\documentclass[preprint,showpacs,preprintnumbers,amsmath,amssymb,superscriptaddress]{revtex4-2}
\usepackage{graphicx}
\usepackage{amsmath}
\usepackage{amssymb}
\usepackage{epsfig}
\usepackage{amsthm}
\usepackage{color}
\usepackage{bm}
\def\be{\begin{equation}}
\def\ee{\end{equation}}
\def\arr{\begin{array}{rll}}
\def\ea{\end{array}}
\def\bea{\begin{eqnarray}}
\def\eea{\end{eqnarray}}

\begin{document}
\title{Theoretical and Numerical Study of Self-Organizing Processes  In a Closed System ``Classical Oscillator + Random Environment"}

\author{Ashot S. Gevorkyan}
\affiliation{Institute for Informatics and Automation Problems NAS of RA,
1, P. Sevak str., Yerevan, 0014, Republic of Armenia}
\affiliation{Institute of Chemical Physics,  NAS of RA, 5/2, P. Sevak str., Yerevan,
 0014, Republic of Armenia}
\author{Aleksander V. Bogdanov}
\affiliation{St. Petersburg State University, 7/9 Universitetskaya nab., St. Petersburg, 199034 Russia}
\affiliation{St. Petersburg State Marine Technical University, Lotsmanskaya D. 3, St. Petersburg,
190121 Russia}
\author{Vladimir V. Mareev}
\affiliation{St. Petersburg State University, 7/9 Universitetskaya nab., St. Petersburg, 199034 Russia}
 \author{Koryun A. Movsesyan}
\affiliation{Institute for Informatics and Automation Problems, NAS of RA,
1, P. Sevak str., Yerevan, 0014, Republic of Armenia}
\date{\today}
\begin{abstract}
A self-organizing joint system ``classical oscillator + random environment" is considered within the
framework of a complex probabilistic process that satisfies a Langevin-type stochastic differential
equation. Various types of randomness generated by the environment are considered. In the limit
of \emph{statistical equilibrium} (SEq), second-order \emph{partial differential equations} (PDE) are
derived that describe the distribution of classical environmental fields. The mathematical expectation
of the oscillator trajectory is constructed in the form of a functional-integral representation, which,
in the SEq limit, is compactified into a two-dimensional integral representation with an integrand - the
solution of the second-order complex PDE. It is proved that the complex PDE in the general case is reduced
to two independent PDEs of the second-order with spatially deviating arguments. The geometric and
topological features of the two-dimensional subspace on which these equations arise are studied in detail.
An algorithm for parallel modeling of the problem has been developed.
\end{abstract}

\maketitle
\section{I\lowercase{ntroduction}}
\label{01}
From ancient sources we know that complex numerical and geometric constructions, as well as complex natural phenomena,
were considered by Pythagoras (l. c. 571 - c. 497 BC) and the students of his school. Nonetheless, the metaphorization
of the word ``system" seems to have been first proposed by Democritus (460-360 BC), which meant the formation
of complex bodies from atoms, similar to the formation of words from syllables and syllables from letters. In
addition, in ancient Greek philosophy, the "system" characterized the orderliness and integrity of natural objects.
Some time later, Plato (427-347 BC) formulated the thesis that \emph{the whole is greater than the sum of its parts}.
Aristotle (384-322 BC), being in a polemic with Plato, formulated the opposite thesis, saying that: \emph{the whole
can be decomposed and studied separately, and then put back together again without losing anything}. It should be
noted that for almost 2500 years, research on the problems of natural science was carried out within the framework
of Aristotle's  concept. This approach, despite its idealized nature, for a long time remained the main method of
cognition in science, stimulated its development and contributed to the creation of a huge number of new
technologies. Nevertheless, the triumph of Aristotle's concept at the beginning of the 20\emph{th}  century was the
creation of a logically perfect theory - classical mechanics of closed systems, which after a short time
experienced a deep crisis, which led to the creation of a more general physical representation of quantum
mechanics.

In the 20\emph{th} century, after Einstein-Smoluchowski created the theory of the Brownian motion of particles
\cite{Ein,Smol}, a new concept of studying nature emerged - the theory of classical and quantum open  systems
\cite{Ingarden}. Recall that an open physical system, interacting with the environment, unlike an isolated system,
exchanges mass, energy, information, etc. with it. At present, the science of open systems is intensively developing
both theoretically and experimentally, especially in the fields of modern quantum physics, chemical physics,
etc. \cite{Zub,Accardi,Taras}. Note that the open systems approach is not equivalent to Plato's concept, but is closer to
it in spirit. Nevertheless, as shown by numerous studies, this approach also has serious difficulties that
do not allow one to reliably describe a number of important phenomena in nonequilibrium thermodynamics,
many-particle quantum systems, etc. Recall that the main drawback of all representations of open systems
is that when describing various physical processes, a certain part of information is inevitably lost, especially
when it comes to physical systems under extreme conditions. This is due to ignoring the influence of the
system on the medium, which excludes the possibility of  formation of a \emph{small environment}
(SE) self-consistent with it, which, most likely, can be considered as its integral part or continuation \cite{Gev}.

The purpose of this study is to combine two opposite concepts of nature cognition, namely - in order
to avoid loss of information, the description of an open system as closed. Just such a statement of the
problem would be completely equivalent to Plato's concept!\\
In this paper, we will demonstrate the implementation of this idea on the example of the problem of
a classical oscillator immersed in a random environment and under the action of an external force.
Recall that this problem has been studied in sufficient detail within the framework of various models
of Brownian motion \cite{Hida}, and its results are widely used in solving a number of important applied
problems (see \cite{Lenzi}). However, in recent decades, a wide range of   phenomena have been discovered,
the descriptions of which cannot be carried out within the framework of the standard representations
of the theory of Brownian motion. In particular, a Brownian particle moving in a viscous
medium exchanges energy and momentum with the environment, which affects the motion of the
particle itself. In other words, such mutual influence leads to the appearance of "memory" in the
Brownian particle, i.e., its behavior becomes dependent on the entire previous history of the process.
However, we know that a Markov model, by definition, cannot describe random processes with
memory, and therefore taking into account the entrainment of particles of the medium
imparts a non-Markov character to the Brownian motion.  To overcome the difficulties of
describing Brownian motion as a non-Markovian process, it is necessary to reformulate the
standard formulation of the problem, introducing significant changes in the mathematical apparatus.
In connection with this, a number of authors have proposed the so-called generalized Langevin
equation, in which instead of a resistance force proportional to velocity, an integral operator of
the convolution type is used \cite{Moroz}. Note that, despite the adaptation of the theory of Brownian
motion to emerging new scientific and technical problems and its development, this theory
describes an exclusively open system with the above limitations and is unsuitable for describing
the properties of systems far from equilibrium or under critical conditions.

Thus, the solution to the problem lies in the development of a fundamentally new mathematical
representation that allows studying the process of self-organization of a  whole system consisting
of finite and infinite subsystems, as a closed  ``classical oscillator + random environment"
(CORE) system.\\
The article is organized as follows:

In section 2, we present statement of the problem and derive complex stochastic differential
equations describing the movement of fields of a random environment for three different cases:
 \begin{itemize}
\item The oscillator frequency is random and the external field is zero,
\item the oscillator frequency is random and the external field is a regular function,
\item the frequency of the oscillator is a regular function and the external force is random.
 \end{itemize}
Note that for all three cases, the corresponding kinetic equations for the distribution of environmental
fields were obtained using a system of stochastic differential equations.

In Section 3, we describe in detail the method for constructing a function space measure; in addition,
a formula for the mathematical expectation of the oscillator trajectory is derived in the form
of a double integral representation from a complex second-order partial differential equation
for the three cases indicated above.

In Section 4, we study in detail the two-dimensional compactified subspace on which complex
second-order partial differential equations are defined. In particular, an algebraic
equation of the 6th order with coefficients depending on two variables is obtained, in the
study of which the possibility of the appearance of topological singularities in a two-dimensional
subspace is proved.

In Section 5, we represent a complex second-order partial differential equation as a
system of two real equations and formulate the Neumann initial-boundary value problem
for this system. We analyze the system of PDEs for the case of symmetry or asymmetry of the
solutions sought and prove that that the system of equations is reduced to two independent PDEs.

  In the case when the solutions do not have a certain symmetry, the PDE
system is again reduced to two independent PDEs, but with a deviation of the arguments.

In Section 6, we present the time-dependent Shannon entropy for a classical oscillator
without taking into account the influence of the particle on the environment. In the same section,
the generalized Shannon entropy for a closed self-organizing “oscillator + random environment”
system is defined.

In Section 7,  we develop numerical algorithms and implement parallel simulation of PDE
with deviating argument, and also present and analyze the results of various numerical experiments.

In Section 8, we discuss the results obtained in this paper and outline directions for future research.


\section{P\lowercase{roblem}}
\label{sec-3}
\subsection{\emph{Statement of the problem}}
The classical action of an one-dimensional oscillator immersed in a random environment can be represented
as  (see \cite{Phy}):
\begin{equation}
\mathbb{S}[x,t_i,t_f)=\int_{t_i}^{t_f}\mathbb{L}\bigl(\dot{x}(t),x(t),t\bigr)dt,
\label{q1.01}
\end{equation}
where $t_i$ and $t_f$ are the moments of time when the interaction of the oscillator with the environment
turns on and off, respectively, in addition,  $\mathbb{L}\bigl(\dot{x}(t),x(t),t\bigr)$ is the Lagrangian
describing the oscillator with a random environment:
\begin{equation}
\mathbb{L}\bigl(\dot{x}(t),x(t),t\bigr)=\frac{1}{2}\dot{x}^2-\frac{1}{2}\Omega^2\bigl(t;\{{\bf f}\}\bigr)x^2
+F\bigl(t;\{{\bf g}\}\bigr)x.
\label{q1.02}
\end{equation}
Recall that $\{{\bf f}\}$ and $\{{\bf g}\}$ are complex probabilistic processes (stochastic sources or
generators), whose properties will be refined below. Obviously, the presence of stochastic generators
in the Lagrangian makes it and, accordingly, the  action stochastic. Despite the fact that the action is
stochastic, it is nevertheless possible to require the fulfillment of the minimum condition:
\begin{equation}
 \delta\mathbb{S}[x,t_i,t_f)=\delta \int_{t_i}^{t_f}\mathbb{L}\bigl(\dot{x}(t),x(t),t\bigr)dt=0.
\label{q1.03}
\end{equation}
Performing the standard procedure for varying the expression (\ref{q1.03}), taking into account the
conditions $\delta x(t_i)=0$ and $\delta x(t_f)=0$ (see \cite{Lan}), we obtain the differential equation
of the second order, which describes the motion of a test particle, i.e. classical oscillator in a random
environment (thermostat):
\begin{equation}
\ddot x +\Omega^2\bigl(t;\{{\bf f}\}\bigr)x=F\bigl(t;\{{\bf g}\}\bigr), \qquad x,t\in (-\infty,+\infty),
\label{1.13}
\end{equation}
where $\dot x=dx/dt$, in addition.

It is important to note that the randomness of the action $\mathbb{S}[x,t_i,t_f)$ and, accordingly, the
Lagrangian $\mathbb{L}\bigl(\dot{x}(t),x(t),t\bigr)$ does not affect the variation procedure, as a result
of which a random equation (\ref{1.13}) is found. However, the equation (\ref{1.13}) in this form is still
not defined, since the stochastic equation must be of the first order.

For definiteness, we will assume that random generators satisfy the \emph{white noise} correlation relations:
\begin{eqnarray}
\mathbb{E}\bigl[f^{(\upsilon)}(t)\bigr] =0,\qquad
\mathbb{E}\bigl[f^{(\upsilon)}(t)f^{(\upsilon)}(t')\bigr]
= 2\epsilon_f^{(\upsilon)}\,\delta (t-t'),\qquad\qquad\qquad\,
\nonumber\\
\mathbb{E}\bigl[g^{(\upsilon)}(t)\bigr] =0,\qquad
\mathbb{E}\bigl[g^{(\upsilon)}(t)g^{(\upsilon)}(t')\bigr]
= 2\epsilon^{(\upsilon)}_g\,\delta (t-t'),\qquad \upsilon=(i,r).
\label{1.07}
\end{eqnarray}
Note that in the expressions (\ref{1.07}), the symbol $\mathbb{E}\bigl[...\bigr]$ denotes the mathematical
expectation of a random variable, in addition, it is assumed that $F\bigl(t;\{{\bf g}\}\bigr)\to 0$ when
$t\to\pm\infty$. Also note that the random forces $f^{(r)}$ and $g^{(r)}$ characterize elastic collisions,
whereas the random forces $f^{(i)}$ and $g^{(i)}$ are responsible for inelastic collisions.

We will consider two different cases:
\begin{enumerate}
\item  When randomness in a joint CORE system generates a complex process $\{{\bf f}\}\neq0$, and the second
source of the random process is absent $\{{\bf g}\}\equiv0$, and, accordingly,
\item  when $\{{\bf f}\}\equiv0$ and randomness in a joint system generates the  generator $\{{\bf g}\}\neq0$,
which has a complex character.
\end{enumerate}

In the case when  the external force  is an arbitrary regular function of time, i.e.  $F_0(t)= F(t;\{\bf g\})
\bigl|_{\{\bf g \}\equiv\,{0}}$, the solution of the equation (\ref{1.13}) can be formally represented as
(see \cite{Baz}):
\begin{equation}
 x(t)=\frac{1}{\sqrt{2\Omega_0^-}}\bigl[\xi(t) d^\ast(t)+\xi^\ast(t)d(t)\bigr],\qquad
 d(t)=\frac{i}{\sqrt{2\Omega_0^-}}\int^t_{-\infty}\xi(t')F_0(t')dt',
 \label{1.n0}
\end{equation}
where the symbol $``^\ast"$ denotes the complex conjugation of a function, $\xi(t)$ is the solution of
the homogeneous equation (\ref{1.13}), i.e. when $F\bigl(t;\{{\bf g}\}\bigr)\equiv0$, in addition, the
following notation are made; $\Omega_0^ -=\lim_{\,t \to-\infty} \Omega\bigl(t,{\bf
\{f\}}\bigr)=const_-$ and $\Omega_0^ +=\lim_{\,t \to+\infty} \Omega\bigl(t,{\bf \{f\}}\bigr)=const_+$.

Note that in the general case, at $t\to-\infty\,\,(in)$ and $t\to+\infty\,\,(out)$ the asymptotic
states can be different and, accordingly, $const_- \neq const_+$. Below, for definiteness, we will
use the regular frequency model $\Omega_0(t)$, which has the following form:
\begin{equation}
\Omega_0(t)=2+\frac{1}{\gamma}\bigl[1+\tanh(\nu t)\bigr],
 \label{q1.n0}
\end{equation}
where $\gamma,\nu>0$ are some constants.

\subsection{\emph{Derivation of environmental field distribution equations}}

{\textbf{Theorem 1.} \emph{If we assume that the equation (\ref{1.13}) for the case
$F(t;\{{\bf g}\})\equiv0$ reduces to a complex Langevin SDE, and the random force
$\{\bf f\}$ is a Gauss-Markovian  process (\ref{1.07}), then the conditional probability
distribution of the environmental fields in the limit of statistical equilibrium will obey
the Fokker-Planck type equation.}}

{\textbf{Proof.}}\\
The solution of the classical oscillator equation (\ref{1.13}) can be represented in the form:
\begin{equation}
\xi(t)= \Biggl\{
\begin{array}{ll}
\,\qquad\xi_{0}(t),\qquad\qquad\qquad\quad\,\, t\leq t_0,
\vspace{0.2 cm}\\
\xi_{0}(t_0)\exp{\bigl\{\int^t_{t_0}\phi(t')\,dt'\bigr\}},\,\,\,\quad t>t_0,
\end{array}
\label{3.01}
\end{equation}
where $\xi_{0}(t)$ is the solution of the classical oscillatory equation (\ref{1.13}) in the
case when the frequency is a regular function of time $\Omega_{0}(t)=\Omega(t;\{\bf f\})
\bigl|_{\{\bf f\}{\equiv\,0}}$ and the external regular force is identically equal to zero
$F_0(t)\equiv0$, in addition, $t_0=0$ denotes the time  of switching on a random
environment. As for the function $\phi(t)$, it denotes a complex probabilistic process.

Substituting (\ref{3.01}) into (\ref{1.13}), taking into account the regular equation:
\begin{equation}
\ddot{\xi}_0+\Omega_0^2(t)\xi_0=0,
\label{3.0z3}
\end{equation}
we obtain the following non-linear \emph{stochastic differential equation} (SDE) of Langevin-type:
\begin{equation}
\label{3.02}
\dot{\phi}+\phi^2+\Omega^2_0(t) +f(t)=0, \qquad \dot{\phi}=d\phi/dt,
\end{equation}
where $\Omega^2(t;\{{\bf f}\})=\Omega_0^2(t)+f(t)$.

For further study, it is convenient to represent the complex probabilistic process $\phi(t)$ as a sum
of fields describing the environment:
\begin{equation}
\label{3.03} \phi(t)=u_{1}(t)+iu_{2}(t).
\end{equation}
Using equation (\ref{3.02}) and representation (\ref{3.03}), we can write the following system of
non-linear SDEs \cite{Gev}:
\begin{equation}
 \Biggl\{
\begin{array}{ll}
 \dot{u}_{1}=[u_{2}]^2-[u_{1}]^2-\Omega_0^2(t)-f^{(r)}(t),\\
\dot{ u}_{2}=-2u_{1}u_{2}-f^{(i)}(t),
\end{array}
\label{3.04}
\end{equation}
where $f(t)=f^{(r)}(t)+if^{(i)}(t)$.\\
Note that the environment fields satisfy the following initial conditions:
$$
\dot{u}_1(t_0)=Re\bigl\{\dot{\xi}_0(t_0)/\xi_0(t_0)\bigr\}=0,\qquad
\dot{u}_2(t_0)=Im\bigl\{\dot{\xi}_0(t_0)/\xi_0(t_0)\bigr\}=\Omega_-.
$$

Let us consider the following functional describing the conditional probability
distribution of fields:
\begin{equation}
P(\textbf{u},t|\textbf{u}',t')=\bigl\langle \delta[\textbf{u}(t)-\textbf{u}(t')]\bigr\rangle,
\qquad \textbf{u}=(u_1,u_2).
\label{3.kn05}
\end{equation}
Differentiating the expression (\ref{3.kn05}) with respect to the time $``t"$, taking into
account the equation (\ref{3.02}), we get:
\begin{eqnarray}
 \partial_t P(\textbf{u},t|\textbf{u}',t')=-\partial_{\textbf{u}}\bigl\langle\textbf{u}_t
 \delta[\textbf{u}(t)-\textbf{u}(t')]\bigr\rangle=\qquad\nonumber\\
\partial_{\textbf{u}}\bigl\{\textbf{K}(\textbf{u},t)P(\textbf{u},t|\textbf{u}',t')
+\bigl\langle{\{\bm f\}}\delta[\textbf{u}(t)-\textbf{u}(t')]\bigr\rangle\bigr\},
\label{3.n01}
\end{eqnarray}
where $\textbf{u}_t=\partial_t\textbf{u}$ and $\textbf{u}'\equiv\textbf{u}(t')$, in addition:
\begin{equation}
 \textbf{K}(\textbf{u},t)=\Biggl\{
\begin{array}{ll}
k_1(u_1,u_2,t)=\bigl[u_{1}\bigr]^2-\bigl[u_{2}\bigr]^2+\Omega_0^2(t),
\nonumber\\
 k_2(u_1,u_2,t)=2u_{1}u_{2}.
\end{array}
\label{3.zn01}
\end{equation}
Using the fact that the vector probabilistic process $\{{\bf f}\}$  satisfy the correlation relations (\ref{1.07}),
we can calculate the second term in the expression (\ref{3.n01}). In particular, using Wick's theorem for an
arbitrary functional $N\bigl(\textbf{u},t;\{{\bm f}\})|\textbf{u}',t'\bigr)$ of argument $\{\bm f\}$,
 we can get (see \cite{Kly}):
\begin{eqnarray}
\bigl\langle\{{\bm f}\}N\bigl(\textbf{u},t;\{{\bm f}\})|\textbf{u}',t'\bigr)\bigr\rangle\,=\,
2\biggl\langle\frac{\delta N\bigl(t;\{{\bm f}\}\bigr)}{\delta f^{(i)}(t)}\biggr\rangle\,+\,
2\biggl\langle\frac{\delta N\bigl(t;\{{\bm f}\}\bigr)}{\delta f^{(r)}(t)}\biggr\rangle\,\,
\nonumber\\
=2\partial_{u_1}\biggl\langle\frac{\delta u_1(t)}{\delta f^{(r)}(t)}\delta[\textbf{u}(t)
-\textbf{u}(t')]\biggr\rangle+2\partial_{u_2}\biggl\langle\frac{\delta u_2(t)}
{\delta f^{(i)}(t)}\delta[\textbf{u}(t)-\textbf{u}(t')]\biggr\rangle.
\label{3.n041}
\end{eqnarray}
Since $u_1(t) $ and $u_2(t)$ are stochastic functions, the corresponding variational
derivatives from $f^{(r)}(t)$ and $f^{(i)}(t)$ are equal:
\begin{equation}
\biggl\langle\frac{\delta u_1(t)}{\delta f^{(r)}(t)}\biggr\rangle=
\epsilon_f^{(r)}sgn(t-t')+O(t-t'),
\quad \biggl\langle\frac{\delta u_2(t)}{\delta f^{(i)}(t)}\biggr\rangle
=\epsilon_f^{(i)}sgn(t-t')+O(t-t').
\label{3.nt07}
\end{equation}
After carrying out the regularization procedure in the sense of the Fourier expansion, we find its value at the time
$t=t'$: $\epsilon_f^{(\upsilon)}sgn(0)=\frac{1}{2}\epsilon_f^{(\upsilon)}.$ Taking into account the equalities
(\ref{3.nt07}) for the conditional probability, the following Fokker-Planck equation can be found:
\begin{equation}
\partial_t P={\hat{L}}({\bf u},t)P,\qquad \partial_t\equiv\partial/\partial t,
\quad {\bf u}={\bf u}(u_1,u_2).
\label{3.05}
\end{equation}
In the equation (\ref{3.05}), the operator ${\hat L}$ has the form:
\begin{eqnarray}
{\hat  L}= \epsilon_f^{(r)}\frac{\partial^{\,2}}{\partial u_1^2}
+\epsilon_f^{(i)}\frac{\partial^{\,2}}{\partial u_2^2}\,+\,
k_1(u_1,u_2,t)\frac{\partial}{\partial u_1}+\,k_2(u_1,u_2,t)\frac{\partial}{\partial u_2}
+k_0(u_1,u_2,t),
\label{3.n05}
\end{eqnarray}
where $k_0=4u_1$, in addition, in the equation (\ref{3.05})-(\ref{3.n05}),
the variables $u_1$ and $u_2$ denote the coordinates of the environment fields'
distribution in the quasi-equilibrium state.

In the case when $t'=t_0$, the conditional probability $P(\textbf{u},t)\equiv P(\textbf{u},t|\textbf{u}_0,t_0)$ describes the distribution of
environmental classical fields  without taking into account the influence of the oscillator. An important issue for the exact formulation of the
problem remains the definition of the type of two-dimensional space on which the equation  for the distribution (\ref{3.05})-(\ref{3.n05})
is given. Recall that the latter implies writing the conditional probability equation (\ref{3.05})-(\ref{3.n05})  in tensor form and studying its
topological and geometric properties in detail. \textbf{Theorem is proved.} $\square$

Recall that for simplicity below we assume that the Fokker-Planck equation (\ref{3.05})-(\ref{3.n05}) is defined on a two-dimensional
Euclidean space and solve this equation as Neumann's initial-boundary value problem  \cite{Gev}.  The numerical study of the free fields
of the environment $P(u_1,u_2,t)$ is carried out using the mathematical algorithm-difference equation (\ref{6.nt03}) developed in \textbf{Listing 1}
 (Section \textbf{VII}). To illustrate the calculations, graphs of the distribution of fields for various media depending on time are plotted
(see FIG 1-3 of  the subsection {\textbf A}).

Now consider the case when the frequency of the oscillator is regular $\Omega_0(t)$, while the external force, on the contrary, is random and
can be represented as the sum:
\begin{equation}
F\bigl(t;\{{\bf g}\}\bigr)=F^{(r)}\bigl(t;\{{\bf g}\}\bigr)+iF^{(i)}\bigl(t;\{{\bf g}\}\bigr)\neq0,
\label{z3.nt07}
\end{equation}
where
$
F^{(r)}\bigl(t;g^{(\upsilon)}(t)\bigr)=F_0(t)+\sqrt{\epsilon_g^{(r)}}{\bar{g}(t)}
$ and $ F^{(i)}\bigl(t;g^{(\upsilon)}(t)\bigr)=
\sqrt{\epsilon_g^{(i)}}{\bar{g}(t)},
$ in addition, $\bar{g}(t)$ is a real Gauss-Markov random process, which will be clearly defined below. In particular, using the
definition (\ref{z3.nt07}) and given that the frequency is regular, the equation (\ref{1.13}) can be written as follows:
\begin{equation}
\ddot x +\Omega^2_0(t)x=F_0(t)+\bigl[\sqrt{\varepsilon^{(r)}}\, +
i\sqrt{\varepsilon^{(i)}}\,\bigr]\bar{g}(t).
\label{n01.13}
\end{equation}

{\textbf{Theorem 2.} \emph{If the oscillator trajectory obeys the equation (\ref{n01.13}), and the random function $\bar{g}(t)$ satisfies
the Gauss-Markov process (\ref{1.07}), then the distribution of fields of the environment in the limit of statistical equilibrium will be
described by PDE of the second order, which, in the $(out)$ asymptotic state or in the limit $t\to+\infty$, transforms into the PDE
of the Fokker-Planck type.}}

{\textbf{Proof.}}\\
The solution to the equation (\ref{n01.13}) can be represented as:
\begin{equation}
x_1(t)= \Biggl\{
\begin{array}{ll}
\,\qquad x_{0}(t),\qquad\qquad\qquad\quad t\leq t_0,
\vspace{0.2 cm}\\
x_{0}(t_0)\exp{\bigl\{\int^t_{t_0}\theta(t')\,dt'\bigr\}},\,\,\quad t>t_0,
\end{array}
\label{4w.n05}
\end{equation}
where $x_0(t)$ is the solution of the regular equation for a classical oscillator with a
time-dependent frequency and under the influence of an external non-stationary force:
\begin{equation}
 \ddot{x}_0+\Omega_0^2(t)x_0=F_0(t).
\label{n07.13}
\end{equation}
Substituting (\ref{4w.n05}) into equation (\ref{n01.13}), setting $\theta(t)=w_1(t)+iw_2(t)$, we obtain
 the following system of stochastic integro-differential equations:
\begin{equation}
 \Biggl\{
\begin{array}{ll}
\dot{w}_{1}=[w_2]^2-[w_1]^2 -\Omega_0^2(t)+F_0(t)e^{-\sigma_1(t)}\cos\sigma_2(t)+\bar{g}(t)\mathcal{A}^+(t),\\
\dot{w}_{2}=-2w_1w_2-  F_0(t)e^{-\sigma_1(t)}\sin\sigma_2(t)+\bar{g}(t)\mathcal{A}^-(t),
\end{array}
\label{4w.n03}
\end{equation}
where
$$
 \mathcal{A}^+(t)=\Bigl[\sqrt{\epsilon_g^{(r)}}\cos\sigma_2(t)+
 \sqrt{\epsilon_g^{(i)}}\sin\sigma_2(t)\Bigr]e^{-\sigma_1(t)},
$$
$$
\mathcal{A}^-(t)=\Bigl[\sqrt{\epsilon_g^{(i)}}\cos\sigma_2(t) -
 \sqrt{\epsilon_g^{(r)}} \sin\sigma_2(t)\Bigr]e^{-\sigma_1(t)}.
$$
As for the functions $\sigma_1(t)$ and $\sigma_2(t)$, they are singly differentiable, i.e.
belong to class $L_1$ and are represented as:
$$
\sigma_1(t)=\int^t_{t_0}w_1(t')dt'+Re[\ln x_0(t_0)],
\qquad
\sigma_2(t)=\int^t_{t_0}w_2(t')dt'+Im[\ln x_0(t_0)].
$$
Assuming that the random function $\bar{g}(t)$ satisfies the white noise correlation relations:
$$
\langle\bar{g}(t)\rangle=0,\qquad \langle\bar{g}(t)\bar{g}(t')\rangle=2\delta(t-t'),
$$
we can use a system of stochastic differential equations  (\ref{4w.n03}) and
obtain the following equation for the conditional probability of the environmental fields:
\begin{eqnarray}
\partial_t\mathcal{P}=\mathcal{\hat L}\bigl({\bf w},t\bigr)\mathcal{P},\qquad
{\bf w}={\bf w}(w_1,w_2),
\label{3v.n05}
\end{eqnarray}
where the evolution operator $\mathcal{\hat L}$ has the form:
\begin{eqnarray}
\mathcal{\hat L}({\bf w},t)= \mathcal{A}^-(t)\frac{\partial^{\,2}}{\partial w_1^2}
+\mathcal{A}^+(t)\frac{\partial^{\,2}}{\partial w_2^2}\,+\,
h_1(w_1,w_2,t)\frac{\partial}{\partial w_1}+\,h_2(w_1,w_2,t)\frac{\partial}{\partial w_2}
+h_0(w_1,w_2,t),
\nonumber\\
\label{3vt.n05}
\end{eqnarray}
in addition:
$$
h_1(w_1,w_2,t)=[w_1]^2-[w_2]^2+\Omega^2_{0}(t)-F_0(t)e^{-\sigma_1(t)}\cos\sigma_2(t),
$$
$$
h_2(w_1,w_2,t)=2w_1w_2+F_0(t)e^{-\sigma_1(t)}\sin\sigma_2(t),\quad h_0(w_1,w_2,t)=4w_1.
$$
It is obvious that if the environment is turned on in the time range $[t_i,t_f]$, then after $t\geq t_f$
the equation (\ref{3v.n05}) turns into the equation of the Fokker-Planck type. \textbf{Theorem is proved}. $\square$

\section{T\lowercase{he mathematical expectation of the trajectory}}

\subsection{The measure of the functional space}
For further analytical constructions of the theory, it is necessary to determine
the distance between functions in the function space $\mathbb{R}_{\{\bm\xi\}}$
or, more precisely, the measure of the function space  (see \cite{Agev,GevA}). Let the conditional probability $P(\textbf{u},t|\textbf{u}',t')$
satisfy the following limiting condition:
\begin{eqnarray}
\lim_{t\to\, t'}P(\textbf{u},t|\textbf{u}',t')=\delta(\textbf{u}-\textbf{u}'),
\qquad t=t'+\Delta t.
\label{t3.05}
\end{eqnarray}
The latter means that for small time intervals, i.e. for $\Delta t=t-t'<<1$, the solution of the equation
(\ref{3.05}) -(\ref{3.n05}) can be represented as:
\begin{eqnarray}
\label{3.b06}
P(\textbf{u},t|\textbf{u}',t')=\frac{1}{2\pi\sqrt{|\det{\bm\epsilon}|}\Delta t}
\exp\Biggl\{-\frac{\bigl[\textbf{u}-\textbf{u}'-\textbf{K}(\textbf{u},t)
\Delta t\bigr]^T{\bm\epsilon}^{-1}\bigl[\textbf{u}-\textbf{u}'-\textbf{K}(\textbf{u},t)
\Delta t\bigr]}{2\Delta t}\Biggr\},
\nonumber\\
\end{eqnarray}
where ${\bm\epsilon}$ denotes a matrix of the second rank with elements; $\epsilon_{11}=\epsilon^{(r)},\,\,\epsilon_{22}
=\epsilon^{(i)}$ and $\epsilon_{12}=\epsilon_{21}=0$, in addition, $[\cdot\cdot\cdot]^T$ is a vector transposition.                                                                                                                                                                                                                                                                                                                                                                                                                                                                                                                                                                                                                                                                                                                                                                                                                                                  The distribution (\ref{3.b06}) can be written explicitly:                                                                                                                                                                                                                                                                                                                                                                                                                                                                                                                                                                                                                                                                                                                                                                                                                                                                                                                                                                                                                                                                                                                                                                                                                                                                                                                                                                                                                                                                                                                                                                                                                                                                                                                                                                                                                                                                                                                                                                                                                                                                                                                                                                                                                                                                                                                                                                                                                                                                                                                                                                                                                                                                                                                                                                                                                                                                                                                                                                                                                                                                                                                                                                                                                                                                                                                                                                                                                                                                                                                                                                                                                                                                                                                                                                                                                                                                                                                                                                                                                                                                                                                                                                                \begin{eqnarray}
\label{4.w06i}
P(u_1,u_2,t|u'_1,u_2',t')=\frac{1}{2\pi\sqrt{\epsilon^{(r)}\epsilon^{(i)}}\Delta t}\,\times
\qquad\qquad\qquad\qquad
\nonumber\\
\exp\Biggl\{-\frac{\bigl[u_1-u_1'- k_1(u_1,u_2,t)\Delta t\bigr]^2}
{2\epsilon^{(r)}\Delta t} -\frac{\bigl[u_2-u_2'-k_2(u_1,u_2,t)\Delta t\bigr]^2}
{2\epsilon^{(i)}\Delta t}\Biggr\},
\end{eqnarray}
where the coefficients $k_1$ and $k_2$ are defined in (\ref{3.zn01}).

In the case when there is no dissipation in the environment, i.e. $\epsilon^{(i)}=0$,
the distribution (\ref{4.w06i}) takes the following form:
\begin{eqnarray}
\label{4.w060i}
P(u_1,u_2,t|u'_1,u_2',t')=\frac{1}{2\pi\sqrt{\epsilon^{(r)}\Delta t}}\,\times
\qquad\qquad\qquad\quad
\nonumber\\
\exp\Biggl\{-\frac{\bigl[u_1-u_1'-k_1(u_1,u_2,t)\Delta t\bigr]^2}
{2\epsilon^{(r)}\Delta t}\Biggr\}\,\delta\bigl [u_2-u_2'+k_2(u_1,u_2,t)\bigr].
\end{eqnarray}

Thus, as can be seen from the expression (\ref{4.w06i}), the evolution of
the system in the functional space $R_{\{\bm\xi\}}$ is characterized by
a regular shift with  a speed $\textbf{K}(\textbf{u},t)$ against the background
of Gaussian fluctuations with the diffusion matrix $\epsilon_{ij}$.
As for the trajectory $\textbf{u}(t)$ in the space $R_{\{\bm\xi\}}$, it is
defined by the following formula (see for example \cite{Gard}):
\begin{equation}
\textbf{u}(t)= \Biggl\{
\begin{array}{ll}
u_1(t+\Delta t)=u_1(t)+ k_1(u_1,u_2,t)\Delta t+f^{(r)}(t){\Delta t}^{1/2},
\vspace{0.1 cm}\\
u_2(t+\Delta t)=u_2(t)+k_2(u_1,u_2,t)\Delta t+f^{(i)}(t){\Delta t}^{1/2}.
\end{array}
\label{4.w760i}
\end{equation}
As can be seen from formula (\ref{4.w760i}), the trajectory is continuous everywhere,
i.e. $\textbf{u}(t+\Delta t)\bigr|_{\Delta t\to\,0}=\textbf{u}(t)$ but nevertheless is
non-differentiable everywhere due to the presence of the $\sim{\Delta t}^{1/2}$ term. If the time interval is represented as
$\Delta t=t/N$, where $N\to\infty$, then expression (\ref{4.w06i}) can be interpreted as the probability of transition from the
vector $\textbf{u}_{\,l}(t)$ to the vector $\textbf{u}_{\,l+1}(t)$ during $\Delta t$ in the Brownian motion model.

Now we can define the Fokker-Planck measure of the functional space:
\begin{eqnarray}
\label{5.w060i}
D\mu(\textbf{u})=d\mu(\textbf{u}_0)\lim_{N\to\,\infty}\Biggl\{
\biggl(\frac{1}{2\pi} \frac{N/t}{\sqrt{\epsilon^{(r)}\epsilon^{(i)}}}\,\biggr)^N\prod_{l=\,0}^N
du_{1(l+1)}du_{2(l+1)}\exp\Biggl[-\frac{N/t}{2\epsilon^{(r)}}\biggl(u_{1(l+1)}-
\nonumber\\
u_{1(l)}-k_{1(l+1)}\frac{t_{l+1}}{N}\biggr)^2-\frac{N/t}{2\epsilon^{(i)}}\biggl(u_{2(l+1)}-
u_{2(l)}-k_{2(l+1)}\frac{t_{l+1}}{N}\biggr)^2\Biggr]\Biggr\},
\end{eqnarray}
where $d\mu(\textbf{u}_0)=\delta(u_{1}-u_{1(0)})\delta(u_{2}-u_{2(0)})du_{1}du_{2}$
denotes the measure of the initial distribution, in addition, the following notations are made:
$$
u_{1(l)}=u_1(t_l),\quad u_{2(l)}=u_2(t_l),\quad k_{1(l)}=k_1(u_{1(l)},u_{2(l)},t_l),
\quad k_{2(l)}=k_2(u_{1(l)},u_{2(l)},t_l).
$$

Note that the measure $D\mu_1({\bf w})$, which describes the probability of a given trajectory
in the functional space $\mathbb{R}_{\{\bm x\}}$, can be constructed in a similar way using the
equation distributions for classical fields of the environment (\ref{3v.n05}).

\subsection{Trajectory calculation}

Now we can rigorously calculate the trajectory of the oscillator for the three cases described above.

\textbf{Definition 1}. \emph{The functional integral along the random trajectory
$\varrho\bigl[u_1(t),u_2(t),t\bigr]$ will be called the mathematical expectation of the trajectory:}
\begin{equation}
\label{i4.0n6i}
\bar{\varrho}(t)= \mathbb{E}[\varrho(t)] =\frac{1}{\alpha(t)}
\int_{\mathbb{R}_{\{\bm\xi\}}}D\mu({\bf u})\varrho\bigl[u_1(t),u_2(t),t\bigr].
\end{equation}
\emph{where}  $\alpha(t)=\int_{\mathbb{R}_{\{\bm\xi\}}}D{\mu(\bf u)}
=\int\int_{\Sigma^{(2)}_{\bf{u}}(t)} P(u_1,u_2,t)$  \emph{is a normalizing constant}.

Let us consider the case when the oscillator is not subjected by a random force, i.e. $F(t;\{{\bf g}\})\equiv0$.
Then the mathematical expectation of the trajectory, taking into account (\ref{3.01}) and (\ref{i4.0n6i}) will
have the following form:
\begin{equation}
\label{4.0n6i}
\bar{\xi}(t)= \mathbb{E}[\xi(t)] =\frac{\xi_0(t_0)}{\alpha(t)}
\int_{\mathbb{R}_{\{\bm\xi\}}}D\mu({\bf u})\exp\biggl\{\int_{t_0}^t\phi(t')dt'\biggr\}.
\end{equation}

Finally, by computing the functional integral (\ref{4.0n6i}) using the generalized Feynman-Kac theorem \cite{GevA},
 one can find the following two-dimensional integral representation for the trajectory expectation:
\begin{equation}
\label{4.0nw6i}
\bar{\xi}(t)=\mathbb{E}[\xi(t)]= \xi_0(t_0)\Lambda_{{Q}}(t),\qquad \Lambda_{{Q}}(t)=\frac{1}{\alpha(t)}
\int\int_{\Sigma^{(2)}_{\bf{u}}(t)}{Q}(u_1,u_2,t)du_1du_2,
\end{equation}
where $\xi_0(t_0)$ is the trajectory of the regular oscillator at time $t_0$ (see equation (\ref{3.0z3})), in addition,
the function ${Q}(u_1,u_2,t)$ is the solution of the following second-order complex PDE:
\begin{equation}
\label{4.0nt6i}
\partial_t Q=\bigl\{\hat{L}({\bf u},t)+u_1+iu_2\bigr\}Q.
\end{equation}
Since $\xi_0(t_0)$ is a constant,  the main role in determining the expectation of the trajectory is played by the function $\Lambda_Q(t)$.

The numerical study of the complex PDE (\ref{4.0nt6i}) is carried out using the developed system of difference equations
(see \textbf{Listing 2}  Section \textbf{VII}). The results of numerical simulation of the function $Q(u_1,u_2,t)$ depending on the
state of the environment and time are shown in FIG 4-6 of subsection  \textbf{B}. The mathematical expectation of the trajectory, depending on
various parameters and time,  was calculated and represented on the graphs (for details see subsection \textbf{C},   FIG 7).

Now let us calculate the mathematical expectation of the trajectory $\bar{x}(t)$ when the oscillator is acted upon by a regular external force $F_0(t)$.

Given the equation (\ref{1.n0}), the trajectory can be formally written as follows:
\begin{equation}
\bar{x}(t)=\mathbb{E}\bigl[x(t)\bigr]=\frac{1}{\sqrt{2\Omega_0}}\bigl[{I}(t)+ {I}^\ast(t)\bigr],
\qquad {I}(t)=\bigl\langle\xi(t)d^\ast(t)\bigr\rangle,
\label{4.0vt6i}
\end{equation}
where the symbol $\bigl\langle\cdot\cdot\cdot\bigr\rangle$ denotes functional integration with respect to the Fokker-Planck measure
 (see expression (\ref{5.w060i})):
\begin{equation}
I(t)=\bigl\langle\xi(t)d^\ast(t)\bigr\rangle=\int_{\mathbb{R}_{\{\bm\xi\}}}D\mu({\bf u})\xi(t)d^\ast(t).
\label{4.0vt7i}
\end{equation}
The functional integral in (\ref{4.0vt7i}) can be calculated and brought to a two-dimensional integral representation, if we use
 the following auxiliary relation:
$$
I(t)=I(m|t)\bigl|_{m=\,0}=\frac{i\xi_0(t_0)}{\sqrt{2\Omega_0}}\partial_\lambda\biggl\langle \int_{t_0}^t
\Bigl[\phi(t')+ \lambda\xi(t')F_0(t')\Bigr]dt'\biggr\rangle\biggl|_{\lambda=\,0}.
$$
The value under the derivative sign can be calculated using the generalized Feynman-Kac theorem:
\begin{equation}
\biggl\langle \int_{t_0}^t \Bigl[\phi(t')+
\lambda\xi(t')F_0(t')\Bigr]dt'\biggr\rangle=\int\int_{\Sigma^{(2)}_{\bf{u}}(t)}Q_\lambda(u_1,u_2,t)du_1du_2,
\label{4.0vt8i}
\end{equation}
where the function $Q_\lambda(u_1,u_2,t)$ is the solution of the following second-order partial differential equation:
\begin{equation}
\partial_t Q_\lambda=\bigl\{\hat{L}+u_1+iu_2+\lambda\xi(t)F_0(t)\bigr\}Q_\lambda.
\label{4.0vt9i}
\end{equation}
Differentiating the equation (\ref{4.0vt9i}) with respect to the parameter $\lambda$ we find a following equation:
$$
\partial_t \mathcal{D}_\lambda= \bigl\{\hat{L}+u_1+iu_2+\lambda\xi(t)F_0(t)\bigr\}\mathcal{D}_\lambda
+\xi(t)F_0(t)Q_\lambda,
$$
where $\mathcal{D}_\lambda(u_1,u_2,t)\equiv\partial_\lambda Q_\lambda(u_1,u_2,t).$

Now, introducing the notation $\mathcal{D}(u_1,u_2,t)=\mathcal{D}_\lambda(u_1,u_2,t)\bigl|_{\lambda=\,0}$,
we obtain the following two-dimensional  integral representation:
$$
I(t)=\frac{i\xi_0(t_0)}{\sqrt{2\Omega_0}}\Lambda_D(t),\qquad
\Lambda_D(t)=\int\int_{\Sigma^{(2)}_{\bf{u}}(t)}\mathcal{D}(u_1,u_2,t)du_1du_2,
$$
where the function $\mathcal{D}(u_1,u_2,t)$ satisfies the following complex second-order PDE:
\begin{equation}
\partial_t \mathcal{D}= \bigl(\hat{L}+u_1+iu_2\bigr)\mathcal{D}+\xi(t)F_0(t)Q,
\qquad Q\equiv Q_\lambda\bigl|_{\lambda=\,0}.
\label{4.0vt10i}
\end{equation}
To solve the equation (\ref{4.0vt9i}), we can require the following initial condition to be met:
\begin{equation}
 \mathcal{D}(u_1,u_2,t)\bigl|_{t=\,t_0}\,=\,0.
\label{4.0vt11i}
\end{equation}
As we can  see, the equation (\ref{4.0vt10i})  includes two parameters, one is the regular $F_0(t)$ (external force), and
the other, respectively, random, the trajectory of the oscillator $\xi(t)\in L_1$, which is generated by stochastic equations (\ref{3.04}).

Integrating the equation (\ref{4.0vt10i}) with respect to the Fokker-Planck measure (\ref{5.w060i}), one can obtain a new
equation:
\begin{equation}
\langle\partial_t{\mathcal{D}}\rangle= \bigl(\hat{L}+u_1+iu_2\bigr)\bar{\mathcal{D}}+\bar{\xi}(t)F_0(t)Q,
\qquad \bar{\mathcal{D}}=\langle{\mathcal{D}}\rangle,
\label{4.0vt12i}
\end{equation}
where $\bar{\xi}(t)$ denotes a regular function representing the mathematical expectation of the oscillator trajectory without
external influence (see (\ref{4.0n6i})). The term $\langle\partial_t{\mathcal{D}}\rangle$ in the equation (\ref{4.0vt12i}) can be
represented as:
\begin{equation}
\langle\partial_t{\mathcal{D}}\rangle=\partial_t{\langle\mathcal{D}\rangle}-
\langle(\partial_t\mathcal{M}(t)){\mathcal{D}}\rangle
=\partial_t{\bar{\mathcal{D}}}-(\partial_t\mathcal{M}(t))\bar{\mathcal{D}},
\label{4.0vt12w}
\end{equation}
where $\mathcal{M}(t)$ denotes the exponent of the Fokker-Planck measure (\ref{5.w060i}) in the limit $N\to\infty$,
 when the sum goes into an integral. It is easy to see that the term $\partial_t\mathcal{M}(t)$ is a random function and,
 accordingly, the new averaging does not allow one to obtain a regular equation for the distribution function of the
environment fields $\bar{\mathcal{D}}$. Nevertheless, since for large times $\partial_t \mathcal{D}\to 0$, then from
the equation (\ref{4.0vt12i}) in the asymptotic $t\to\infty$, one can obtain the following regular stationary equation:
\begin{equation}
\bigl(\hat{L}+u_1+iu_2\bigr)\bar{\mathcal{D}}+\bar{\xi}(t)F_0(t)Q=0,
\label{4.0vtn3i}
\end{equation}
Finally, taking into account (\ref{4.0vtn3i}), we can write the total expectation of the oscillator trajectory in the external
 field in the asymptotic channel $(out)$:
\begin{equation}
 \bar{x}(t)\,=\frac{1}{\sqrt{2\Omega_0}}\bigl\{\bar{I}(t)+\bar{I}^\ast(t)\bigr\},\qquad
\bar{I}(t)=\frac{i\xi_0(t_0)}{\sqrt{2\Omega_0}}\int\int_{\sum^{(2)}_{\bf{u}}(t)}\bar{\mathcal{D}}(u_1,u_2,t)du_1du_2.
\label{4.0vt13i}
\end{equation}
Taking into account (\ref{4.0vtn3i}) for large times, i.e. for $t\to\infty$, the expectation of the trajectory $\bar{x}(t)$ is a regular function.

In the end, we will calculate the mathematical expectation of the trajectory of the oscillator, which is under the action of an
external random force $F(t;\{\bf g\})$. Carrying out similar reasoning, we can write the following functional
integral for the mathematical expectation of the trajectory:
\begin{equation}
\bar{x}_1(t)=\mathbb{E}\bigl[x_1(t)\bigr]=\frac{x_0(t_0)}{\alpha(t)}
\int_{\mathbb{R}_{\{\bm x\}}}D\mu_1({\bf w)}\exp\biggl\{\int_{t_0}^t\theta(t')dt'\biggr\}.
\label{7.0vtn3i}
\end{equation}
Doing a similar calculation in the functional integral (\ref{7.0vtn3i}), we get:
$$
\bar{x}_1(t)= x_0(t_0)\Lambda_{Q_1}(t),\qquad \Lambda_{Q_1}(t)=\frac{1}{\alpha(t)}
\int\int_{\Sigma^{(2)}_{\bf w}(t)}Q_1(w_1,w_2,t)dw_1dw_2,
$$
where $\Sigma^{(2)}_{\bf w(1)}(t)$ is a two-dimensional manifold, the geometric and topological features of which
must be studied specially, in addition, the function $x_0(t_0)$ is the solution of the regular equation (\ref{n07.13}),
 in addition, the function $Q_1(w_1,w_2,t)$ is the solution of the following complex PDE:
\begin{equation}
\partial_t Q_1=\bigl\{\mathcal{\hat L}({\bf w},t)+w_1+iw_2\bigr\}Q_1.
\label{7.0vtw3i}
\end{equation}
As we can see, the equation (\ref{7.0vtw3i}) differs significantly from the parabolic complex PDE (\ref{4.0nt6i}), and it
can turn into an ordinary equation, i.e. complex PDE, in the limit of $t\to \infty$ when statistical equilibrium occurs in the joint system.

\section{G\lowercase{eometric and topological features of a compactified space} }
As we saw in the previous section, in the limit of statistical equilibrium, the functional space $\mathbb{R}_{\{\bm\xi\}}$
compactifies into the two-dimensional manifold. In particular, for a random frequency and no external force, the functional
space $\mathbb{R}_{\{\bm\xi\}}$ compactifies into a two-dimensional manifold $\Sigma^{(2)}_{\bf{u}}(t)$, and for a
random external force and a regular frequency, the functional space, respectively, is compactified into another two-dimensional
 manifold; $\mathbb{R}_{\{\bm\xi\}}\to \Sigma^{(2)}_{\bf{w}}(t)$.

Thus, it is obvious that in this case the joint CORE system in the limit of statistical equilibrium is described in three-dimensional space;
$
\mathbb{R}^{3}_\bullet\cong{\mathbb{R}^1}\otimes\Sigma^{(2)}_{\bf{u}}(t)
$
or
$
\mathbb{R}^{3}_\bullet\cong{\mathbb{R}^1}\otimes\Sigma^{(2)}_{\bf{w}}(t),
$
where $\mathbb{R}^1$ is a one-dimensional Euclidean subspace, and  $\Sigma^{(2)}_{\bf{u}}(t)$ and $\Sigma^{(2)}_{\bf{w}}(t)$
are a two-dimensional manifolds whose topological and geometric features will be studied in detail below.

\subsection{Geometry of two-dimensional subspace $\Sigma^{(2)}_{\bf{u}}(t)$}
For definiteness, below we will study the property of the subspace
$\Sigma^{(2)}_{\bf{u}}(t)$.

\textbf{Definition 2.}  \emph{A generalized Riemannian (or pseudo-Riemannian) space is a smooth manifold \,\,$\Sigma^{(2)}_{\bf{u}}(t)$
with a doubly covariant tensor $g_{\mu\nu}$ defined on it, which we will call the generalized metric tensor}.

\textbf{Theorem 3.} \emph{If the motion of a dynamical system is described by stochastic differential equations of the Langevin
type (\ref{3.04}),  then in the limit of statistical equilibrium these equations generate a two-dimensional space with an antisymmetric
metric (Riemann–Cartan manifold).}

\textbf{Proof.}\\
Let us represent the equation (\ref{3.05})-(\ref{3.n05}) in tensor form (see for example \cite{Jost}):
\begin{equation}
\partial_t P= \nabla^2P+k_0(u^1,u^2,t)P,\qquad \nabla^2=\frac{1}{\sqrt{|g|}}
\sum^2_{i,j=1}\frac{\partial}{\partial u^i}\biggl(\sqrt{|g|}g^{ij}\frac{\partial}{\partial u^j}\biggr),
\label{nw3.01}
\end{equation}
where the following notation are made; $u_1=u^1$ and $u_2=u^2$.

To find the elements of the metric tensor, we write the two-dimensional Laplace-Beltrami operator $\nabla^2$ in explicit form:
\begin{eqnarray}
\nabla^2=g^{11}\frac{\partial^2}{\partial u_1^2} +
\frac{1}{\sqrt{|g|}}\biggl[\frac{\partial}{\partial u_1}\Bigl(\sqrt{|g|}g^{11}\Bigr)+\frac{\partial}{\partial u_2}
\Bigl(\sqrt{|g|}g^{21}\Bigr)\biggr]\frac{\partial}{\partial u_1}
+g^{12}\frac{\partial^2}{\partial u_1\partial u_2}\,\,
\nonumber\\
\quad\quad+\,\,g^{22}\frac{\partial^2}{\partial u_2^2}
+\frac{1}{\sqrt{|g|}}\biggl[\frac{\partial}{\partial u_2}\Bigl(\sqrt{|g|}g^{22}\Bigr)+\frac{\partial}{\partial u_1}
\Bigl(\sqrt{|g|}g^{12}\Bigr)\biggr]\frac{\partial}{\partial u_2}
+g^{21}\frac{\partial^2}{\partial u_2\partial u_1}.
\label{nw3.02}
\end{eqnarray}
Comparing (\ref{nw3.02}) with (\ref{3.n05}) and requiring the corresponding terms in the equations to be equal, we find:
\begin{equation}
g^{11}=\epsilon^{(r)},\qquad g^{22}=\epsilon^{(i)},\qquad g^{12}=-g^{21},
\qquad g=g^{11}g^{22}-g^{12}g^{21}=\epsilon^{(r)}\epsilon^{(i)}+\bigl(g^{12}\bigr)^2.
\label{qnw3.01}
\end{equation}
As can be seen from (\ref{qnw3.01}), the metric tensor of the subspace $\Sigma^{(2)}_{\bf{u}}(t)$ is antisymmetric
and, therefore, the geometry  it describes is non-commutative. Note that such spaces usually arise both in mathematics
and quantum physics and naturally correspond to non-commutative algebras \cite{Connes}.

Before proceeding to the study of various properties of the $\Sigma^{(2)}_{\bf{u}}(t)$ subspace, we perform the following
coordinate scaling transformation:
\begin{equation}
u_1\to\bar{u}_1=u_1/\sqrt{\varepsilon^{(r)}/\lambda},\qquad
u_2\to\bar{u}_2=u_2/\sqrt{\varepsilon^{(i)}/\lambda},
\label{w3.t01}
\end{equation}
where $\lambda>0$ is some constant.\\
In this case, the metric tensor $\bar{g}^{ij}$ in an orthogonal basis can be represented as the following
sum; $\bar{g}^{ij}=\lambda^{ij}+\bar{y}^{ij},\,(\lambda^{ij}=\lambda^{ji},\,\,\bar{y}^{ij}=-\bar{y}^{ji})$
or in the following explicit form:
\begin{equation}
\bar{g}^{ij}=\lambda\Biggl(
\begin{array}{cc}
1& 0 \\
0 & 1\\
\end{array}
  \Biggr)+
\bar{y} \Biggl(
\begin{array}{cc}
    0 & 1 \\
    -1 & 0\\
    \end{array}
 \Biggr),
\label{7w3.t01}
\end{equation}
where
$y(u_1,u_2,t)=g^{12}(u_1,u_2,t)\mapsto\bar{g}^{12}(\bar{u}_1,\bar{u}_2,t)=\bar{y}(\bar{u}_1,\bar{u}_2,t)$.

The first feature of the generalized metric is that the non-symmetric part does not contribute to the definition of the length,
since the sum $\bar{y}^{ij}v_iv_j=0$ and, therefore:
$$
|{\bf v}|=\sqrt{\bar{g}^{ij}v_iv_j}=\sqrt{\lambda^{ij}v_iv_j+\bar{y}^{ij}v_iv_j}=\sqrt{\lambda^{ij}v_iv_j}.
$$
Recall that the tensor $\lambda^{ij}$ defines the Euclidean geometry of the plane tangent to the manifold
$\textbf{v}\in \Sigma^{(2)}_{\bf{u}}(t)$ at a given point. In this "symmetric"  space $\lambda^{ij}=\lambda^{ji}$
angular measure and coordinates of the unit vector $\textbf{v}$ are defined and equal to $\textbf{v}=(\cos{\vartheta},\,\sin{\vartheta})$,
respectively, where $\vartheta$ is the Euclidean angle between the vector $\textbf{v}$ and the first basis vector.

Obviously, the angle between two unit vectors $\textbf{v}_1=(\cos{\vartheta_1},\,\sin{\vartheta_1})$ and
$\textbf{v}_2=(\cos{\vartheta_2},\,\sin{\vartheta_2})$ is equal to their scalar product. Based on this definition
and the antisymmetry of the off-diagonal element of the metric tensor $\bar{g}^{12}(u_1,u_2,t)=-\bar{g}^{21}(u_1,u_2,t)$,
it is easy to obtain two expressions for the cosine of the angle \cite{Bur}:
\begin{equation}
\Biggl\{
\begin{array}{ll}
\cos(\widehat{\textbf{v}_1,\textbf{v}_2})=\frac{\lambda^{ij}{(v_1)}_i{(v_2)}_j\,+\,\bar{y}^{ij}{(v_1)}_i{(v_2)}_j}
{\sqrt{\lambda^{ij}{(v_1)}_i{(v_1)}_j}\sqrt{\lambda^{ij}{(v_2)}_i{(v_2)}_j}},\\
\cos(\widehat{\textbf{v}_2,\textbf{v}_1})=\frac{\lambda^{ij}{(v_1)}_i{(v_2)}_j\,-\,\bar{y}^{ij}{(v_1)}_i{(v_2)}_j}
{\sqrt{\lambda^{ij}{(v_1)}_i{(v_1)}_j}\sqrt{\lambda^{ij}{(v_2)}_i{(v_2)}_j}},
\end{array}
\label{4aw.n0q3}
\end{equation}
where ${(v_1)}_i$ denotes the projection of the $\textbf{v}_1$ vector onto the $u_i$ axis (see (\ref{3.kn05})).

By doing simple calculations we find:
\begin{equation}
 \Biggl\{
\begin{array}{ll}
 \cos\psi^+=\cos(\widehat{\textbf{v}_1,\textbf{v}_2})= \sqrt{1\,+(\bar{y}/\lambda)^2}\cos(\Delta\vartheta+\delta),\\
\cos\psi^-=\cos(\widehat{\textbf{v}_2,\textbf{v}_1})= \sqrt{1\,+(\bar{y}/\lambda)^2} \cos(\Delta\vartheta-\delta).
\end{array}
\label{4w.n0q3}
\end{equation}
where $\Delta\vartheta=\vartheta_2-\vartheta_1$ is the Euclidean angle between the vectors $\textbf{v}_1$ and $\textbf{v}_2$,
 in addition, the angle $\delta$ is determined from the following relations:
$$
\frac{\lambda}{\sqrt{\lambda^2+\bar{y}^2}}=\cos\delta,\qquad \frac{\bar{y}}{\sqrt{\lambda^2+\bar{y}^2}}=\sin\delta.
$$
From (\ref{4w.n0q3}) also follows the important conditions for the Euclidean angles $\vartheta^+=\Delta\vartheta+\delta$ and
$\vartheta^-=\Delta\vartheta-\delta$. In particular, it follows from the definition of Euclidean geometry that the angles must satisfy
the following constraint conditions:
\begin{equation}
\cos\vartheta^+\leq\frac{1}{\sqrt{1\,+(\bar{y}/\lambda)^2}},\qquad \cos\vartheta^-\leq\frac{1}{\sqrt{1\,-(\bar{y}/\lambda)^2}}.
\label{5wl.n0z3}
\end{equation}

Recall that two different values of the angle $\psi^+$ and $\psi^-$ between the vectors $\textbf{v}_1=\textbf{v}_1(u_1,u_2)$ and
$\textbf{v}_2=\textbf{v}_2(u_1,u_2)$ (see (\ref{4w.n0q3})) depending on the direction of rotation - to the right or to the left -
is a characteristic peculiarity of Kozyrev's theory \cite{Koz}.

Taking into account the antisymmetry of the metric of the two-dimensional space $\Sigma^{(2)}_{\bf{u}}(t)$, it is easy to prove that its
 Gaussian curvature is equal to zero. However, following Cartan \cite{Car,Car1}, one can introduce a generalized linear connection:
\begin{equation}
G^i_{jk} =\Gamma^i_{jk}+K^i_{jk},\qquad i,j,k=1,2,
\label{4wl.n0q3}
\end{equation}
where $\Gamma^i_{jk}=\frac{1}{2}{\lambda}^{il}\bigl(\lambda_{lj;k}
+\lambda_{lk;j}-\lambda_{jk;l}\bigr),\,\,(\lambda_{lj;k}=
\partial \lambda_{lj}/\partial\bar{u}_k )$ is the Christoffel symbol and $K^i_{jk}$ denotes the contortion tensor generated by the
interaction of the oscillator with a random environment. This tensor can be defined as follows:
$$
K_{ijk}(\bar{u}_1,\bar{u}_2,t;\bar{y})=\frac{1}{2}\Bigl(\frac{\partial w_i}{\partial \bar{u}^k}-
\frac{\partial w_k}{\partial \bar{u}^i}\Bigr)\bar{u}_j,\qquad
w_i= \bar{y}\delta_{ii},
$$
where $\delta_{ij}$ denotes the Kronecker symbol.

Note that since the $K_{ijk}$ tensor is antisymmetric with respect to the first pair of indices, the $G_{ijk}$ connection is consistent with the metric.

Now we can write the equation of motion of a quasi-particle or excitation of an environment, which, taking into account the representation
 (\ref{7w3.t01}), will have the following form:
\begin{equation}
\frac{\partial^2 {\bar{u}^i}}{\partial s^2} + K^i_{jk}(\bar{u}_1,\bar{u}_2,t;\bar{y})\dot{\bar{u}}^j\dot{\bar{u}}^k=0,\qquad
\Gamma^i_{jk}\equiv0,\quad i,j,k=1,2,
\label{5wl.n03}
\end{equation}
where $s=\int\sqrt{\lambda^{ij}du_i du_j}=\int\sqrt{d\bar{u}_i d\bar{u}^i}$ and  $\dot{\bar{u}}^i=\partial{\bar{u}}^i/\partial s$.

Note that equation (\ref{5wl.n03}) can be considered as the equation of a geodesic line in a space with connection $K^i_{jk}$. To solve the
equation (\ref{5wl.n03}), it is necessary to know the form of the contortion tensor $K^i_{jk}$ as a function of coordinates and time.
\textbf{Theorem is proved}. $\square$

\subsection{Topology of two-dimensional subspace $\Sigma^{(2)}_{\bf{u}}(t)$}
\textbf{Theorem 4.} \emph{If the two-dimensional Riemann–Cartan space $\Sigma^{(2)}_{\bf{u}}(t)$ is characterized by the antisymmetric metric
$\bar{g}^{ij}$ (see (\ref{7w3.t01})), then this manifold is topological, the type of which is determined from the algebraic equation of the fourth
degree for the off-diagonal element of the metric.}

\textbf{Proof.}

Comparing the operators (\ref{3.n05}) and (\ref{nw3.02}) taking into account the transformations of coordinate (\ref{w3.t01}), for determine
the antisymmetric element of the tensor ${g}^{12}={y}$ one can obtain the following partial differential equations of the first order:
\begin{eqnarray}
\varepsilon^{(r)}\chi\frac{\partial{y}}{\partial{u}_1}-\bigl(1+
{y}\chi\bigr)\frac{\partial{y}}{\partial{u}_2}={k}_1({u}_1,{u}_2,t),
\nonumber\\
\varepsilon^{(i)}\chi\frac{\partial{y}}{\partial{u}_2}\,+\bigl(1+
{y}\chi\bigr)\frac{\partial{y}}{\partial{u}_1}={k}_2({u}_1,{u}_2,t),
\label{nw3.03}
\end{eqnarray}
where $\chi={y}/(a+{y}^2)$ and $a=\epsilon^{(r)}\epsilon^{(i)}$.\\
From the equations (\ref{nw3.03}) it is easy to find expressions for two different derivatives of the off-diagonal component of the metric tensor:
\begin{equation}
{y}_{1}=\frac{\partial{y}}{\partial{u}_1}=\frac{\varepsilon^{(i)}{k}_1\chi+{k}_2(1+{y}\chi)}{a\chi^2+(1+{y}\chi)^2},
\qquad
{y}_{2}=\frac{\partial{y}}{\partial{u}_2}=\frac{\varepsilon^{(r)}{k}_2\chi -{k}_1(1+{y}\chi)}{a\chi^2+(1+{y}\chi)^2}.
\label{nw3.0f3}
\end{equation}

By virtue of the Schwarz theorem (see for example \cite{Rudin}), we can require that the second derivatives to be symmetric, that is:
 $$
{y}_{12}=\frac{\partial^2\,{y}}{\partial{u}_1\partial{u}_2}={y}_{21}
=\frac{\partial^2\,{y}}{\partial{u}_2\partial{u}_1}.
 $$
If we write this equality explicitly, it will look like this:
$$
2\big[\varepsilon^{(r)}{k}_2\chi-\,{k}_1(1+{y}\chi)\bigr]
\frac{a\chi\chi_{;1}+(1+{y}\chi)({y}_1\chi+y\chi_1)}{a\chi^2+(1+{y}\chi)^2}
+\,({k}_1+\,{k}_2{y})\chi_{;2}
$$
$$
=2\big[\varepsilon^{(i)}{k}_1\chi+{k}_2(1+{y}\chi)\bigr]\frac{a\chi\chi_{;2}+
(1+{y}\chi)({y}_2\chi+y\chi_{;2})}{a\chi^2+(1+{y}\chi)^2}+ ({k}_2-{k}_1{y})
\chi_{;1}
$$
\begin{equation}
-(k_{1;1}+{k}_{2;2})(1+{y}\chi)-({k}_1{y}_1+k_2y_2+\varepsilon^{(i)}{k}_{1;2}-\varepsilon^{(r)}{k}_{2;1})\chi,
\label{nw3.0zf3}
\end{equation}
where $\chi_{;j}=\partial\chi/\partial{u}_j$ and ${k}_{i;j}=\partial{k}_i/\partial{u}_j,
\,\,\,( i,j=1,2).$\\
Finally, given (\ref{nw3.0f3}) from (\ref{nw3.0zf3}), we obtain the following 4\emph{th} degree algebraic equation:
\begin{eqnarray}
\sum_{n=0}^4 A_n(u_1,u_2,t)y^n=0,
\label{nw3z.0zf3}
\end{eqnarray}
where the coefficients of the algebraic equation $A_n(u_1,u_2,t)$ are defined by the expressions:
$$
A_0=a\bigl\{4au_1-4\varepsilon^{(r)}u_1^2u_2^2 -\varepsilon^{(i)}[u_1^2-u_2^2+\Omega^2_0(t)]^2\bigr\},\quad
 A_1=-2au_2(\varepsilon^{(r)}+\varepsilon^{(i)}),\,\,\,A_2=
$$
$$
\,\, 24au_1+\,8\varepsilon^{(r)}u_1^2u_2^2\,+\,2\varepsilon^{(i)}[u_1^2-u_2^2+\,\Omega^2_0(t)]^2,\quad
 A_3=-8u_2(\varepsilon^{(r)}+\varepsilon^{(i)}),\quad A_4=32 u_1.
$$
The equation of the 4\emph{th} degree (\ref{nw3z.0zf3}) of general form is solved exactly by the Ferrari method and has four
 solutions, some of which may be complex \cite{Artin}. Since the coefficients of the equation (\ref{nw3z.0zf3}) are functions of
two coordinates $(u_1,u_2)$ and time, the solutions must form a continuum of sets in two-dimensional Euclidean space. However,
we will be interested in those sets of solutions that are complex. In particular, it is obvious that if we cut out and remove from the
Euclidean space all the regions on which the solution of an algebraic equation (\ref{nw3z.0zf3}) is complex, then the remaining
space will be a topological space. As the numerical solution of equation (\ref{nw3z.0zf3}) shows, depending on the parameters
$\varepsilon^{(r)}$ and $\varepsilon^{(i)}$, the topological space $\Sigma^{(2)}_{\bf{u}}(t)$ can be an $n$-connected space,
where in the case under consideration $n \leq 4.$ \textbf{Theorem is proved}. $\square$

To illustrate the above, below are the results of visualization of a series of calculations (see FIG 8-11), which allow obtaining a
detailed idea of the topological features of the two-dimensional manifold $\Sigma^{(2 )}_{\ bf {u}}(t)$ , which arises after the
compactification of the function space $\mathbb { R}_{\{\bm x\}}$ (see subsection {\textbf D} of section { \textbf VII} for details).
It is also important to note that a similar analysis for the complex equation (\ref{4.0nt6i}) describing the solution $Q(u_1,u_2,t)$
 proves that the exact geometry for solving this problem is also the manifold $\Sigma^{(2)}_{\bf {u}}(t)$.

\section{S\lowercase{tatement of the initial-boundary value problem for the complex} PDF}
For definiteness, let us study the expectation of the trajectory of an oscillator immersed in a thermostat when it is not subjected
by an external field. Recall that it is described by a two-dimensional integral representation (\ref{4.0nw6i}), where the function
$Q(u_1,u_2,t)$ is the solution of a complex second-order partial differential equation (\ref{4.0nt6i}). For simplicity, we will consider
the case when the subspace $\Sigma^{(2)}_{\bf{u}}(t)$ is a two-dimensional Euclidean space, i.e. $\Sigma^{(2)}_{\bf{u}}(t)\cong \mathbb{R}^2
\equiv (-\infty,\,+\infty)\times (-\infty,\,+\infty)$.

{\textbf{Theorem 5.} \emph{If a two-dimensional complex PDE has the form (\ref{4.0nt6i}), then using the internal symmetry of the
equation it can be reduced to two independent PDFs belonging to the class of PDFs with a deviated argument given by affine
 transformations such as reflection.}

{\textbf{Proof}}.\\
Representing the solution of equation (\ref{4.0nt6i}) as a sum of real and imaginary parts:
\begin{equation}
Q(u_1,u_2,t)=Q^{(r)}(u_1,u_2,t)+iQ^{(i)}(u_1,u_2,t),
\label{5.n01}
\end{equation}
one can obtain the following system of differential equations:
\begin{equation}
 \Biggl\{
\begin{array}{ll}
 \partial_t Q^{(i)}(u_1,u_2,t)=\bigl\{\hat{L}+u_1\bigr\}Q^{(i)}(u_1,u_2,t)+u_2Q^{(r)}(u_1,u_2,t),\\
 \partial_t Q^{(r)}(u_1,u_2,t)=\bigl\{\hat{L}+u_1\bigr\}Q^{(r)}(u_1,u_2,t)-u_2Q^{(i)}(u_1,u_2,t),
\end{array}
\label{4w.n03}
\end{equation}
where the functions $Q^{(i)}(u_1,u_2,t)$ and $Q^{(r)}(u_1,u_2,t)$ can be normalized and given the meaning of the probability density:
\begin{eqnarray}
 \bar{Q}^{(i)}(u_1,u_2,t)=\beta^{-1}(t)Q^{(i)}(u_1,u_2,t),\qquad
 \bar{Q}^{(r)}(u_1,u_2,t)=\beta^{-1}(t)Q^{(r)}(u_1,u_2,t),\nonumber\\
\beta(t)=\int\int_{\Sigma^{(2)}_{\bf{u}}(t)}\bigl[Q^{(i)}(u_1,u_2,t)+Q^{(r)}(u_1,u_2,t)\bigr]du_1du_2.
\qquad\qquad
\label{5.n02}
\end{eqnarray}
Obviously, in this case, the probability is normalized to unity and has the form:
$$\int\int_{\Sigma^{(2)}_{\bf{u}}(t)}\bigl[\bar{Q}^{(i)}(u_1,u_2,t)
+\bar{Q}^{(r)}(u_1,u_2,t)\bigr]du_1du_2=1.$$

It is easy to see that when changing the coordinates $(u_1,u_2)\to(u_1,-u_2)$, the system of equations (\ref{4w.n03}) becomes:
\begin{equation}
 \Biggl\{
\begin{array}{ll}
 \partial_t Q^{(i)}(u_1,-u_2,t)=\bigl\{\hat{L}+u_1\bigr\}Q^{(i)}(u_1,-u_2,t)-u_2Q^{(r)}(u_1,-u_2,t),\\
 \partial_t Q^{(r)}(u_1,-u_2,t)=\bigl\{\hat{L}+u_1\bigr\}Q^{(r)}(u_1,-u_2,t)+u_2Q^{(i)}(u_1,-u_2,t).
\end{array}
\label{5.n03}
\end{equation}
If we assume that; $Q^{(i)}(u_1,-u_2,t)=Q^{(r)}(u_1,u_2,t)$ and, accordingly, $Q^{(r)}(u_1 ,-u_2 ,t)=Q^{(i)}(u_1,u_2,t)$, then the
system of equations (\ref{5.n03}) takes the original form (\ref{4w.n03}), i.e. the first equation goes over into the second, the second
into the first. Using this obvious symmetry properties, we can write the system of coupled PDEs (\ref{5.n03}) as two independent PDEs:
\begin{equation}
 \Biggl\{
\begin{array}{ll}
 \partial_t Q^{(i)}(u_1,u_2,t)=\bigl\{\hat{L}+u_1\bigr\}Q^{(i)}(u_1,u_2,t)\,-u_2Q^{(i)}(u_1,-u_2,t),\\
 \partial_t Q^{(r)}(u_1,u_2,t)=\bigl\{\hat{L}+u_1\bigr\}Q^{(r)}(u_1,u_2,t)+u_2Q^{(r)}(u_1,-u_2,t).
\end{array}
\label{5.n04}
\end{equation}
As we can see in the system (\ref{5.n04}), the equations are independent, and each of them belongs to the PDE class with a spatially
deviated argument given by affine transformations such as reflection. By solving one of the equations, we can get the solution of the
second one using a 180 degree rotation in the two-dimensional space $\Sigma^{(2)}_{\bf{u}}(t)$. Before proceeding to the numerical
solution of these PDEs, we consider three possible scenarios:

a)  when the solutions of the equations (\ref{5.n04}) are even functions with respect to the $u_2$ coordinate, i.e.
 $ Q^{(i)}(u_1,-u_2,t)= Q^{(i)}(u_1,u_2,t)$ and $Q^{(r)}(u_1,-u_2,t)= Q^{(r)}(u_1,u_2,t)$;

b) when the solutions of the equations (\ref{5.n04}) are odd functions with respect to the coordinate $u_2$, i.e.
$Q^{(i)}(u_1,-u_2,t)=- Q^{(i)}(u_1,u_2,t)$ and  $Q^{(r)}(u_1,-u_2,t)=- Q^{(r)}(u_1,u_2,t)$, and, accordingly,
the case;

c) when the indicated functions do not have definite parity.\\
 In the first  a)  case from (\ref{5.n04}) we get two unrelated PDEs:
\begin{equation}
 \Biggl\{
\begin{array}{ll}
 \partial_t Q^{(i)}(u_1,u_2,t)=\bigl\{\hat{L}+u_1\bigr\}Q^{(i)}(u_1,u_2,t)-u_2Q^{(i)}(u_1,u_2,t),\\
 \partial_t Q^{(r)}(u_1,u_2,t)=\bigl\{\hat{L}+u_1\bigr\}Q^{(r)}(u_1,u_2,t)+u_2Q^{(r)}(u_1,u_2,t).
\end{array}
\label{5.n14}
\end{equation}
In the second b) case, we again obtain a system of uncoupled equations, only in this case it is necessary to replace;
$Q^{(i)}(u_1,u_2,t)\to Q^{( r)}(u_1,u_2,t)$  in the first equation, and in the second one; $Q^{(r)}(u_1,u_2,t)\to Q^{(i)}(u_1,u_2,t)$,
respectively.

In the third c) case, the   functions $Q^{(i)}(u_1,u_2,t)$ and $Q^{(r)}(u_1,u_2,t)$ are described by the system of equations (\ref{5.n04}).
 Note that this is the most general and difficult case for numerical simulation, which will be considered in detail below.
\textbf{Theorem is proved}. $\square$

Below we will consider a more complicated case where a PDE with a deviating argument. Our task will be to formulate
an initial-boundary value problem for solving one of the PDE of the system of equations (\ref{5.n04}).

For definiteness, let us consider the second equation in  (\ref{5.n04}), which can be written as the following system:
\begin{equation}
 \Biggl\{
\begin{array}{ll}
\,\partial_t \,Q^{(r)}(u_1,u_2,t)\,\,=\,\bigl\{\hat{L}+u_1\bigr\}Q^{(r)}(u_1,u_2,t)+u_2Q^{(r)}(u_1,-u_2,t),\\
 \partial_t Q^{(r)}(u_1,-u_2,t)=\bigl\{\hat{L}+u_1\bigr\}Q^{(r)}(u_1,-u_2,t)+u_2Q^{(r)}(u_1,u_2,t),
\end{array}
\label{5.n15}
\end{equation}
Recall that the second equation in  (\ref{5.n15}) is obtained from the first one by replacing $u_2\to-u_2$.

As an initial condition, we assume that the probability distribution of the environmental fields $Q^{(r)}(u_1,u_2,t)$ at time $t_0$
 is described by the Dirac delta function:
\begin{equation}
 {Q}^{(r)}(u_1,u_2,t_0)=\prod_{j=1}^2\delta(u_j-u_{0j}),\qquad u_{0j}=u_j(t_0).
\label{5.n05}
\end{equation}
As for the boundary conditions, since the classical oscillator in the environment experiences both elastic and inelastic interactions, it
 is useful to define the Neumann boundary conditions respectively on the perpendicular axes $u_1$ and $u_2$:
\begin{equation}
 \frac{\partial}{\partial u_2} {Q}^{(r)}(u_1,u_2,t)\bigl|_{u_2=\,0}= {Q}^{(r)}_{\,;2}(u_1,0,t),
\qquad
 \frac{\partial}{\partial u_1} {Q}^{(r)}(u_1,u_2,t)\bigl|_{u_1=\,0}= {Q}^{(r)}_{\,;1}(0,u_2,t),
\label{5.n06}
\end{equation}
where ${Q}^{(r)}_{;\,i}=\partial {Q}^{(r)}/\partial u_i, \quad (i=1,2)$ we will derive from a series of physical considerations. As we will
see below, the conditions (\ref{5.n06}) lead to two different equations.

First, consider the behaviour of the second equation in the (\ref{5.n04}) near the $u_1\in (-\infty,+\infty)$ axis. In particular the solution
of this equation near the $u_1$ axis can be represented as:
\begin{equation}
Q^{(r)}(u_1,u_2,t)\bigl|_{u_2\,\sim\,0}=(a_0+a_1u_2+a_2u_2^2+\cdot\cdot\cdot)e^{-u_2^2/2}
\mathbb{Q}^{(r)}_1(u_1,t),
\label{5.n07}
\end{equation}
where $a_0, a_1$ and $a_2$ are some unknown constants that will be determined based on physical considerations.

 Substituting (\ref{5.n07}) into the second equation in (\ref{5.n04}), in the
limit of $ u_2\to 0$, one obtains the following second-order partial differential equation:
\begin{equation}
\frac{\partial }{\partial t}\mathbb{Q}^{(r)}_1=\biggl\{\epsilon^{(r)}\frac{\partial^2}{\partial
 u_1^2}+\bigl(u_1^2+\Omega^2_0(t)\bigr)\frac{\partial}{\partial u_1}+\biggl[5u_1-\epsilon^{(i)}
 \biggl(\frac{2a_2}{a_0}-1\biggr)\biggr]\biggr\}\mathbb{Q}^{(r)}_1.
\label{5.n08}
\end{equation}
In particular, for the first Neumann condition we obtain:
\begin{equation}
\frac{\partial}{\partial u_2}{Q}^{(r)}(u_1,u_2,t)\bigl|_{u_2=\,0}={Q}^{(r)}_{\,;2}(u_1,0,t)=a_1
\mathbb{Q}^{(r)}_1(u_1,t).
\label{5.nt08}
\end{equation}
Since the equation for the function ${Q}^{(r)}(u_1,u_2,t)$ is not symmetric with respect to the change $u_2\to-u_2$, the first
Neumann boundary condition cannot be equal to zero and, accordingly, $a_1\neq0 $. Since the function
 ${Q}^{(r)}_{\,;2}(u_1,t)\sim \mathbb{Q}^{(r)}_1(u_1,t)$ has the meaning of the probability
density on the axis, we can normalize it to one. The latter is equivalent to setting $a_1=1$ and normalizing the
solution of the equation $\mathbb{Q}^{(r)}_1(u_1,t)$ to unity, i.e.:
$$\bar{\mathbb{Q}}^{(r)}_1(u_1,t)=c^{-1}_0(t)\mathbb{Q}^{(r)}_1(u_1,t),\qquad
c_0(t)=\int_{-\infty}^{+\infty}\mathbb{Q}^{(r)}_1(u_1,t)du_1,
$$
where $c_0(t)$ is a normalizing constant.

Proceeding from the fact that the boundary conditions on the perpendicular axes $u_1$ and $u_2$ describe the probability
distributions of elastic and inelastic collisions independent of each other,  it is natural to assume that the term of inelastic
collision in equation (\ref{5.n08}) should be identically equal to zero. In other words, we can require that the equality
$2a_2/a_0-1=0$  gets satisfied, and thus the equation (\ref{5.n08}) can be written as:
\begin{equation}
\frac{\partial }{\partial t}\mathbb{Q}^{(r)}_1=\biggl\{\epsilon^{(r)}\frac{\partial^2}{\partial u_1^2}+\bigl(u_1^2+
\Omega^2_0(t)\bigr)\frac{\partial}{\partial u_1}+5u_1\biggr\}\mathbb{Q}^{(r)}_1.
\label{5.n09}
\end{equation}
To solve this equation, it is necessary to formulate the following initial-boundary value problem:
\begin{equation}
\mathbb{Q}^{(r)}_1(u_1,t)\bigl|_{t=t_0}=\delta(u_1-u_{01}),
\label{5.n10}
\end{equation}
and, correspondingly;
\begin{equation}
\mathbb{Q}^{(r)}_1(u_1,t)\bigl|_{u_1=\,s_1}=0,\qquad \mathbb{Q}^{(r)}_1(u_1,t)\bigl|_{u_1=\,s_2}=0.
\label{5.n011}
\end{equation}
where $s_1$ and $s_2$ denote points far enough from the origin $0$, which are located to the left and right, respectively.

Now, to establish the second Neumann boundary condition, we consider the solution ${Q}^{(r)}(u_1,u_2,t)$ near the axis
$u_2\in (-\infty,+\infty)$. To do this, represent the solution in the form:
\begin{equation}
{Q}^{(r)}(u_1,u_2,t)\bigl{|}_{u_1\,\sim\,0} =(b_0+b_1u_1+b_2u_1^2+\cdot\cdot\cdot)e^{-u_1^2/2}
\mathbb{Q}^{(r)}_2(u_2,t),
\label{5.n012}
\end{equation}
where $b_0,\,b_1$ and $b_2$ are some constants that we will define below.

Substituting (\ref{5.n012}) into (\ref{5.n04}), in the limit of $u_1\to0$ we get the following equation:
\begin{equation}
\frac{\partial }{\partial t}\mathbb{Q}^{(r)}_2(u_2,t)=\biggl\{\epsilon^{(i)}\frac{\partial^2}{\partial u_2^2}+
\biggl[\epsilon^{(r)}\Bigl(\frac{2b_2}{b_0}-1\Bigr)+\frac{b_1}{b_0}\Bigl(\Omega_0^2(t)-u_2^2\Bigr)\biggr]\biggr\}
\mathbb{Q}^{(r)}_2(u_2,t)+u_2\mathbb{Q}^{(r)}_2(-u_2,t).
\label{5.n013}
\end{equation}
Since only inelastic processes are taken into account on the $u_2$ axis, we can require that the following condition gets fulfilled;
${2b_2}/{b_ 0}-1=0$, in addition, for definiteness, we can set the constant $b_0=1$  and $b_1=1/2$. Taking into account the
above clarifications,  the equation (\ref{5.n013}) can be simplified by presenting it in the form:
\begin{equation}
\frac{\partial }{\partial t}\mathbb{Q}^{(r)}_2(u_2,t)=\biggl\{\epsilon^{(i)}\frac{\partial^2}{\partial u_2^2}+
\frac{1}{2}\bigl(\Omega_0^2(t)-u_2^2\bigr)\biggr\}\mathbb{Q}^{(r)}_2(u_2,t)+u_2\mathbb{Q}^{(r)}_2(-u_2,t) .
\label{5.n013}
\end{equation}
Due to the fact that (\ref{5.n013}) is a second-order PDE with a deviating argument, it can be solved together with the same equation, but
after changing the argument $u_2\to-u_2.$ For the equation (\ref{5.n013}), we can formulate the following initial-boundary conditions:
$$
\mathbb{Q}^{(r)}_2(u_2,t)\bigl|_{t=t_0}=\delta(u_2-u_{02}),
$$
and, correspondingly;
$$
\mathbb{Q}^{(r)}_2(u_2,t)\bigl|_{u_2=\,\pm\, s}=0,\qquad \qquad |s|>>1.
$$

Taking into account the above, for the second  boundary condition, we can write the following expression:
\begin{equation}
\frac{\partial}{\partial u_1} {Q}^{(r)}(u_1,u_2,t)\bigl|_{u_1=\,0}= {Q}^{(r)}_{2;1}(0,u_2,t)=
b_1\mathbb{Q}^{(r)}_2(u_2,t)=\frac{1}{2}\mathbb{Q}^{(r)}_2(u_2,t).
\label{5.nw08}
\end{equation}
Recall that, as in the case of the solution $\mathbb{Q}^{(r)}_1(u_1,t)$, one can also normalize the solution $\mathbb{Q}^{(r)}_2( u_2 ,t )$
as a probability density per unit.

\section{E\lowercase{ntropy of a self-organizing system}}
As  known, for a classical dynamical system an important characteristic is the non-stationary Shannon entropy \cite{Shann}. In particular, the
entropy production rate is a quantitative measure of a non-equilibrium process, and knowledge of its value indicates information about the
dissipated heat \cite{Tak,Shoi}, the difference in free energy between two equilibrium states \cite{Jarz,Gavin}, and also about the efficiency,
 if the considered nonequilibrium system is an engine \cite{Gatien,Gat,Sreekanth}. It should be noted that the rate of entropy production provides
important information for systems with hidden degrees of freedom \cite{Massimiliano,Kyogo}, as well as for interacting subsystems, where the
amount of information plays a key role \cite{Takah,Sos,Jordan,Juan}.

In philosophy, physics and mathematics, the term negentropy is often used, which has the opposite meaning of entropy. Recall that if entropy
characterizes the measure of orderliness and organization of the system, then negentropy is the possibility of reducing entropy or moving
towards order. Note that this concept was first proposed by Schr\"{o}dinger \cite{Sh} when explaining the behaviour of living systems:
\emph{in order not to die, a living system struggles with the surrounding chaos and with the entropy it produces, organizing and ordering the
 latter by introducing negentropy}. This, in particular, explains the behaviour of self-organizing systems.

For definiteness, let us consider the question of the change in the entropy of a classical oscillator in the case when its frequency has a random
component. We first calculate the dynamics of the oscillator entropy without taking into account its influence on the random environment, when,
as in the second case, we will take this influence into account consistently and strictly. In particular, in the first case, the non-stationary entropy
of the oscillator is determined by the standard expression of the form:
\begin{equation}
\mathcal{S}(t)= -\int\int_{\Sigma^{(2)}_{\bf{u}}(t)}\bar{P}(u_1,u_2,t)\ln \bar{P}(u_1,u_2,t)du_1du_2.
\label{6.n01}
\end{equation}
It is often important to know the change in entropy or production of entropy over a given period of time, which can be determined by the
expression:
$$
\Delta{\mathcal{S}}(t_1,t_2)\equiv\mathcal{S}(t_2)-\mathcal{S}(t_1).
$$
Since processes of a fundamentally different nature occur in the problem under consideration, we can also introduce the concept of partial
entropy, which characterizes the production of entropy of a particular process:
\begin{equation}
\mathcal{S}^{(\sigma)}_{par}(t)=-\int\int_{\Sigma^{(2)}_{\bf{u}}(t)}\bar{Q}^{(\sigma)}(u_1,u_2,t)\ln
 \bar{Q}^{(\sigma)}(u_1,u_2,t)du_1du_2,\qquad \sigma=i,\,r.
\label{6.nt03}
\end{equation}
Recall that the partial entropies $\mathcal{S}^{(r)}_{par}(t)$ and $\mathcal{S}^{(i)}_{par}(t)$ are in any case related by the total probability
normalization condition (\ref{5.n02}) and, accordingly, will influence each other in the course of evolution. It is important to note that the partial
entropy $\mathcal{S}^{(r)}_{par}(t)$  characterizes the processes of elastic collisions of the oscillator with the environment, while
$\mathcal{S}^{(i)}_{par}(t)$ describes the processes of inelastic collisions of the oscillator with the environment. The study of partial entropies
will obviously provide additional important information about a dynamical system immersed in a thermostat.

Finally, it is very important to determine the Shannon entropy for the case, when the ``oscillator + thermostat" is considered as a closed
self-organizing system. In this case, the expression for the generalized entropy  $\mathcal{S}_ {gen}(t)$ can be represented as:
\begin{equation}
\mathcal{S}_{gen}(t)=-\int\int_{\Sigma^{(2)}_{\bf{u}}(t)}\Bigl(\sum_{\sigma=i,\,r}\bar{Q}^{(\sigma)}(u_1,u_2,t)\Bigr)
\ln\Bigl(\sum_{\sigma=i,\,r}\bar{Q}^{(\sigma)}(u_1,u_2,t)\Bigr)du_1du_2.
\label{6.n03}
\end{equation}
As can be seen from the graphs (see FIG 11), at intermediate times, the behavior of the entropies calculated by the formulas (\ref{6.n01})
and (\ref{6.n03}) differ greatly in value and in character. Moreover, in some time intervals, the generalized entropy (\ref{6.n03}), regardless
of the intensity of the processes in the environment, takes a negative value, which is quite natural for a self-organizing system. Finally, as
follows from the simulation and the corresponding visualizations, the behavior of both ordinary and generalized entropy in the limit of long
times is established and tends to a constant value. Details of entropy calculations are discussed in subsection \textbf{E} of Section \textbf{VII}.

\section{N\lowercase{umerical methods for solving the problem}}
Numerical simulation of self-organization processes in the joint system ``Classical oscillator + thermostat"
even for a simple case, i.e. in the absence of an external field $F\bigl(t;\{{\bf g}\}\bigr) \equiv0$, requires
a lot of computational time. This is due to the fact that complex probabilistic processes are described by three
key PDEs, the equation (\ref{3.05})-(\ref{3.n05}), as well as the system of equations  (\ref{4w.n03}), turn out
to be quite complicated in numerical implementation. Recall that in \cite{Gev} we have already considered PDEs
of this type. Various finite-difference methods of solution were also considered there. Taking into account the
performed analysis and test calculations, for the numerical solution, an explicit finite-difference scheme of the
second order of accuracy in coordinates and the first order in time was chosen. Despite the external simplicity
of this method, we believe that this scheme satisfies the goals of our work in terms of efficiency and accuracy.
Equations of the form  (\ref{3.05})-(\ref{3.n05}) and (\ref{4w.n03}) are a second-order partial differential
equations of the parabolic type with a source term. The main difficulty in the calculations arises in connection
with the convective transfer, which must be taken into account when increasing the size of the computational domain.

Below we consider a numerical algorithm for solving the initial boundary value problems for these PDEs
in the two-dimensional Euclidean space $\mathbb{R}^2$, shown in \textbf{Listings 1} and \textbf{2}.

\textbf{Listing 1}. Numerical algorithm for solving PDE (\ref{3.05})-(\ref{3.n05}) after coordinate transformations
(\ref{w3.t01}).
\begin{enumerate}
\item The continuous area $\mathbb{R}^2$ for the equation (\ref{3.05})-(\ref{3.n05}) is replaced by the calculated
area; $[(\bar{u}_1)_{\min},(\bar{u}_1)_{\max}]\times[(\bar{u}_2)_{\min},(\bar{u}_2)_{\max}]\times[0,T]$. In the
computational domain, a uniform difference grid is set in time $t$ and in spatial coordinates $\bar{u}_1$ and
$\bar{u}_2$:
$$
(\bar{u}_1)_j=j\Delta\bar{u}_1,\qquad j\in[1,M],\qquad (\bar{u}_1)_{\min}=(\bar{u})_{j=1},\qquad (\bar{u}_1)_{\max}
=(\bar{u})_{j=M},
$$
$$
(\bar{u}_2)_k=k\Delta\bar{u}_2,\qquad k\in[1,L],\qquad (\bar{u}_2)_{\min}=(\bar{u})_{k=1},\qquad (\bar{u}_2)_{\max}
=(\bar{u})_{k=L},
$$
$$
t_n=n\Delta t,\qquad n=0,1,2,...T/\Delta t-1,
$$
where $\Delta\bar{u}_1$ and $\Delta\bar{u}_2$  are steps in spatial coordinates, $\Delta t$ is a step in time.
\item In these notations for the constructed grid, we have the following difference equation for equations  (\ref{3.05})-(\ref{3.n05}) the
following difference equation:
\begin{eqnarray}
P^{n+1}_{j,k}= P^{n}_{j,\,k}+r_1\bigl[P^{n}_{j+1,k}-2P^{n}_{j,k}+P^{n}_{j-1,k}\bigr]+r_2
\bigl[P^{n}_{j,k+1}-2P^{n}_{j,k}+P^{n}_{j,k-1}\bigr]+4\Delta t P^{n}_{j,k}+
\nonumber\\
\frac{\Delta t}{2\Delta\bar{u}_1}\bigl[(\bar{u}_1^2)_j-(\bar{u}_2^2)_k+\Omega_{0(n)}^2\bigr]\bigl(P^{n}_{j+1,k}
-P^{n}_{j-1,k}\bigr)+\frac{\Delta t}{\Delta\bar{u}_2}(\bar{u}_1)_j(\bar{u}_2)_k\bigl(P^{n}_{j,k+1}-P^{n}_{j,k-1}\bigr),\quad\,\,
\label{6.nt03}
\end{eqnarray}
where the following notation are introduced:
$$
P^{n}_{j,\,k}=P(j\Delta\bar{u}_1,k\Delta\bar{u}_2;n\Delta t),\quad \Omega^2_{0(n)}=\Omega_0^2(t_n),\quad r_1=
\varepsilon^{(r)}\frac{\Delta t}{(\Delta\bar{u}_1)^2},\quad r_2= \varepsilon^{(i)}\frac{\Delta t}{(\Delta\bar{u}_2)^2}.
$$
\item
To calculate the equation (\ref{3.05})-(\ref{3.n05}), it is also necessary to set two boundary conditions in
the form of difference equations on the coordinate axes $\bar{u}_1$ and
$\bar{u}_2$ respectively, which
can be easily found by approximating the equation (\ref{6.nt03}). Note that these difference equations must
be solved taking into account the Dirichlet boundary conditions;
$\mathbb{P}_l(x)|_{x=\,\pm\,|s|}=0,\,(|s|>>1)$,
where the index $(l=1,2)$ denotes the first and second boundary conditions, respectively.
 \item The condition is set at the center of the coordinate axes:
 $$P(\bar{u}_1,\bar{u}_2=0,t)\bigl|_{\bar{u}_1=0}=P(\bar{u}_1=0,\bar{u}_2,t)\bigl|_{\bar{u}_2=0}.$$
In addition, the Dirichlet condition $P(\bar{u}_1,\bar{u}_2,t)\bigl|_{\partial\mathcal{G}}=0$
is specified on the boundaries of the computational domain, where $\partial\mathcal{G}$
denotes the boundary.
\item As an initial condition, instead of the Dirac delta function, the Gaussian distribution was considered:
$$
P(\bar{u}_1,\bar{u}_2,t)\bigl|_{t=0}\approx\sigma \exp\biggl\{-\omega\Bigl(\Bigl[\frac{\bar{u}_1}{a}\Bigr]^2+
\Bigl[\frac{\bar{u}_2}{b}\Bigr]^2\Bigr)\biggr\}.
$$
Note that the parameters $\sigma=500,\,\omega=\pi\sigma a^2$ and $a=b=1/2$ included in the function were
chosen in such a way that the condition of normalizing the initial distribution to unit.
\end{enumerate}

\textbf{Listing 2}. Numerical algorithm for solving the system of PDEs (\ref{w3.t01}) after coordinate transformation.
\begin{enumerate}
\item
The continuous region $\mathbb{R}^2$ for the PDE system (\ref{4w.n03}) is replaced by a discrete grid, as
described in the \textbf{Listing 1}.
\item Using the PDE system (\ref{4w.n03}), we can derive the following system of difference equations on the
constructed grid:
\begin{eqnarray}
\bigl[Q^{(i)}\bigr]^{n+1}_{j,\,k}=\,\bigl[Q^{(i)}\bigr]^{n}_{j,\,k}+\,r_1\Bigl\{\bigl[Q^{(i)}\bigr]^{n}_{j+1,\,k}\,-
2\bigl[Q^{(i)}\bigr]^{n}_{j,\,k}+\bigl[Q^{(i)}\bigr]^{n+1}_{j-1,\,k}\Bigr\}+r_2\Bigl\{\bigl[Q^{(i)}\bigr]^{n}_{j,\,k+1}-
\nonumber\\
2\bigl[Q^{(i)}\bigr]^{n}_{j,\,k}\,+\,\bigl[Q^{(i)}\bigr]^{n+1}_{j,\,k-1}\Bigr\}\,+\,\frac{\Delta t}{2\Delta\bar{u}_1}
\bigl[(\bar{u}_1^2)_j\,-\,(\bar{u}_2^2)_k\,+\Omega_{0(n)}^2\bigr]\Bigl\{\bigl[Q^{(i)}\bigr]^{n}_{j+1,\,k}-
\bigl[Q^{(i)}\bigr]^{n}_{j-1,\,k}\Bigr\}
\nonumber\\
+\frac{\Delta t}{\Delta\bar{u}_2}(\bar{u}_1)_j(\bar{u}_2)_k \Bigl\{\bigl[Q^{(i)}\bigr]^{n}_{j,\,k+1}\,-\,
\bigl[Q^{(i)}\bigr]^{n}_{j,\,k-1}\Bigr\}\,+\,\Delta t\Bigl\{5 (\bar{u}_1)_j\bigl[Q^{(i)}\bigr]^{n}_{j,\,k}+
(\bar{u}_2)_k \bigl[Q^{(r)}\bigr]^{n}_{j,\,k}\Bigr\},
\nonumber\\
\bigl[Q^{(r)}\bigr]^{n+1}_{j,\,k}=\bigl[Q^{(r)}\bigr]^{n}_{j,\,k}+r_1\Bigl\{\bigl[Q^{(r)}\bigr]^{n}_{j+1,\,k}-
2\bigl[Q^{(r)}\bigr]^{n}_{j,\,k}+\bigl[Q^{(r)}\bigr]^{n+1}_{j-1,\,k}\Bigr\}+r_2\Bigl\{\bigl[Q^{(r)}\bigr]^{n}_{j,\,k+1}-
\nonumber\\
2\bigl[Q^{(r)}\bigr]^{n}_{j,\,k}\,+\,\bigl[Q^{(r)}\bigr]^{n+1}_{j,\,k-1}\Bigr\}\,+\,\frac{\Delta t}{2\Delta\bar{u}_1}
\bigl[(\bar{u}_1^2)_j-(\bar{u}_2^2)_k+\Omega_{0(n)}^2\bigr]\Bigl\{\bigl[Q^{(r)}\bigr]^{n}_{j+1,\,k}-
\bigl[Q^{(r)}\bigr]^{n}_{j-1,\,k}\Bigr\}
\nonumber\\
+\,\frac{\Delta t}{\Delta\bar{u}_2}(\bar{u}_1)_j(\bar{u}_2)_k \Bigl\{\bigl[Q^{(r)}\bigr]^{n}_{j,\,k+1}\,-
\bigl[Q^{(r)}\bigr]^{n}_{j,\,k-1}\Bigr\}\,+\,\Delta t\Bigl\{5(\bar{u}_1)_j\bigl[Q^{(r)}\bigr]^{n}_{j,\,k}-
(\bar{u}_2)_k \bigl[Q^{(i)}\bigr]^{n}_{j,\,k}\Bigr\},\nonumber\\
\label{6.nt01}
\end{eqnarray}
where $\bigl[Q^{(\upsilon)}\bigr]^{n}_{j,\,k}=Q^{(\upsilon)}(j\Delta\bar{u}_1,k\Delta\bar{u}_2;n\Delta t)$, in addition,
$\upsilon=i,\,r$.
\item Similarly, as in \textbf{Listing 1}, for the solutions $Q^{(i)}\bar{u}_1,\bar{u}_2,t)$ and
$Q^{(r)}(\bar{u}_1,\bar{u}_2,t)$, boundary conditions are set on the $\bar{u}_1$ and $\bar{u}_2$
axes in the form of difference equations, which can be obtained by approximating the equations (\ref{6.nt01})
on the same axes. It should be noted that each equation obtained for the boundary conditions is solved as an
interior Dirichlet problem with a zero value of the solution at the boundary.
\item
As in the case of the probability density $P(\bar{u}_1,\bar{u}_2,t)$ (see \textbf{Listing 1}), the solutions
$Q^{(i)}(\bar{u}_1,\bar{u}_2,t)$ and $Q^{(r)}(\bar{u}_1,\bar{u}_2,t)$ are subject to similar conditions at
the center of the coordinate axes:
$$
Q^{(i)}(\bar{u}_1,\bar{u}_2=0,t)\bigl|_{\bar{u}_1=0}=Q^{(i)}(\bar{u}_1=0,\bar{u}_2,t)\bigl|_{\bar{u}_2=0},
$$
$$
Q^{(r)}(\bar{u}_1,\bar{u}_2=0,t)\bigl|_{\bar{u}_1=0}=Q^{(r)}(\bar{u}_1=0,\bar{u}_2,t)\bigl|_{\bar{u}_2=0}.
$$
In addition, we assume that at the boundary of the computational domain:
$$
Q^{(i)}(\bar{u}_1,\bar{u}_2,t)\bigl|_{\partial\mathcal{G}}=0,\qquad Q^{(r)}(\bar{u}_1,\bar{u}_2,t)
\bigl|_{\partial\mathcal{G}}=0.
$$
\item Finally, as an initial condition for solving the system of equations (\ref{6.nt01}) for both solutions
$Q^{(i)}(\bar{u}_1,\bar{u}_2,t)$ and $Q ^{(r)}(\bar{u}_1,\bar{u}_2,t)$ Gaussian distribution with
parameters as in \textbf{Listing 1} is chosen.
\end{enumerate}
For the system of equations (\ref{6.nt01}) with the corresponding conditions, on the coordinate axes, the finite difference
scheme is similar to \textbf{Listing 1}.

In view of the symmetry of the equations (\ref{6.nt03}) and (\ref{6.nt01}) with respect to the coordinate $\bar{u}_2$, the
calculation was carried out for the upper half-plane, i.e. for the $\bar{u}_2\geq0$ for a square grid of $600\times400$
nodes in $\bar{u}_1\times\bar{u}_2$, respectively. Then the results were recalculated for the entire region $\mathbb{R}^2$,
that is on the $600\times800$ grid for all solutions $P(\bar{u}_1,\bar{u}_2,t)$, $Q^{(r)}(\bar{u}_1,\bar{u}_2,t)$ and
$Q^{(i)}(\bar{u}_1,\bar{u}_2,t)$. Space steps; $\Delta\bar{u}_1=\Delta\bar{u}_2=0.02$, time step for the first option
$\Delta t=10^{-5}$; for the second and third options $\Delta t=2\cdot10^{-5}$.

\subsection{Distributions of the free  environmental fields }
To calculate the distribution equation for the free fields of the environment  (\ref{3.05})-(\ref{3.n05}), we use the difference equation
(\ref{6.nt03}), which is described in detail in \textbf {Listing 1}. To implement the numerical simulation of the problem, we used
the data presented in the {\textbf{Table}} below.
\begin{center}
\medskip
\textbf{Table}\qquad\qquad
\begin{tabular}{c c c c }
 \hline
$\nu=0.5$;& $\gamma=2$;& $\varepsilon^{(r)}=0.01$;& $\varepsilon^{(i)}=0.01$;\\
$\nu=0.5$;& $\gamma=2$;& $\varepsilon^{(r)}=1.00$;& $\varepsilon^{(i)}=0.01$;\\
$\nu=0.5$;& $\gamma=2$;& $\varepsilon^{(r)}=1.00$;& $\varepsilon^{(i)}=0.50$.\\
\end{tabular}
\end{center}
Note that each line of this table characterizes different state of the environment:

a. when the processes going on in the environment, both elastic and inelastic, are weak,

b. when elastic processes are strong and inelastic processes are weak,

c. when both elastic and inelastic processes are strong in the environment.
\begin{figure}[ht!]
\centering
\includegraphics[width=52mm]{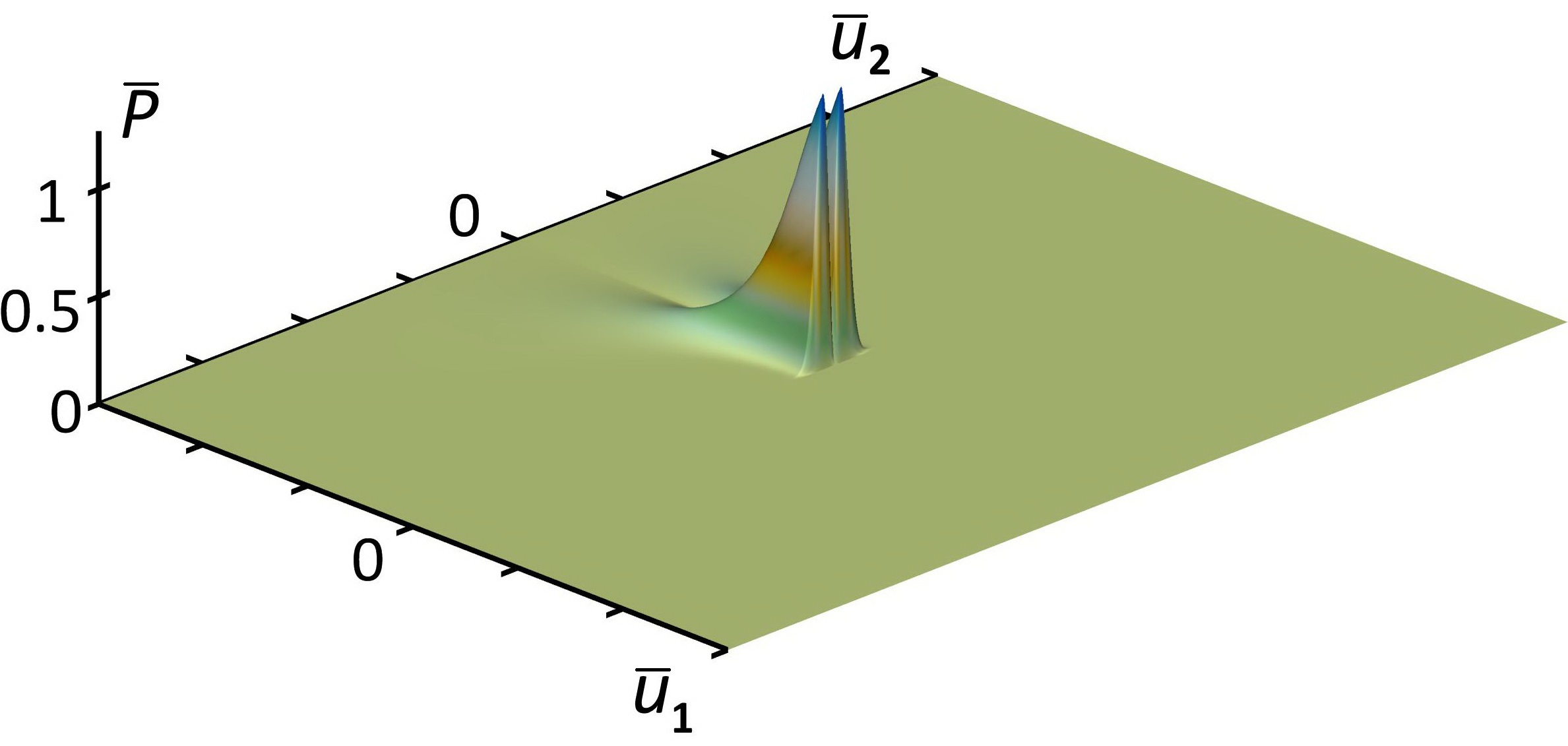}
\,\includegraphics[width=52mm]{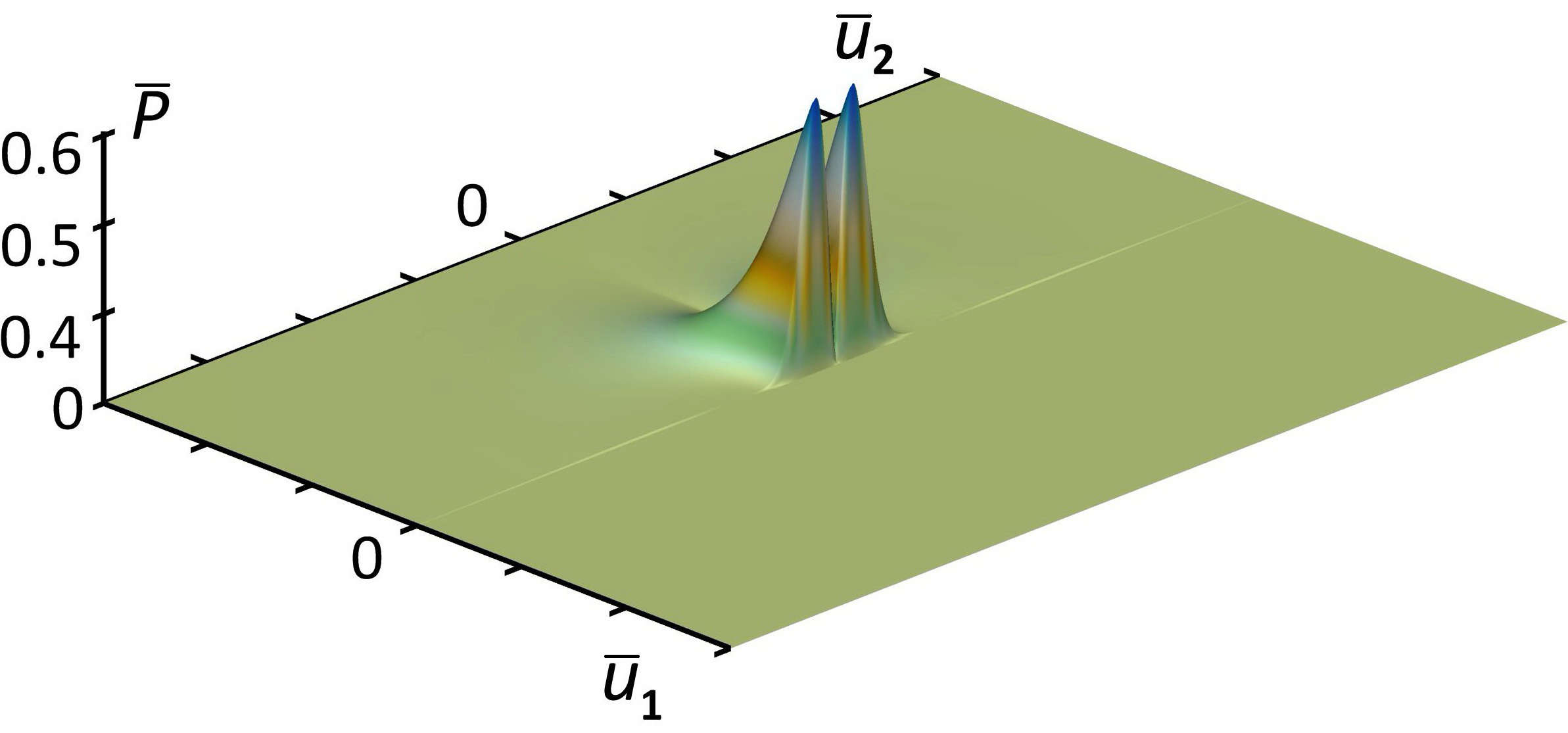}
\, \includegraphics[width=52mm]{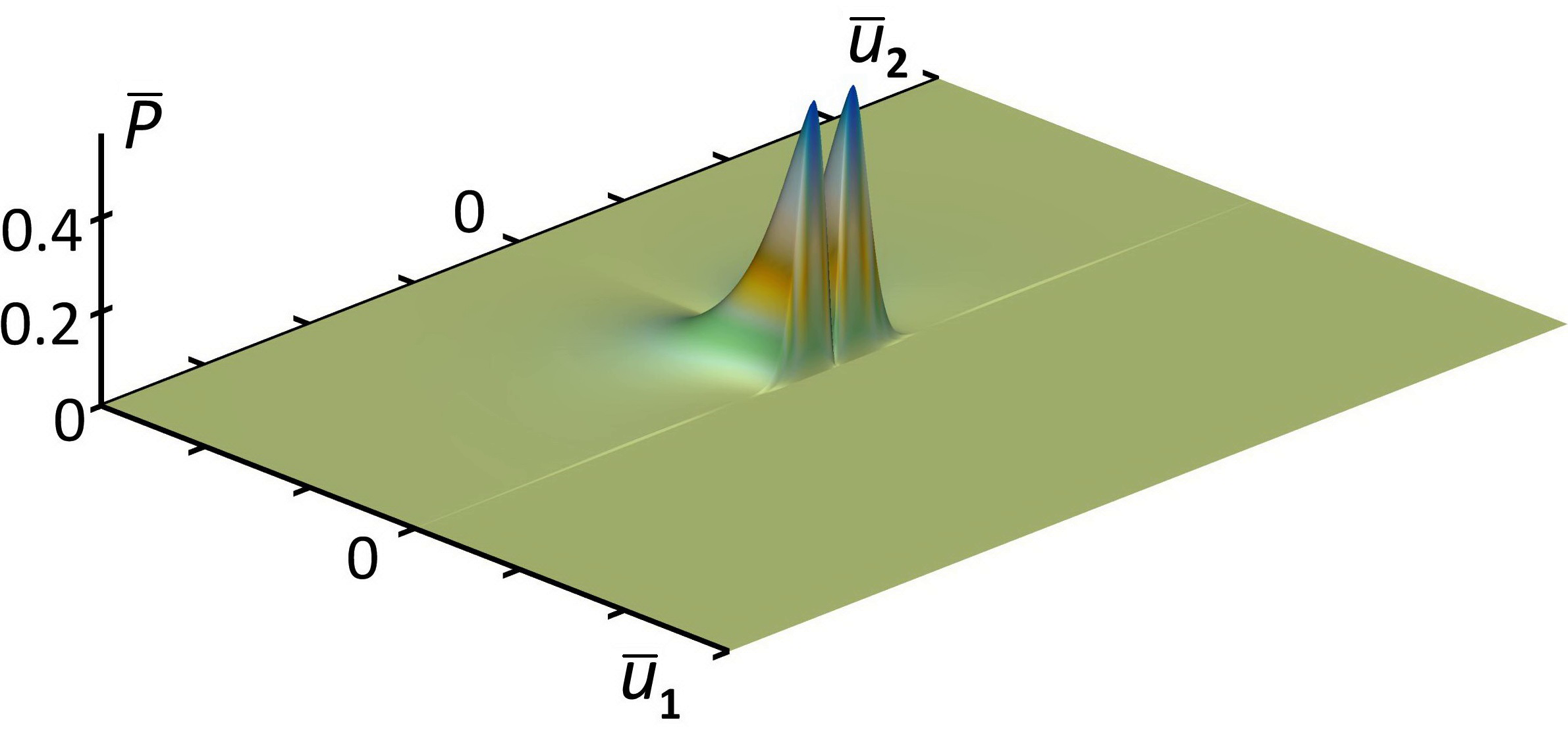}
\caption{\emph{The  different stages of the evolution of the free fields of the environment, respectively, at $t_1=1.5$,
$t_2=10$ and $t_3=20$. Note that these distributions were calculated using the data from the first row of the \textbf{Table}, which corresponds to weak
elastic and inelastic processes in the environment. Comparing the distributions at different times, it is easy to see that, as $t\sim 10$, the distribution
 $\bar{P}(u_1,u_2,t)$ normalized to unity is established or tends to its stationary limit.}
\label{overflow}}
\end{figure} \\
\begin{figure}[ht!]
\centering
\includegraphics[width=52mm]{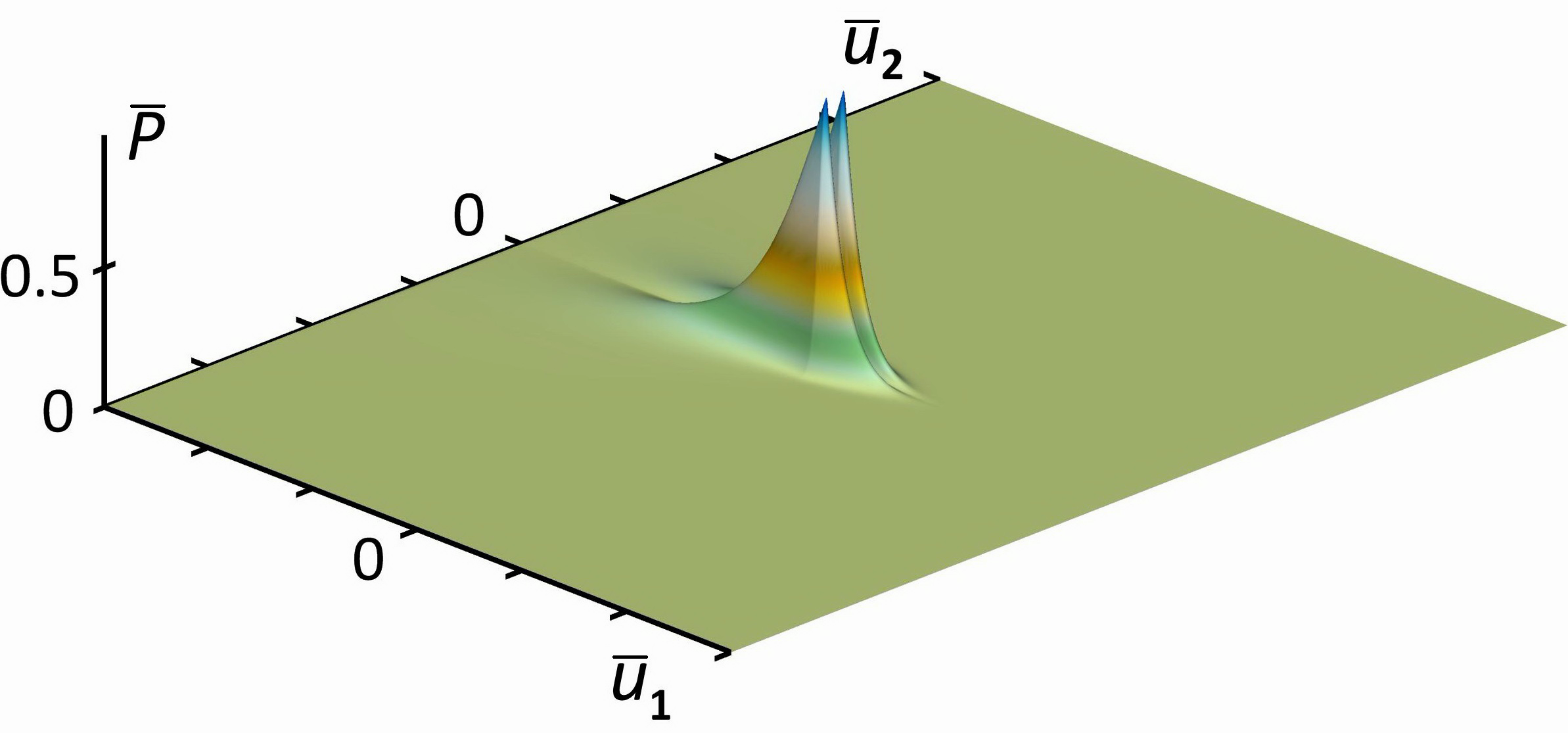}
\, \includegraphics[width=52mm]{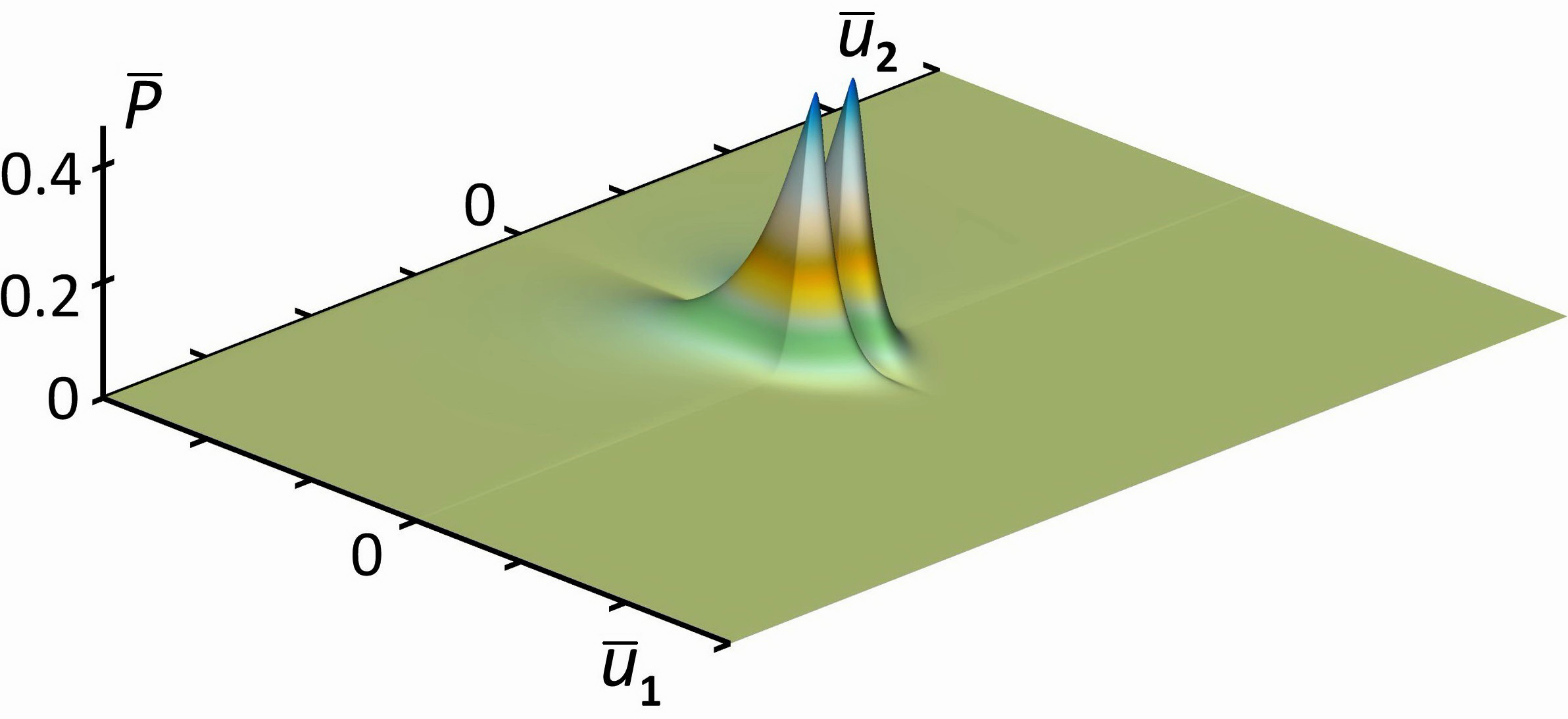}
\, \includegraphics[width=52mm]{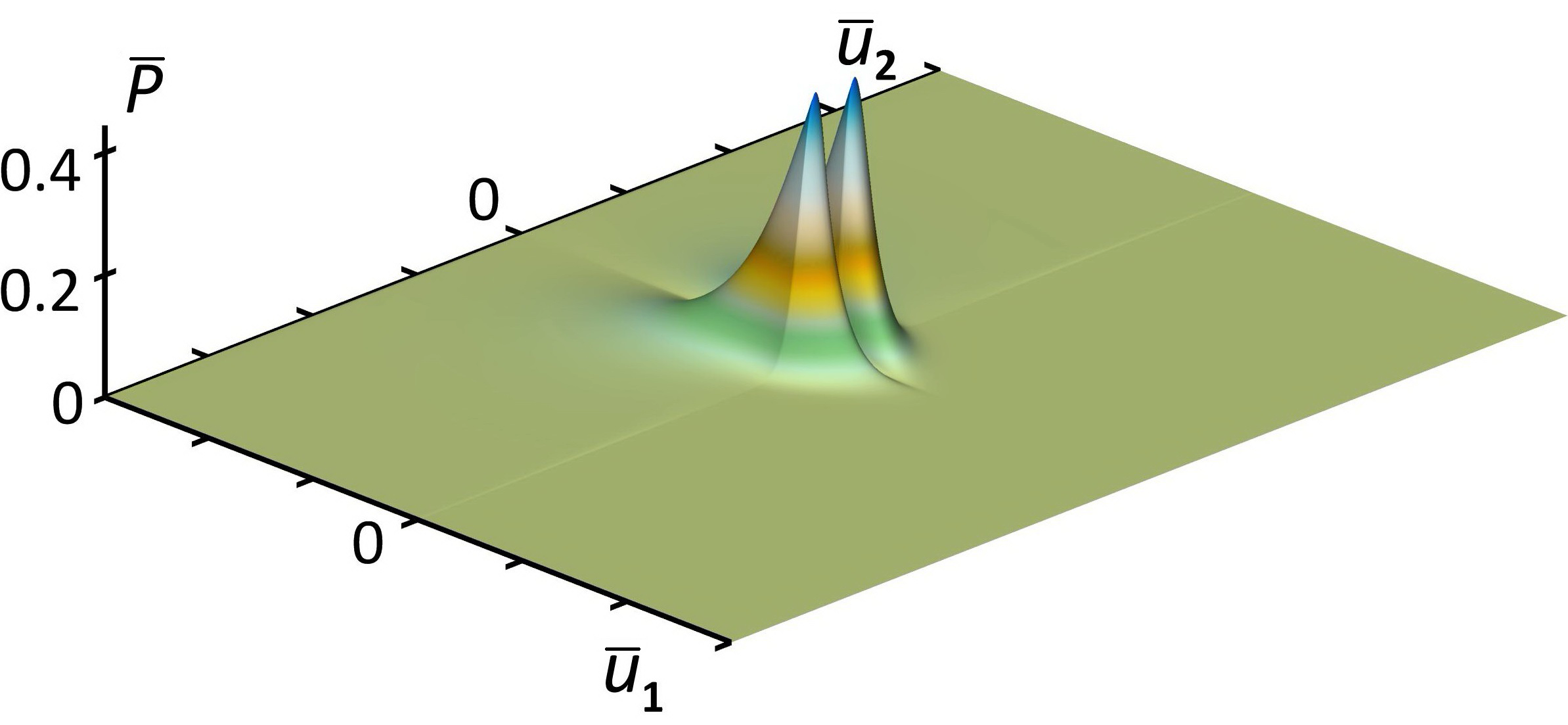}
\caption{\emph{A series of graphs of the distribution of free fields of the environment at time points $t_1=1.5,\,t_2=10,\,\,t_3=20$. Recall that we used
the data of the second line \textbf{Table}, which characterizes strongly elastic and weakly inelastic processes occurring in the environment.
An analysis of the graphs shows that the distribution tends to the stationary limit already at $t\sim 7. $}
\label{overflow}}
\end{figure}
\begin{figure}[ht!]
\centering
\includegraphics[width=52mm]{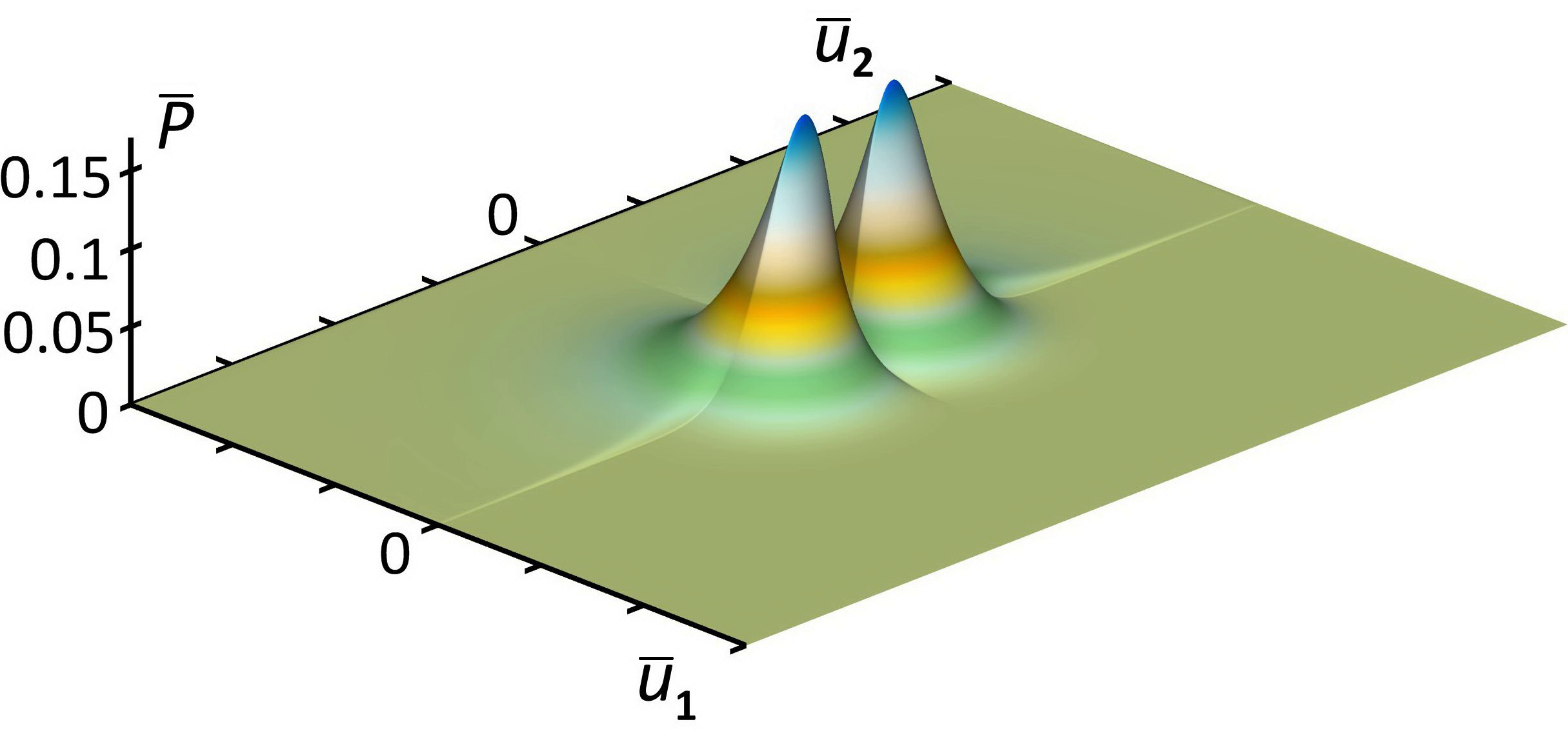}
\, \includegraphics[width=52mm]{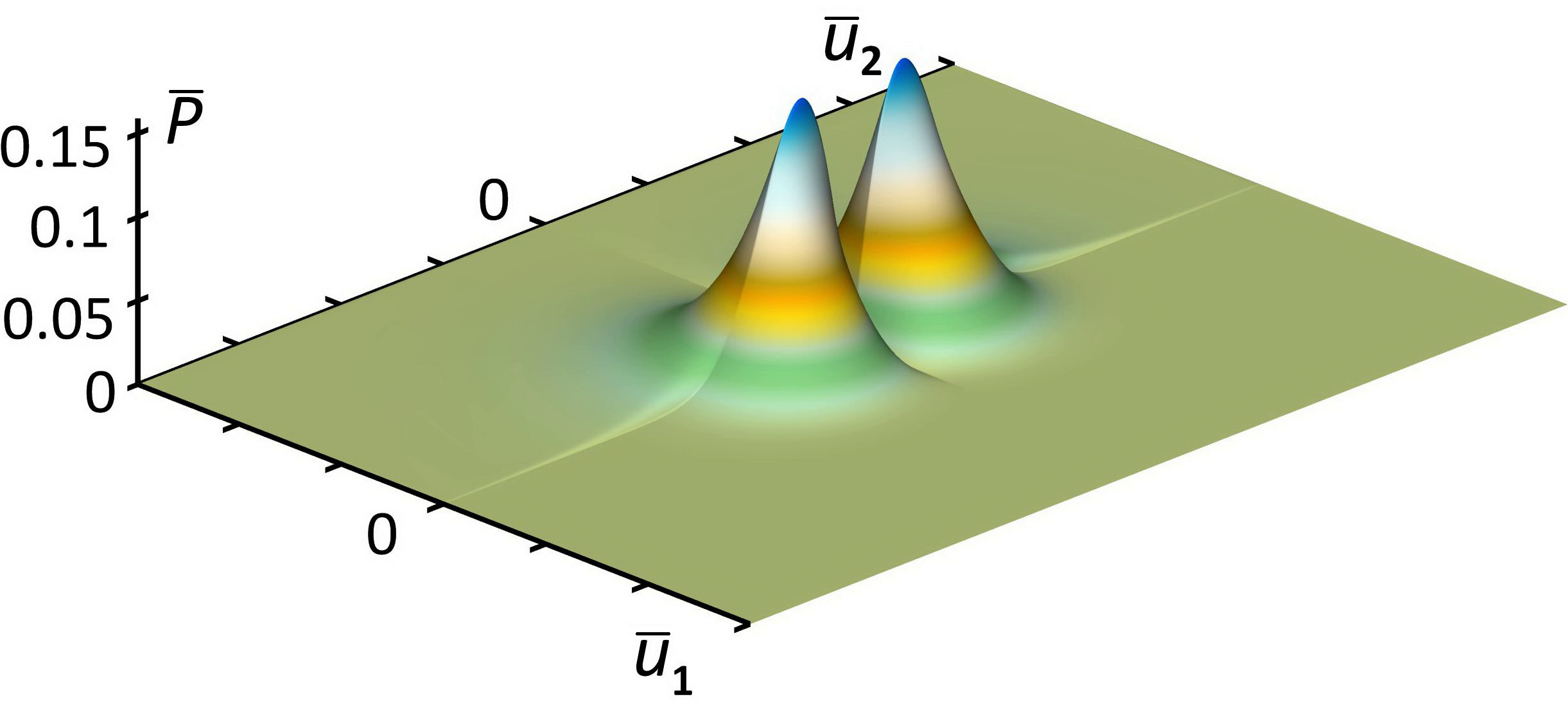}
\, \includegraphics[width=52mm]{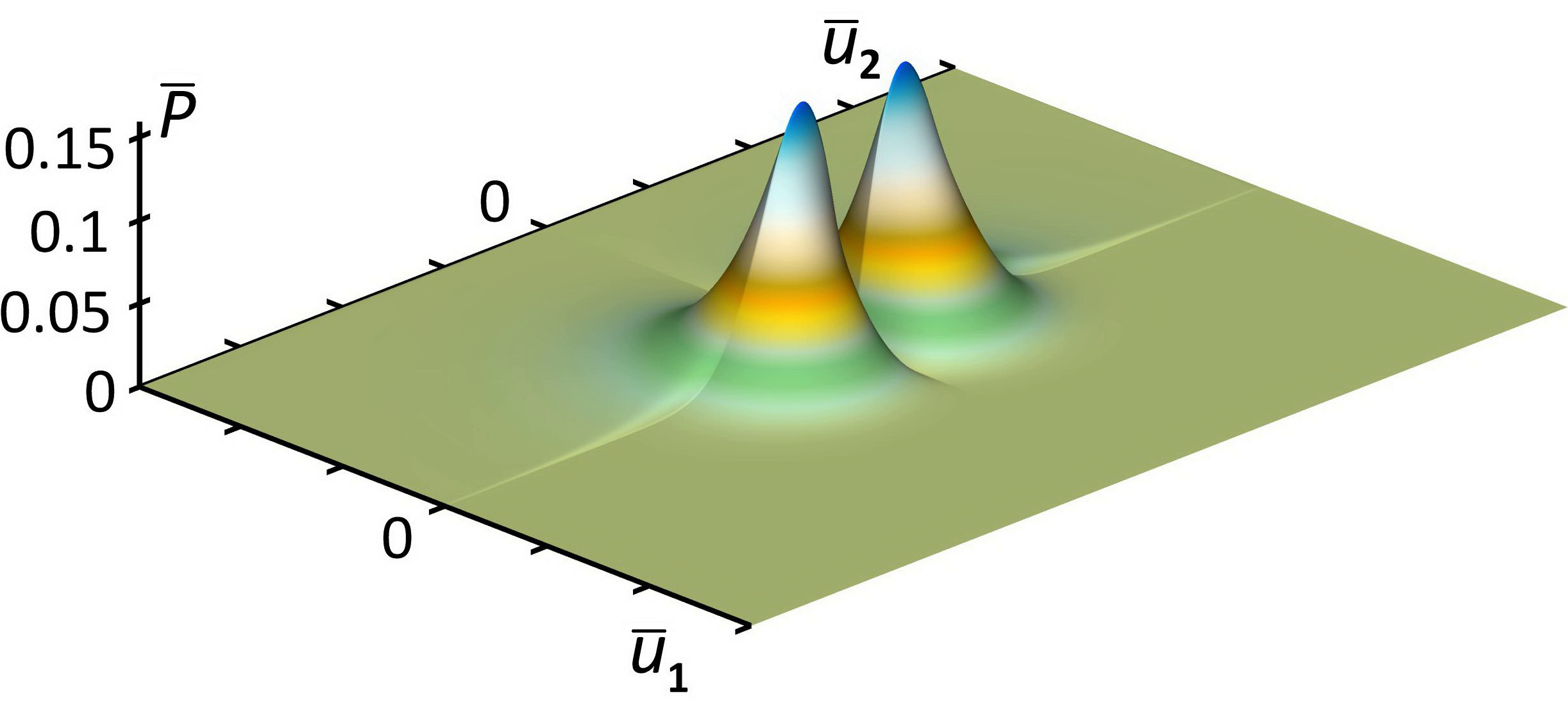}
\caption{\emph{Distributions of free  fields of the environment, respectively, at the time points $t_1=1.5,\,\,t_2=3,\,\,t_3=10$ and $t_4=20$. The data
of the third row \textbf{Table}, corresponding to strong elastic and inelastic processes occurring in the environment, were used for the calculation. As
can be seen from the figures, the greater the constants that determine the powers of elastic and inelastic processes, the faster the distribution of
environmental fields is established.}
\label{overflow}}
\end{figure}

\subsection{Distributions of environmental fields taking into account the influence of the oscillator}
We now present figures illustrating the evolution of the normalized functions $\bar{Q}^{(r)}(u_1,u_2,t)$ and $\bar{Q}^{(i)}(u_1,u_2,t)$
depending on time. Using the algorithm developed in \textbf{Listing 2}, we can calculate and visualize all of these solutions. Recall that it
is these functions that are responsible for the release of a small environment from the thermostat, which is a structural formation
self-consistent with the oscillator. Below are represented distributions for three different cases.
\begin{figure}[ht!]
\centering
\includegraphics[width=52mm]{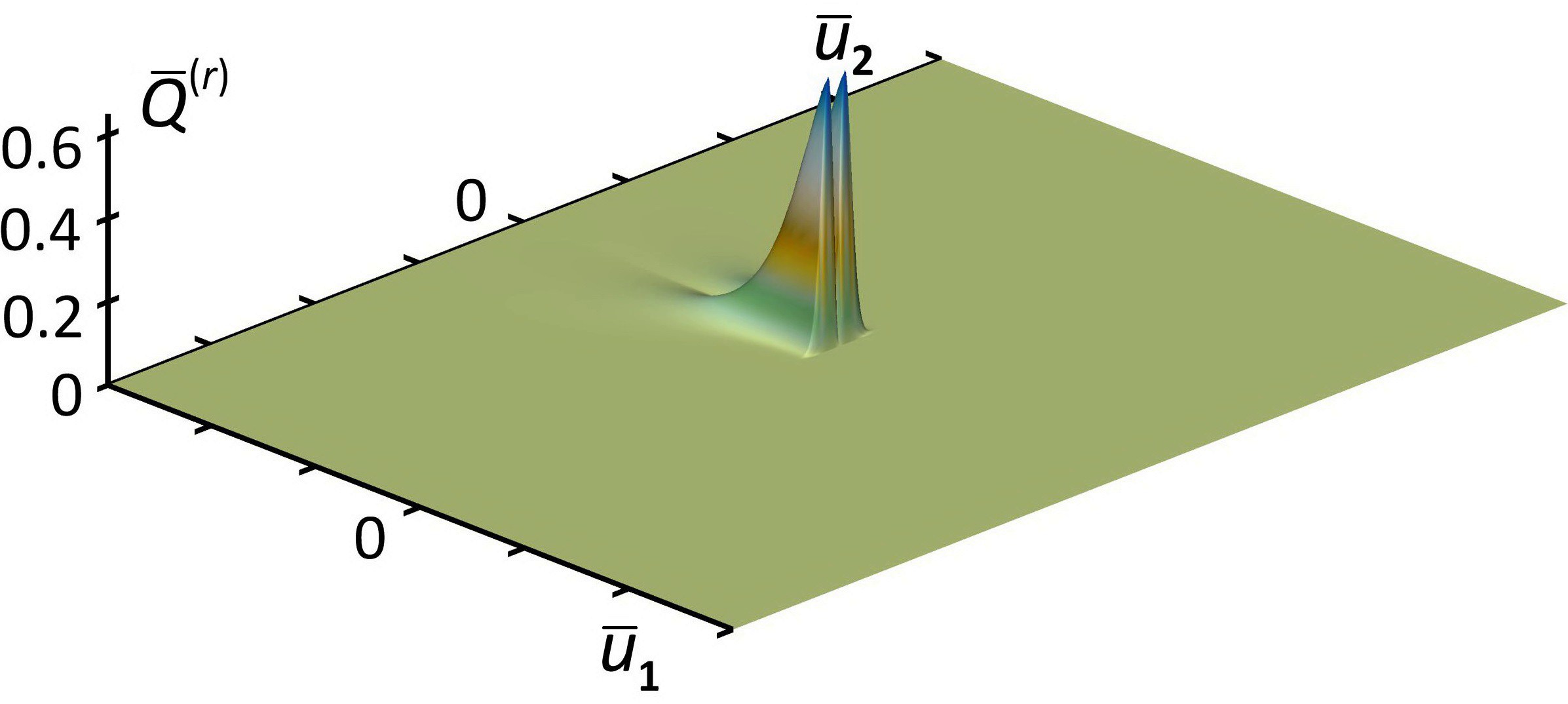}
\,\includegraphics[width=52mm]{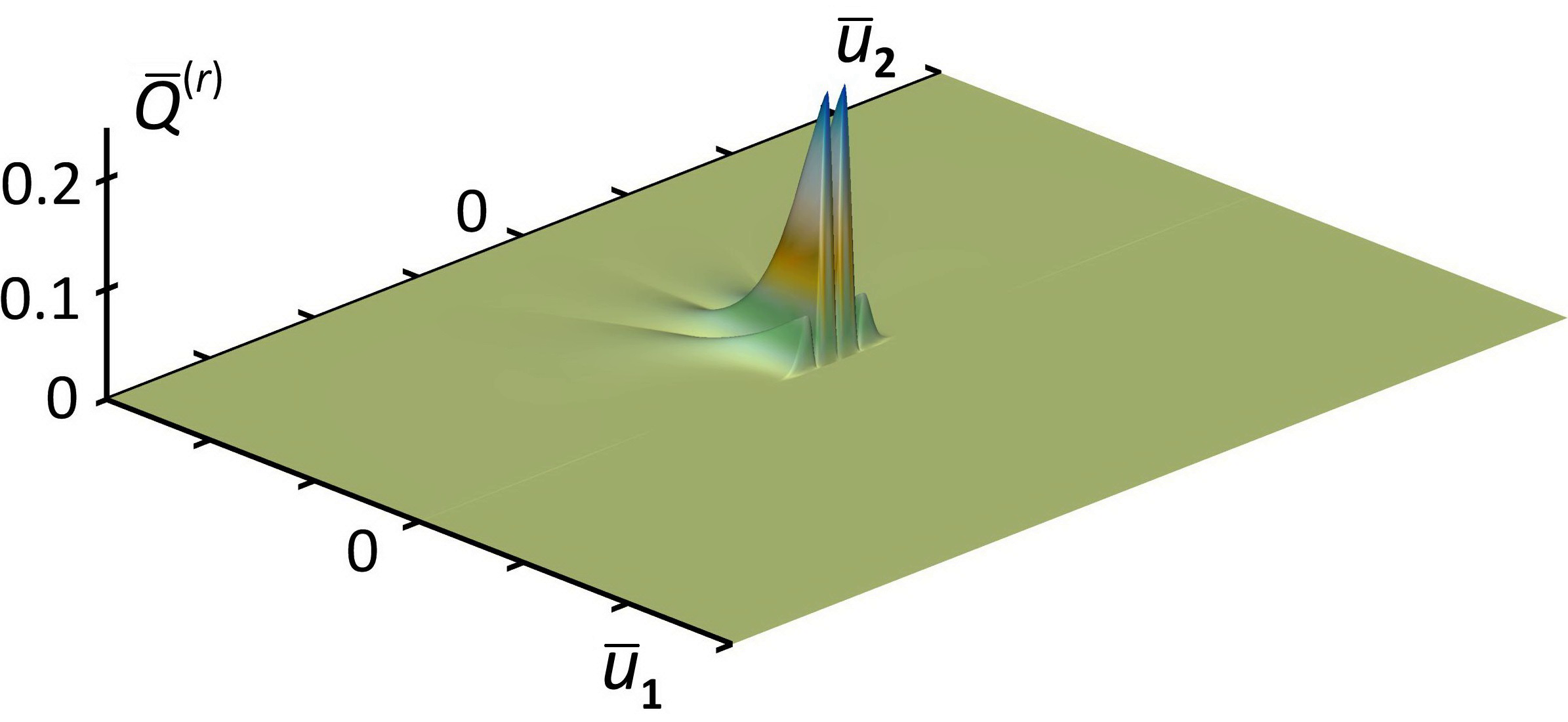}
\,\includegraphics[width=52mm]{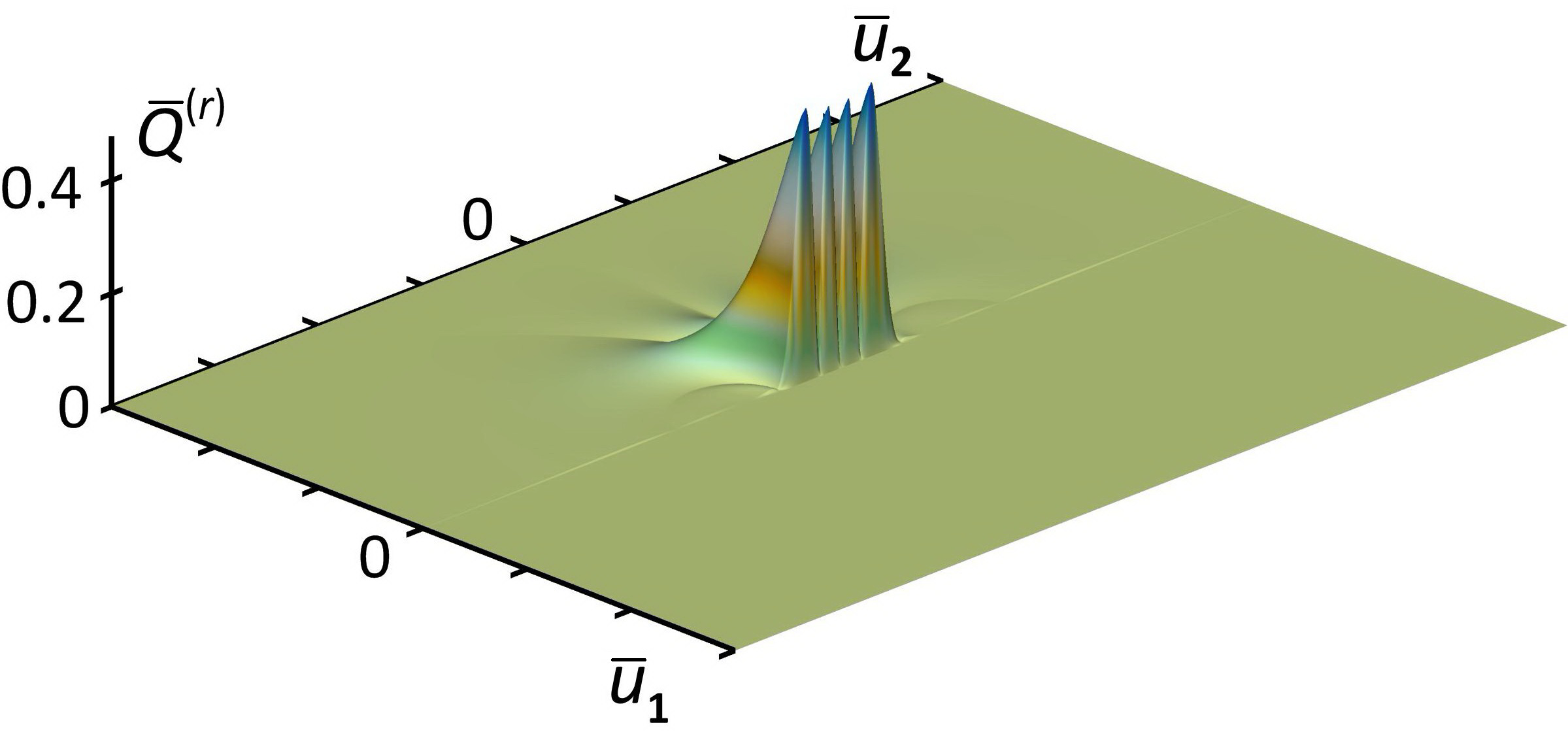}
\, \includegraphics[width=52mm]{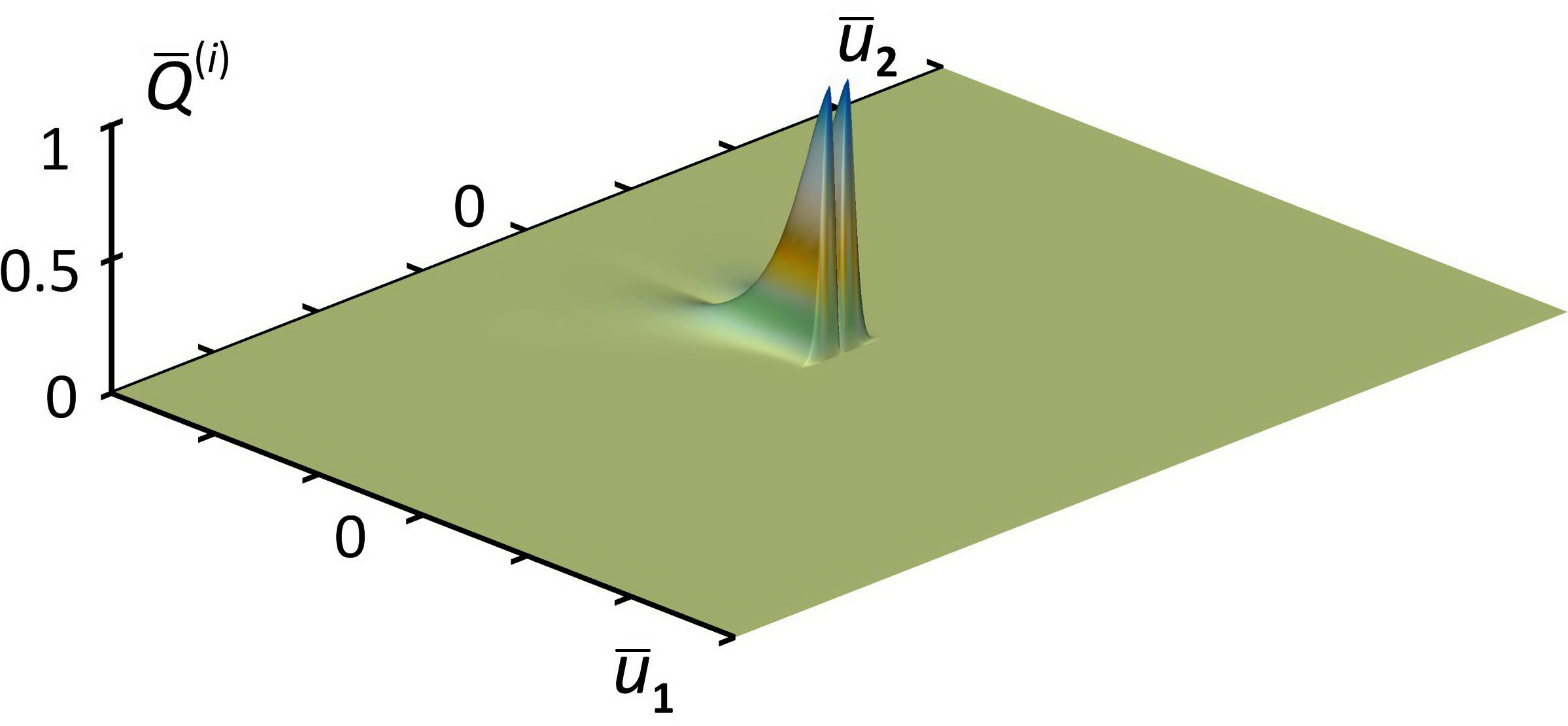}
\, \includegraphics[width=52mm]{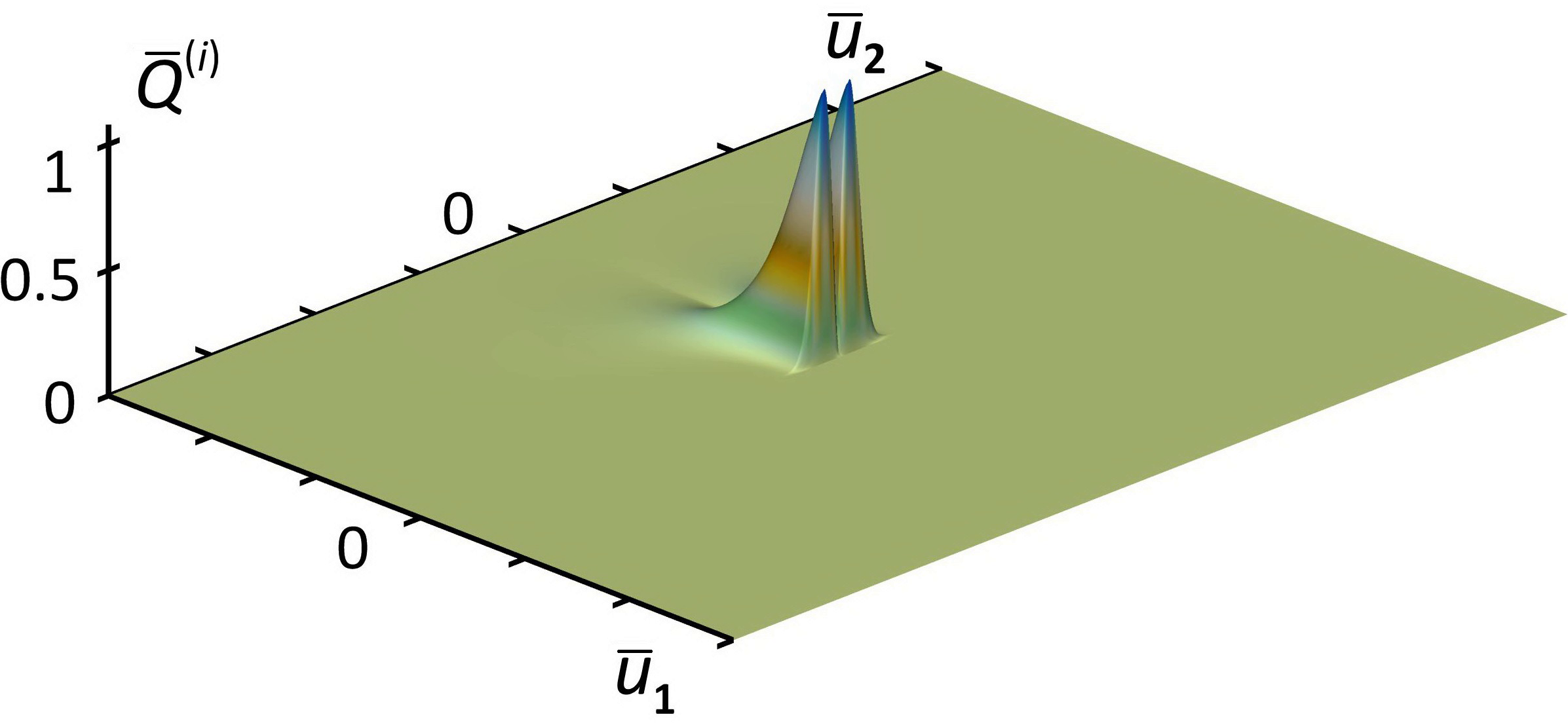}
\, \includegraphics[width=52mm]{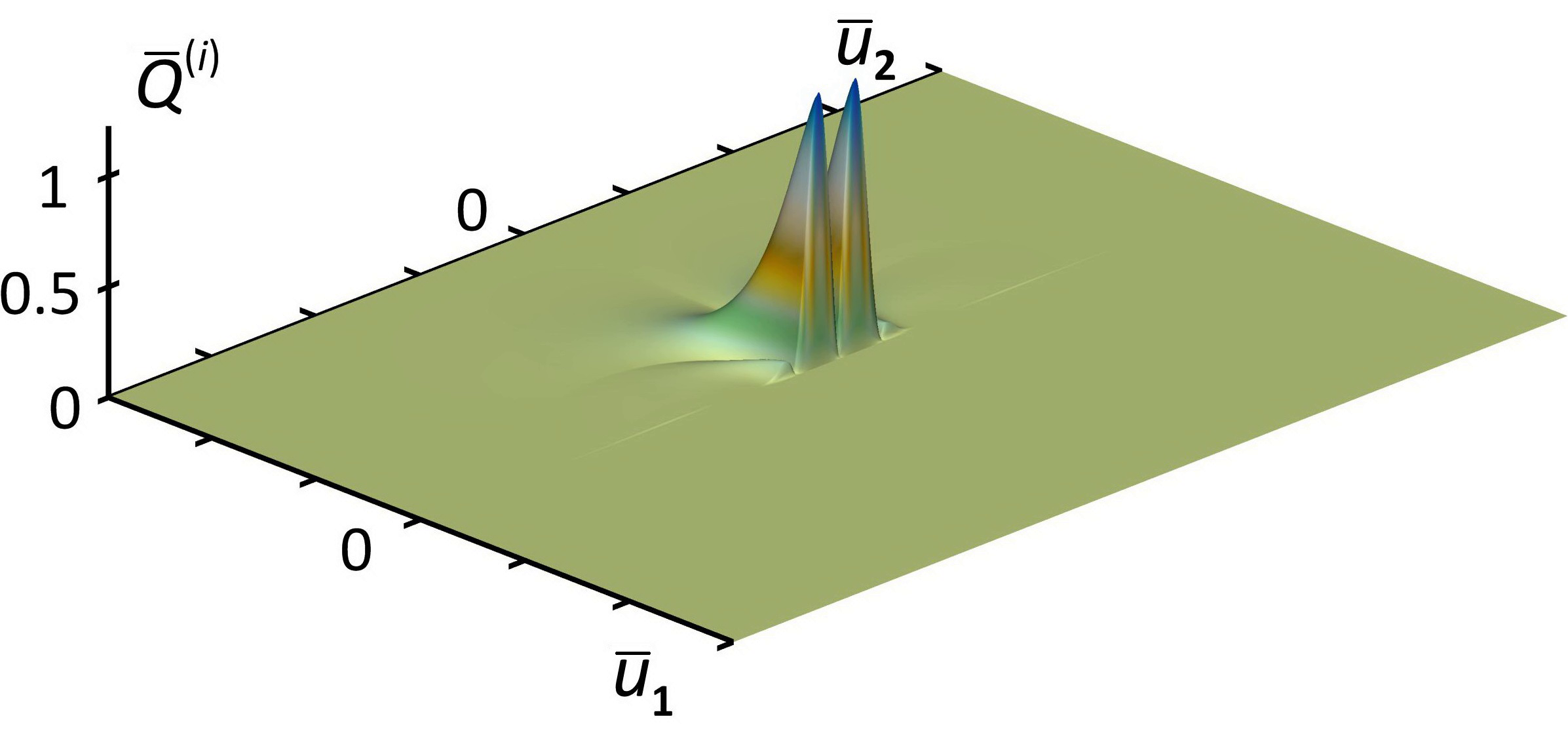}
\caption{\emph{Distributions of the environment fields in the process of evolution, respectively, at the time points $t_1=1.5,\,t_2=10$
and $t_3=20$, calculated from the data of the first line \textbf{Table}.}
\label{overflow}}
\end{figure}
 \begin{figure}[ht!]
\centering
\includegraphics[width=52mm]{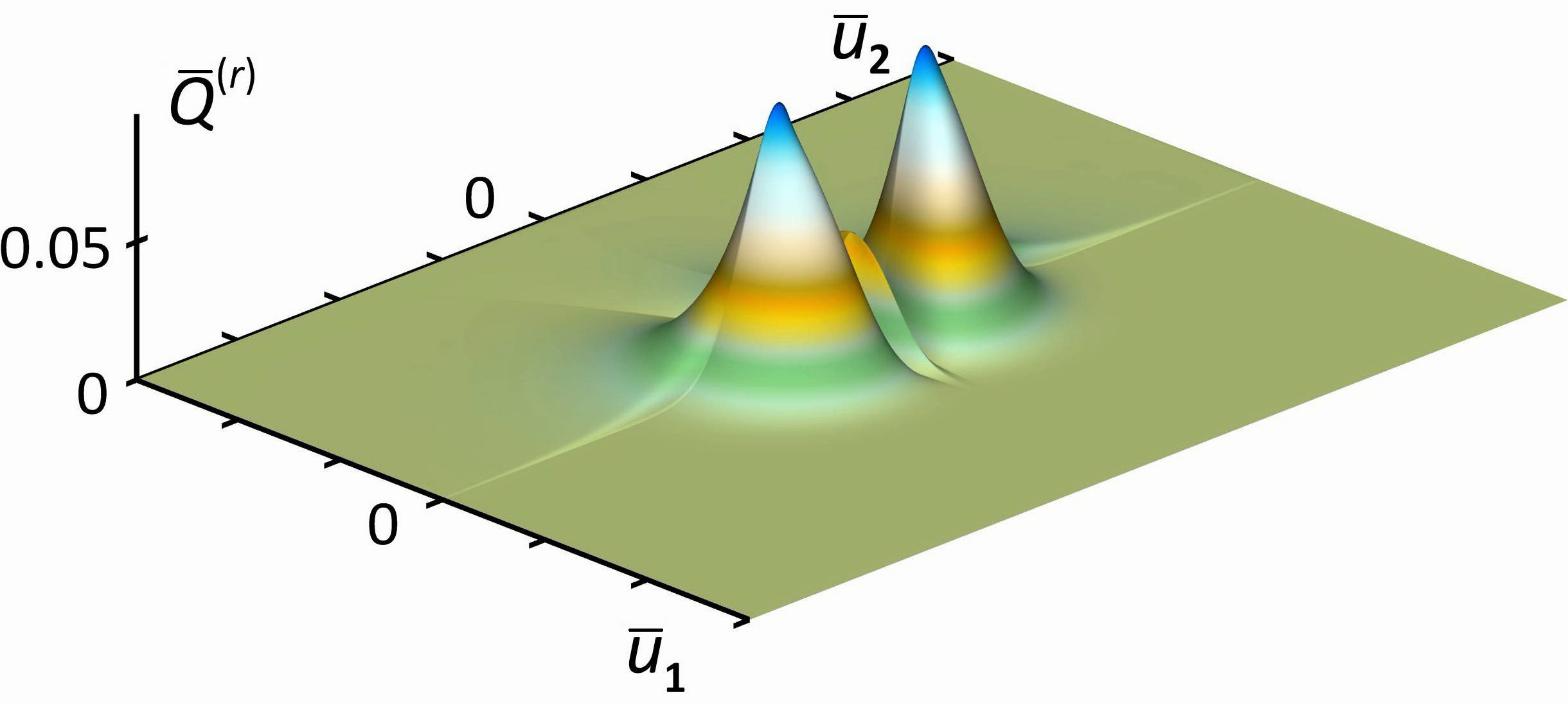}
\,\includegraphics[width=52mm]{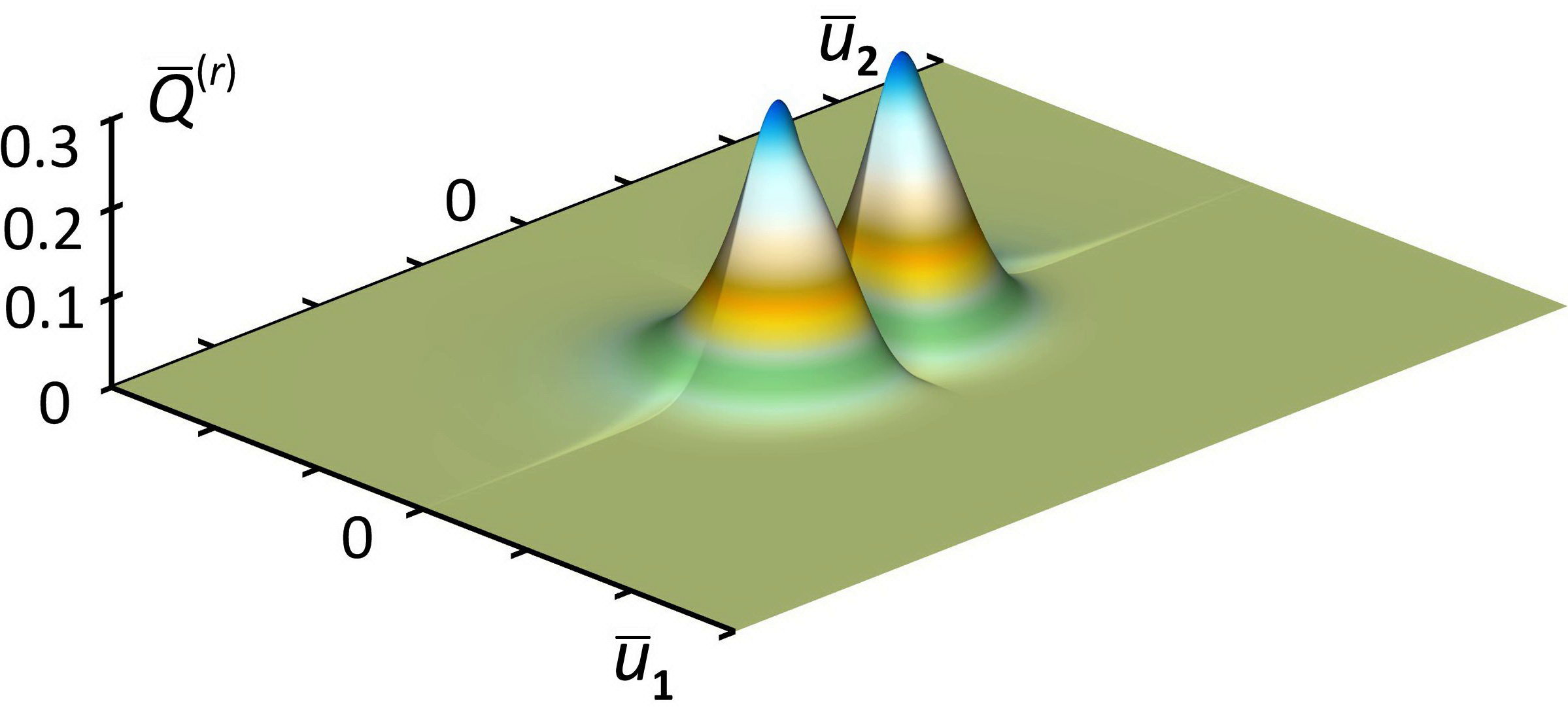}
\,\includegraphics[width=52mm]{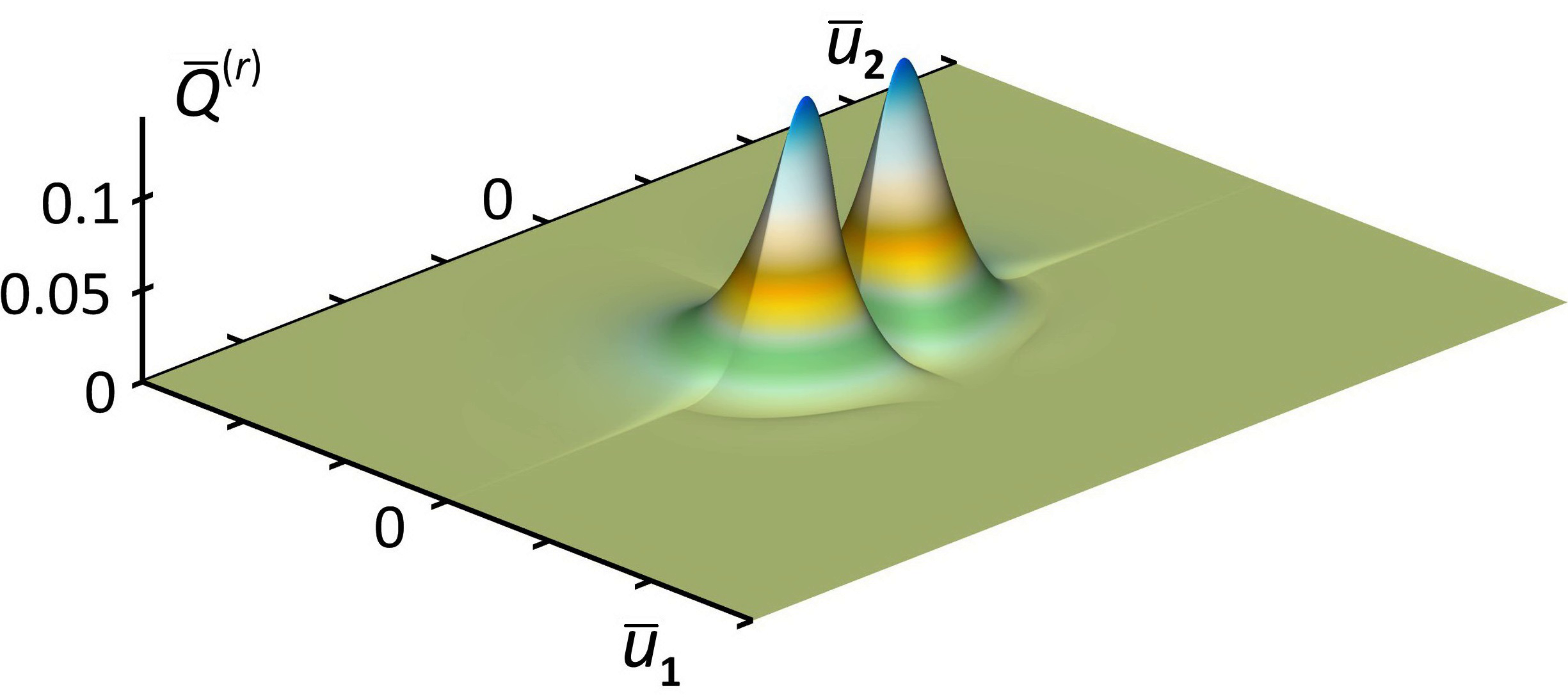}
\,\includegraphics[width=52mm]{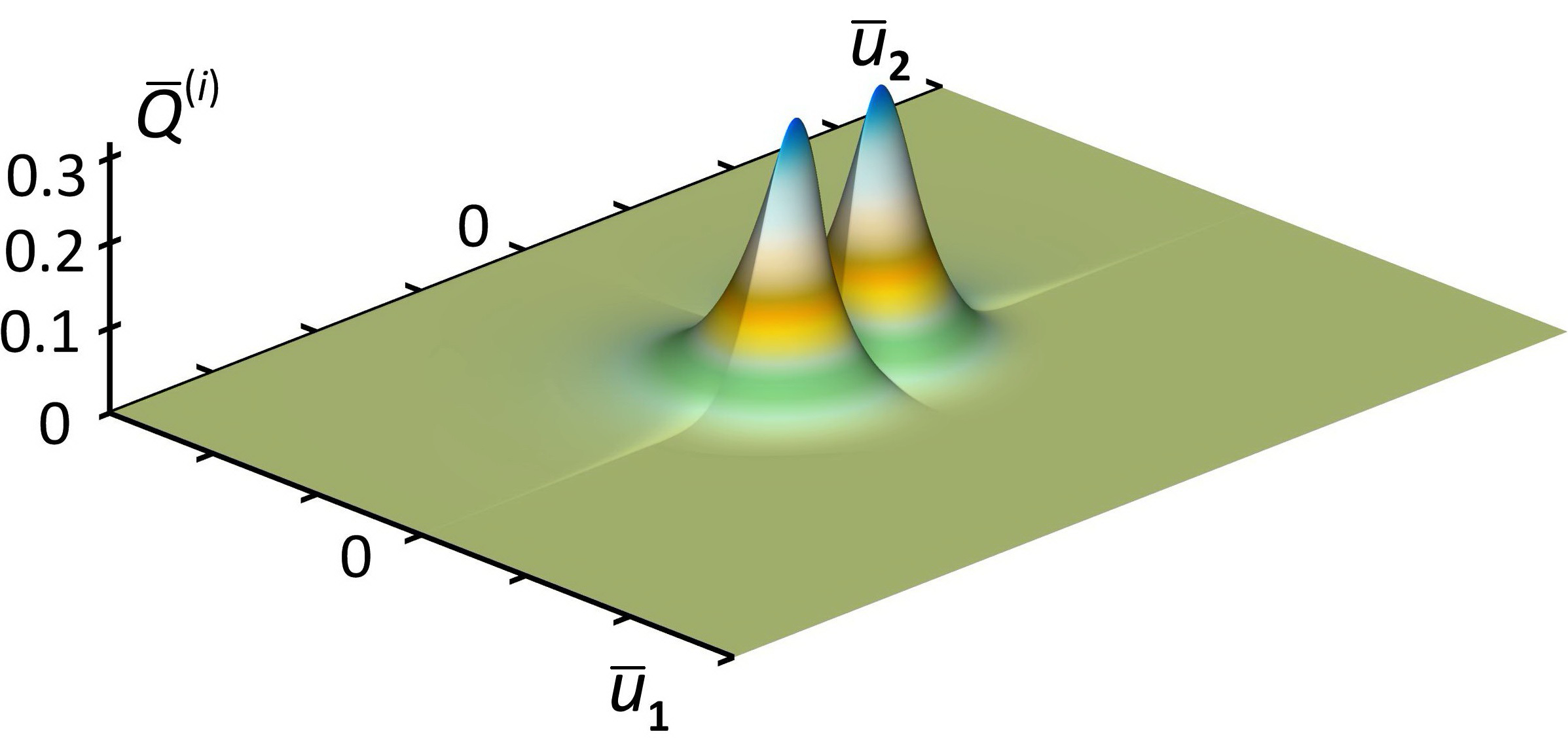}
\,\includegraphics[width=52mm]{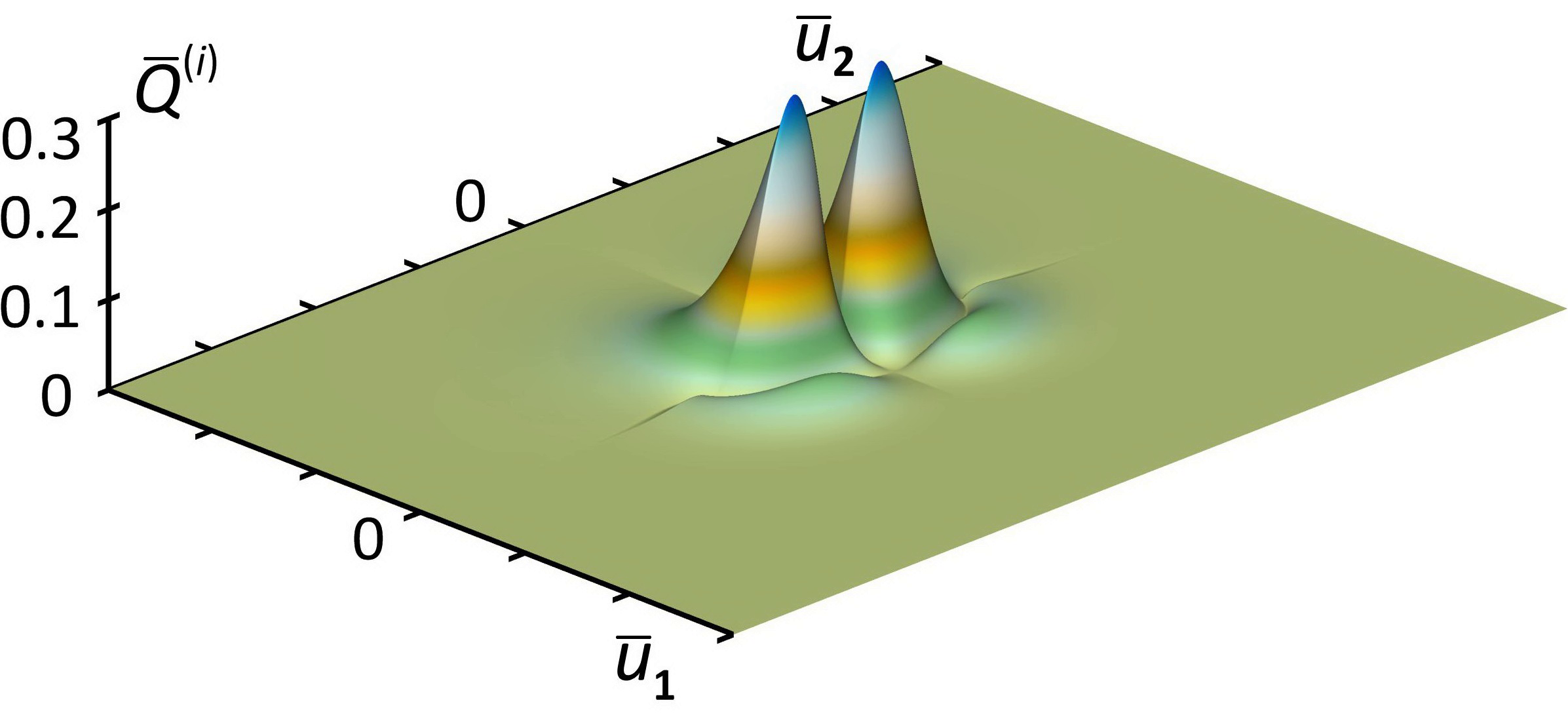}
\,\includegraphics[width=52mm]{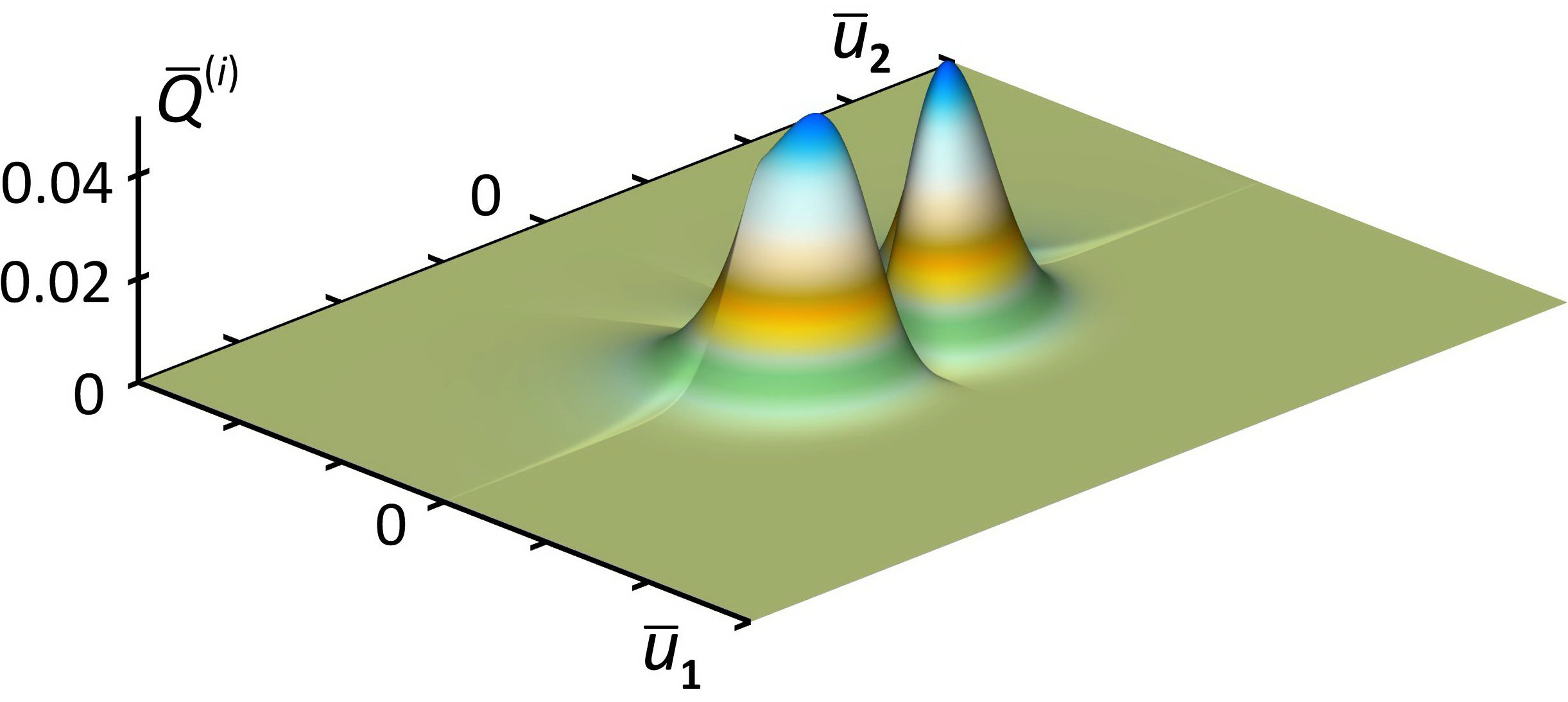}
\caption{\emph{Distributions of the fields of the environment in the course of evolution, respectively, at the time points $t_1=1.5,\,t_2=10$
and $t_3=20$, calculated using the third line of the {\textbf{Table}}. }
\label{overflow}}
\end{figure}
\begin{figure}[ht!]
\centering
\includegraphics[width=52mm]{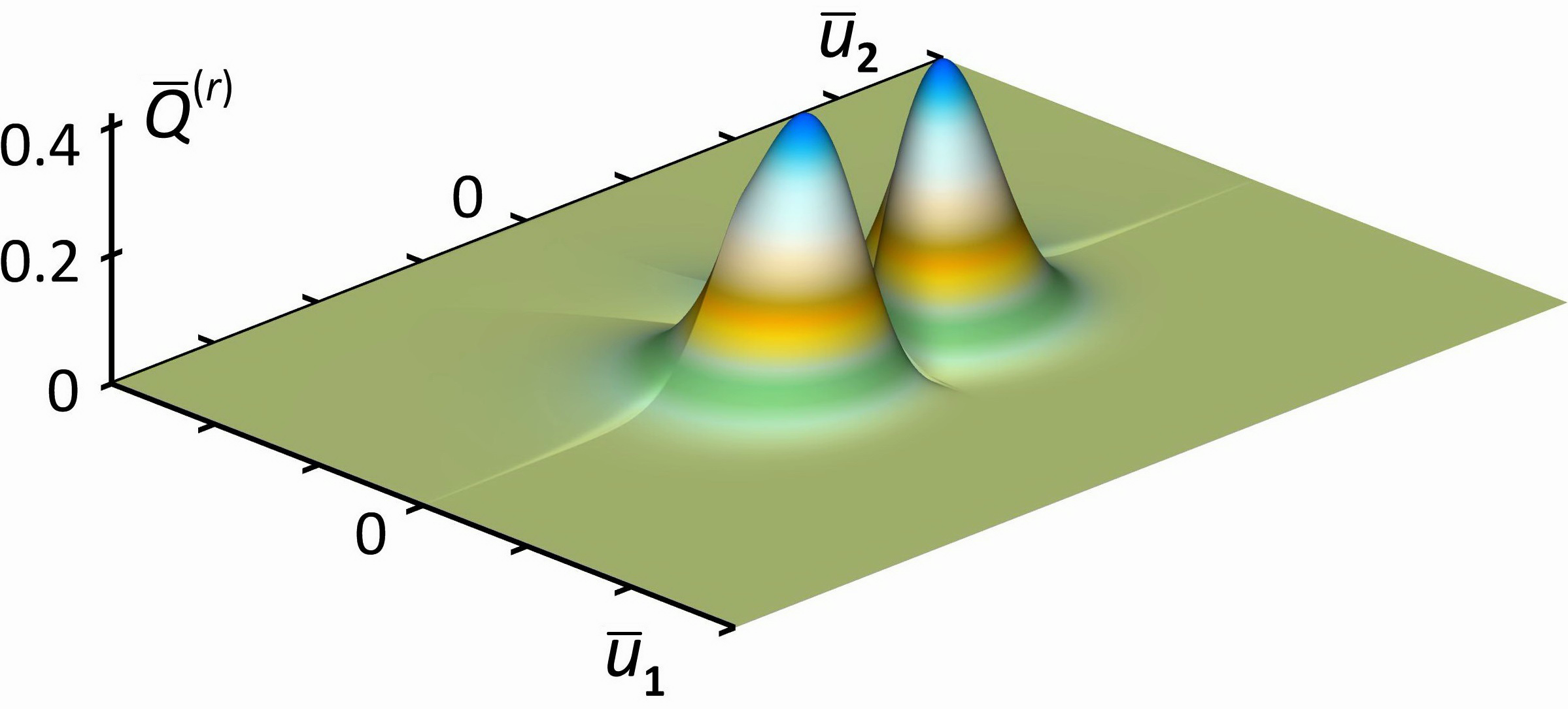}
\,\includegraphics[width=52mm]{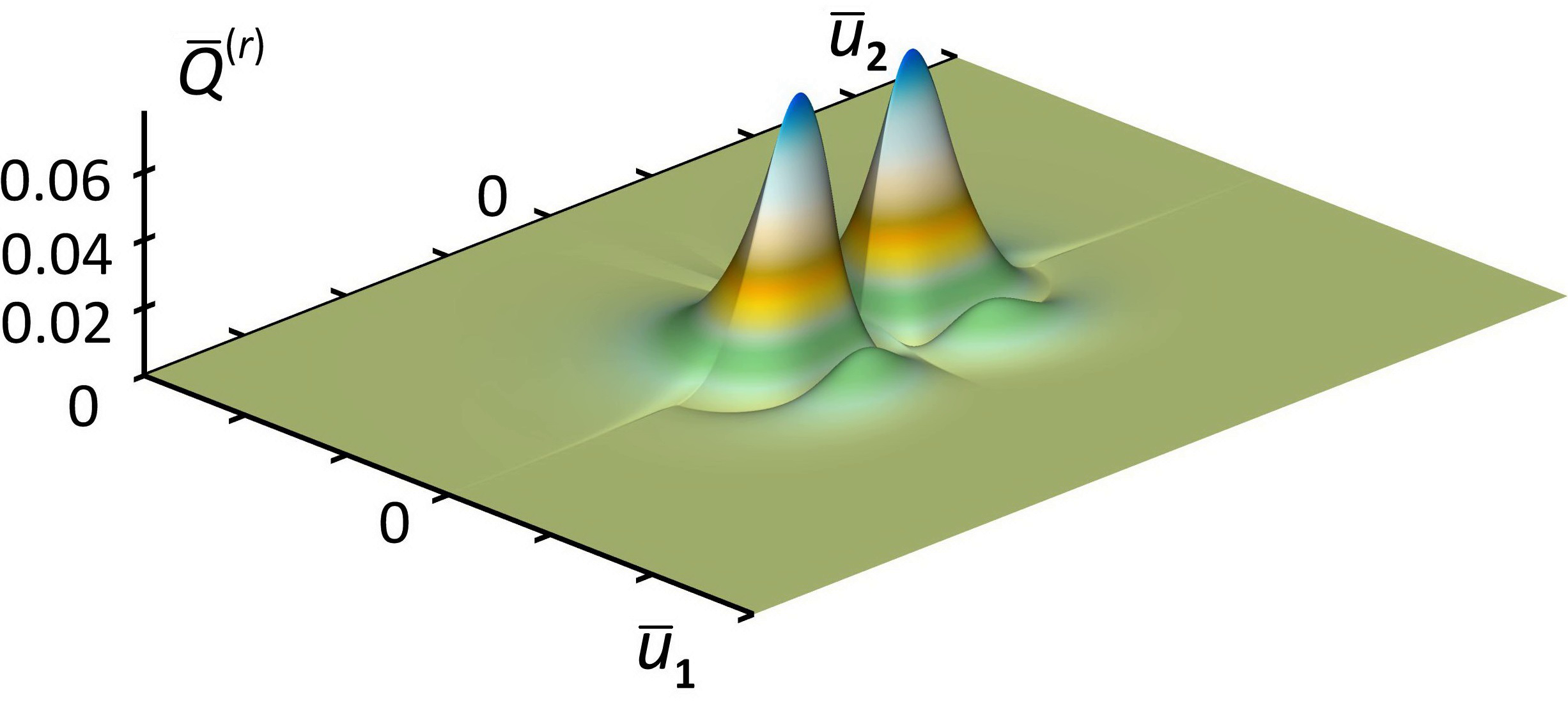}
\,\includegraphics[width=52mm]{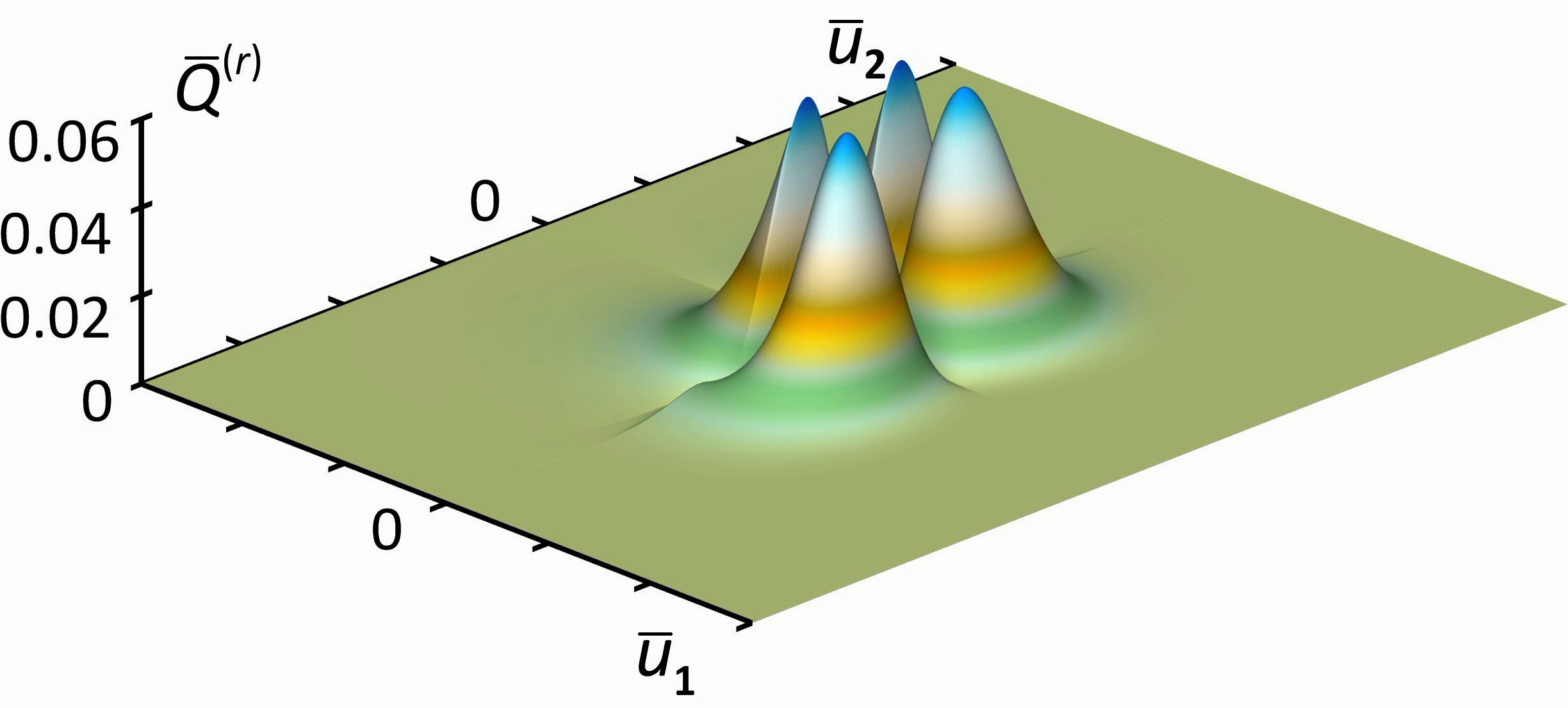}
\,\includegraphics[width=52mm]{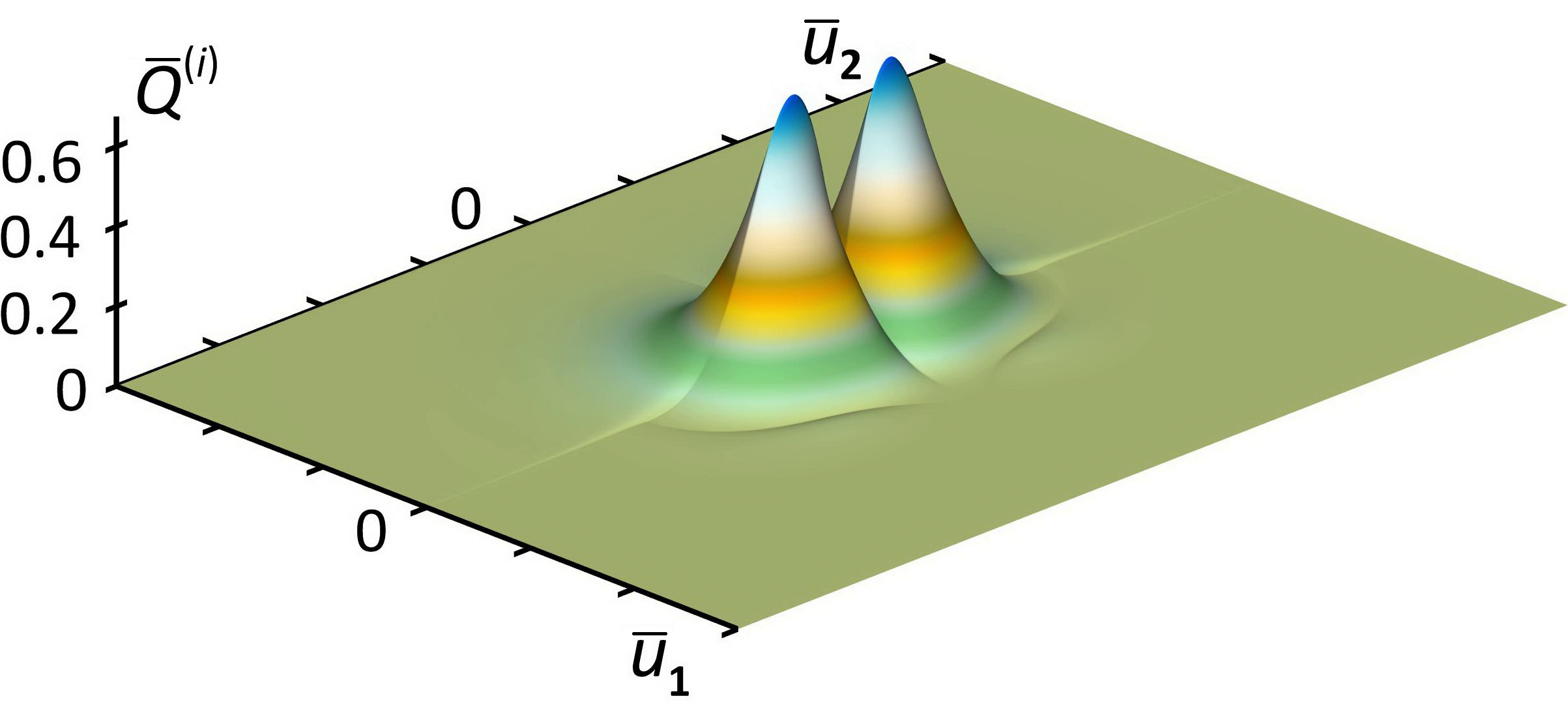}
\,\includegraphics[width=52mm]{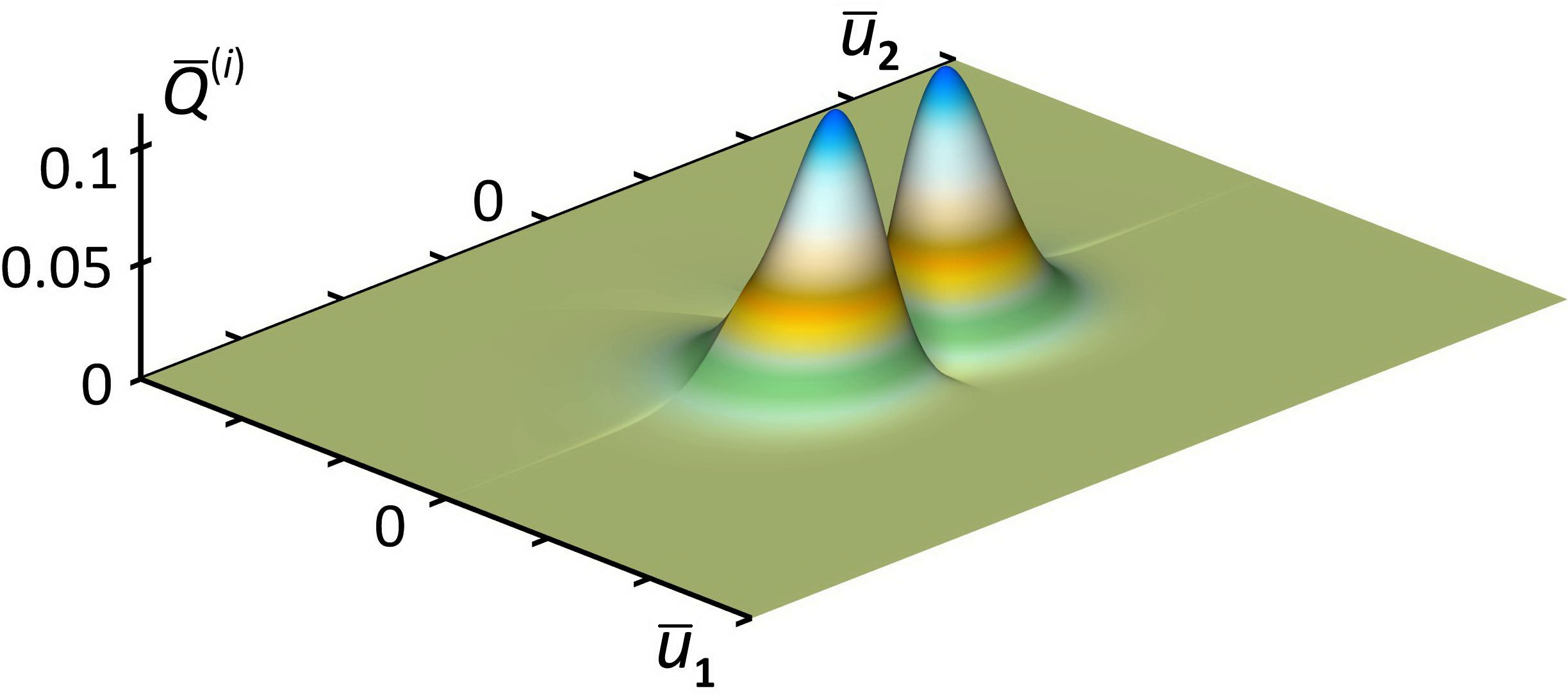}
\,\includegraphics[width=52mm]{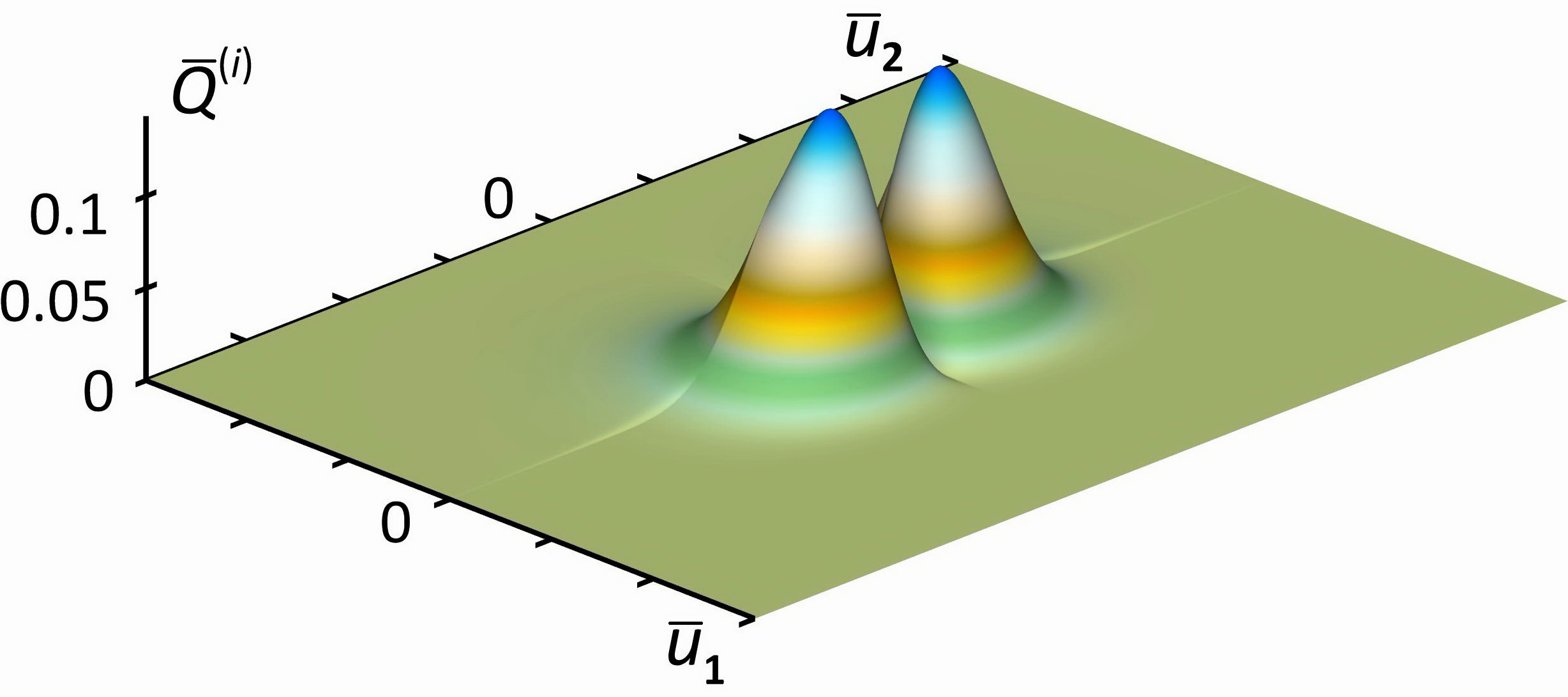}
\caption{\emph{Distributions of environmental fields in the process of evolution, respectively, at time points $t_1=1.5,\, t_2=10$ and $t_2=20$.
The calculations are performed for the case $ \varepsilon^{(r )}=\varepsilon^{(i)}=1$, when processes, both elastic and inelastic,
are strongly developed.}
\label{overflow}}
\end{figure}
In particular, as the analysis of these figures shows, with an increase in the constants of interaction with the environment, the
time for the establishment of distributions is steadily reduced.

 \subsection{Mathematical expectation of the oscillator trajectory}
Using all the above calculations, one can simulate the expected value of the oscillator trajectory in the absence of an external field based on
the equation (\ref{4.0nw6i}). We modeled the expected oscillator trajectory for both its real and imaginary parts for five different
environmental states and visualized their behavior as a function of time (see FIG 7). In particular, as can be seen from the figures, in all
the cases under consideration, the mathematical expectation of the trajectory not only has a non-trivial oscillatory character, but it decreases
with time, taking both positive and negative values.
\begin{figure}[ht!]
\centering
\includegraphics[width=50mm]{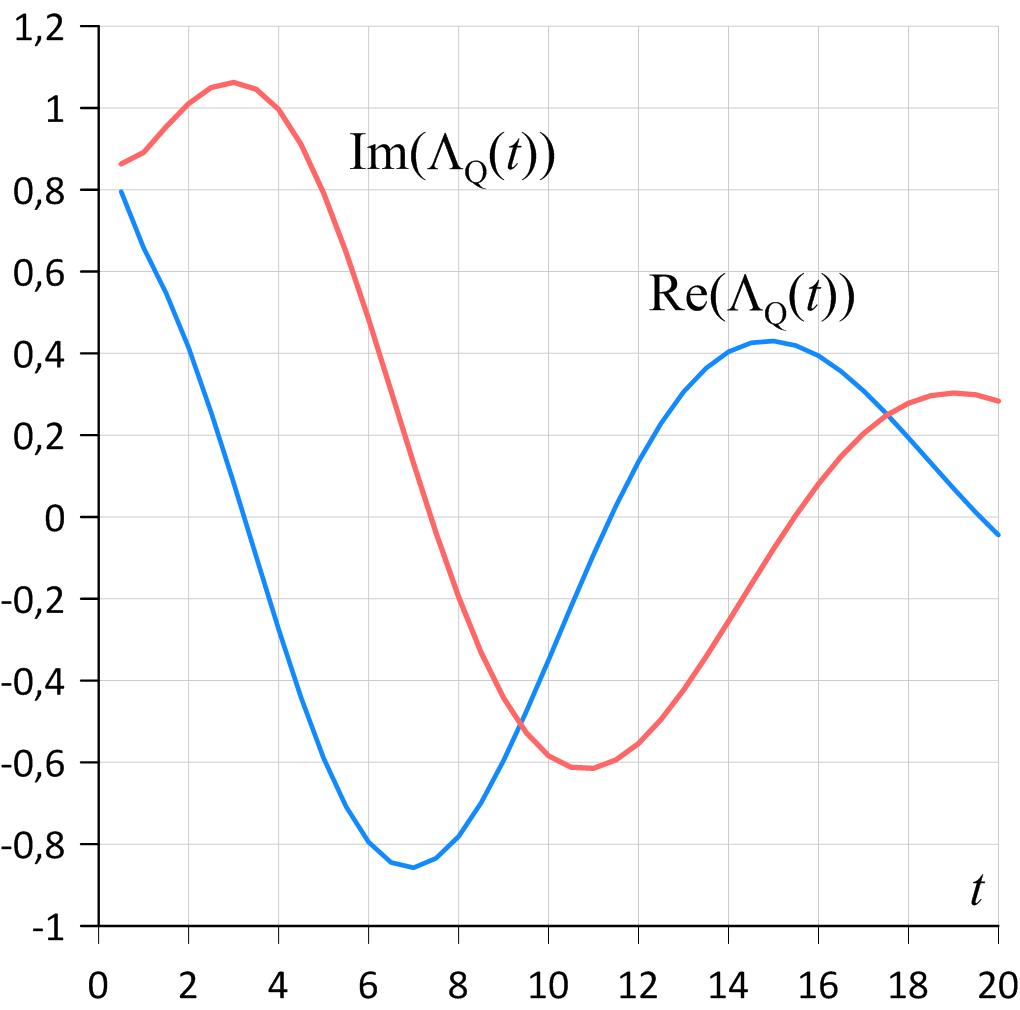}
\,\,\,
 \includegraphics[width=50mm]{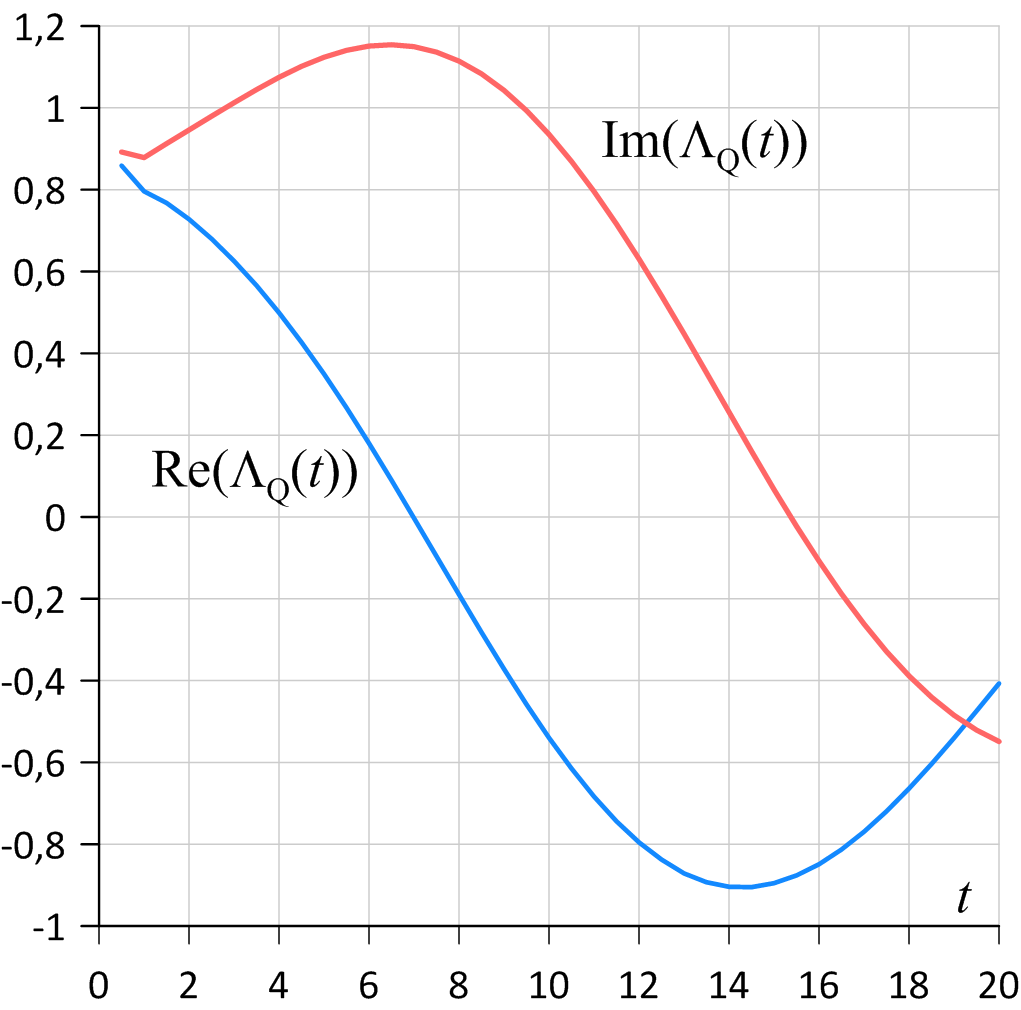}
\,\,\,
 \includegraphics[width=50mm]{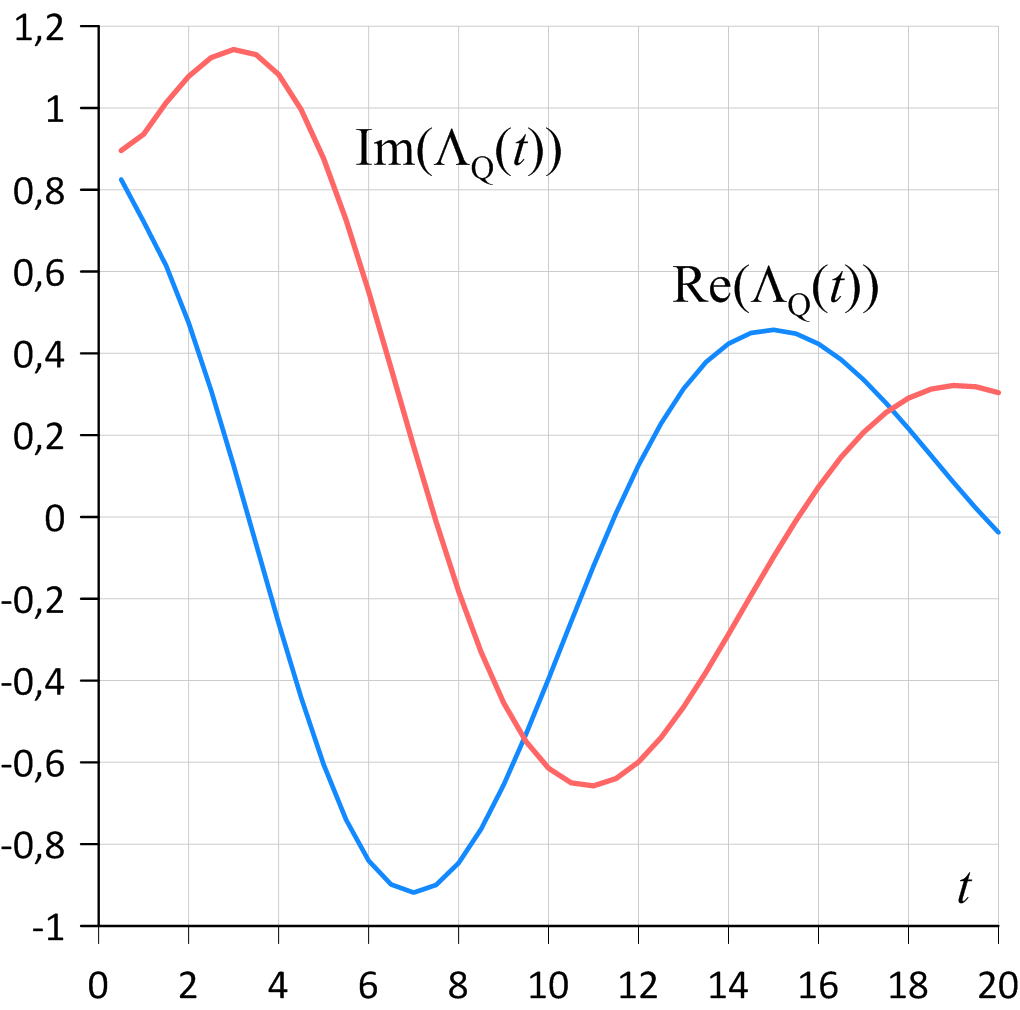}
\,\,\,
\includegraphics[width=50mm]{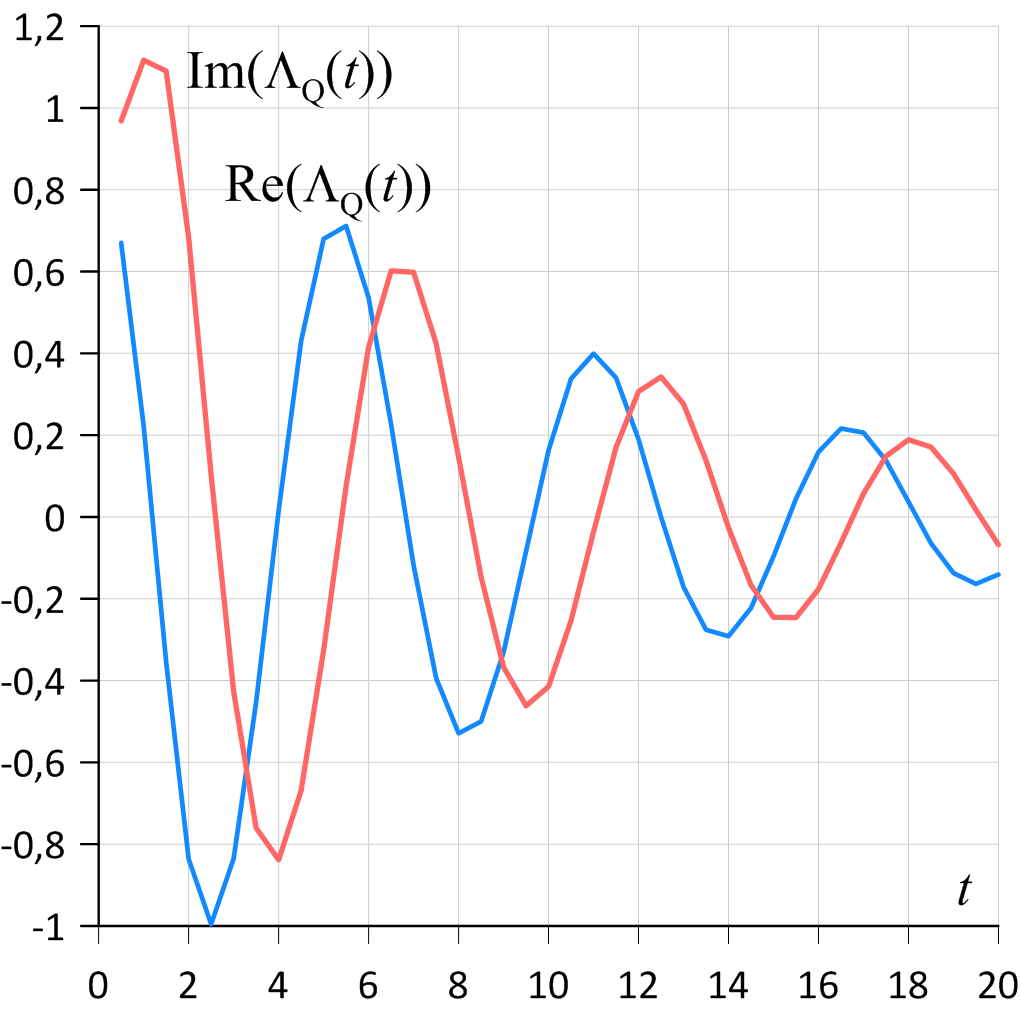}
\qquad
 \includegraphics[width=50mm]{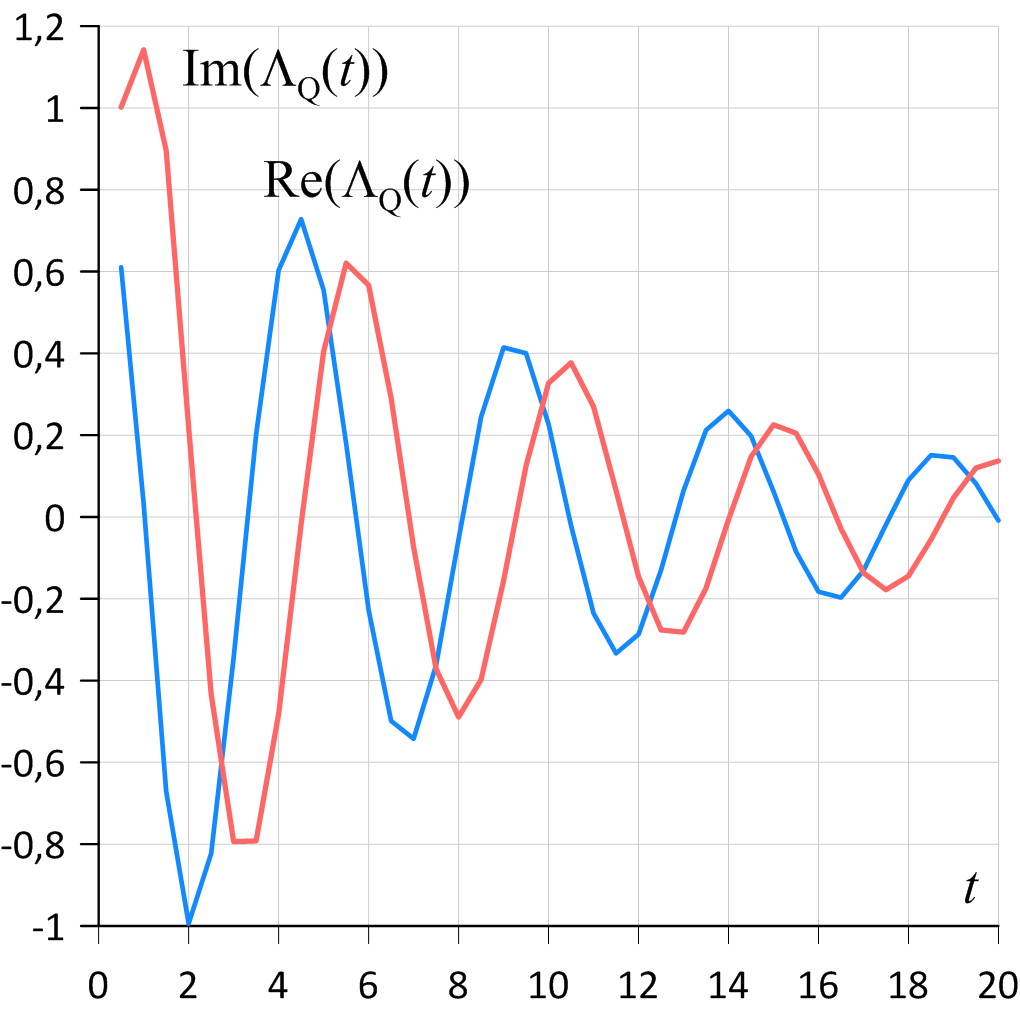}
\caption{\emph{Mathematical expectations of the real (blue) and imaginary (red) parts of the oscillator trajectory for five sets of parameters.
Left to right, front row; $(\varepsilon^{(r)}=\varepsilon^{(i)}=0.01),\,(\varepsilon^{(r)}=1,\,\, \varepsilon^{(i) }= 0.001)$ and $(\varepsilon^{(r)}=1,\,\,
\varepsilon^{(i)}=0,01)$, in the second row; $(\varepsilon^{(r)}=1,\, \,\varepsilon^{(i)}=0.5)$ and $(\varepsilon^{(r)}= \varepsilon^{(i) }=1)$.}
\label{overflow}}
\end{figure}

\subsection{Calculation of topological and geometric features of the manifold $\Sigma^{(2)}_{\bf{u}}(t)$}

As we saw above, the off-diagonal term of the metric tensor $y=g^{12}(u_1,u_2,t)$ of a manifold $\Sigma^{(2)}_{\bf{u}}(t)$ is
determined by the algebraic equation of the 4\emph{th} degree (\ref{nw3z.0zf3}) with coordinates- and time-dependent coefficients.
Computing this equation with the Mathematics - Wolfram solver for three sets of parameters (see data {\textbf{Tables}}), we obtain   sets of surfaces for
the off-diagonal term of the metric tensor (see FIG 8-11), which allows us to study and understand geometric and topological features of the manifold
$\Sigma^{(2)}_{\bf{u}}(t)$.  An analysis of the surfaces (see FIG 8) shows that when the oscillator is immersed in an environment with weak elastic and inelastic processes, these surfaces do not have interesting topological features.  However, there is an obvious  singularity at the point $ u_1=0-\epsilon,\,
\,(u_1\in(-\infty,0]\setminus 0,\,\epsilon>0)$,  because as $\epsilon\to0$ the term of $g^{12}=y(u_1,u_2,t)$ tends to plus infinity in the upper half-space
 when as in the lower half-space the term $g^{21}=-y(u_1,u_2,t)$ tends to minus infinity. It should be noted that in the case under consideration, the
evolution of the environment characteristically does not change the geometry of space, but only shifts the minimum point of the surface.
\begin{figure}[ht!]
\centering
\includegraphics[width=65mm]{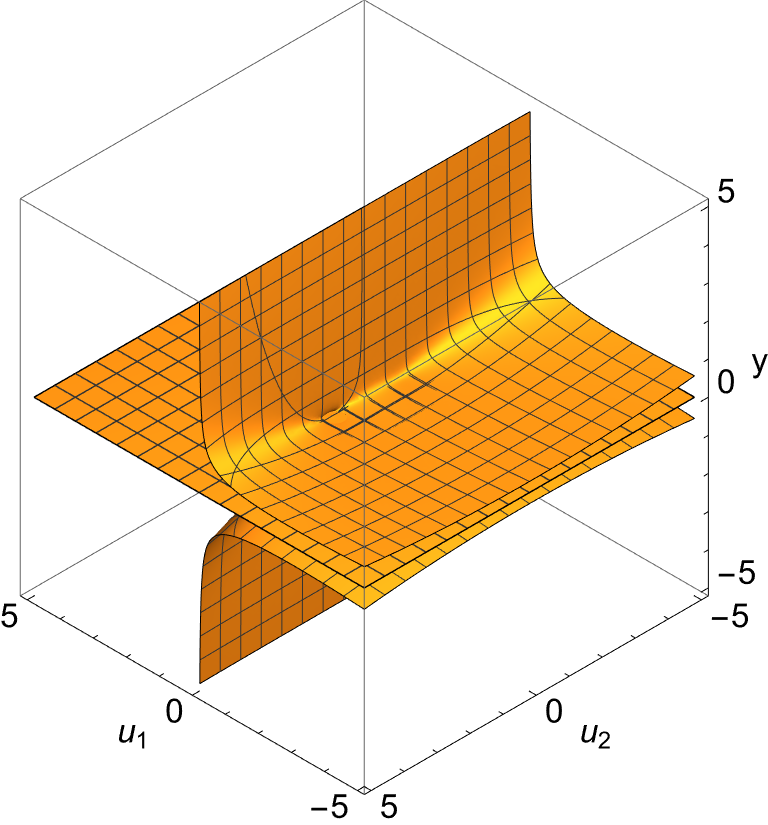}
\quad\,\,\includegraphics[width=65mm]{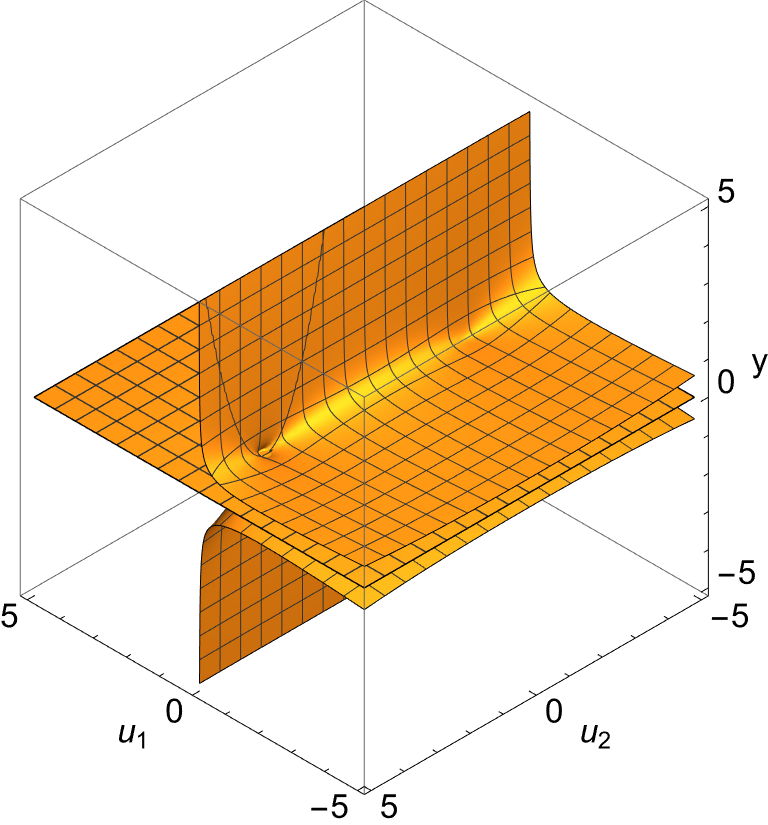}
\caption{\emph{The two figures (from left to right) show the evolution of the space $\Sigma^{(2)}_{\bf{u}}(t)$ from the asymptotic state $(in)$ with
 the environment data $ \bigl(\varepsilon^{( r)}=\varepsilon^{(i)}=0.01\bigr)$ and frequency $\Omega_0^-=1$, to the asymptotic state $(out)$ with
frequency $\Omega_0 ^+=3$. }
\label{overflow}}
\end{figure}
\begin{figure}[ht!]
\centering
\includegraphics[width=65mm]{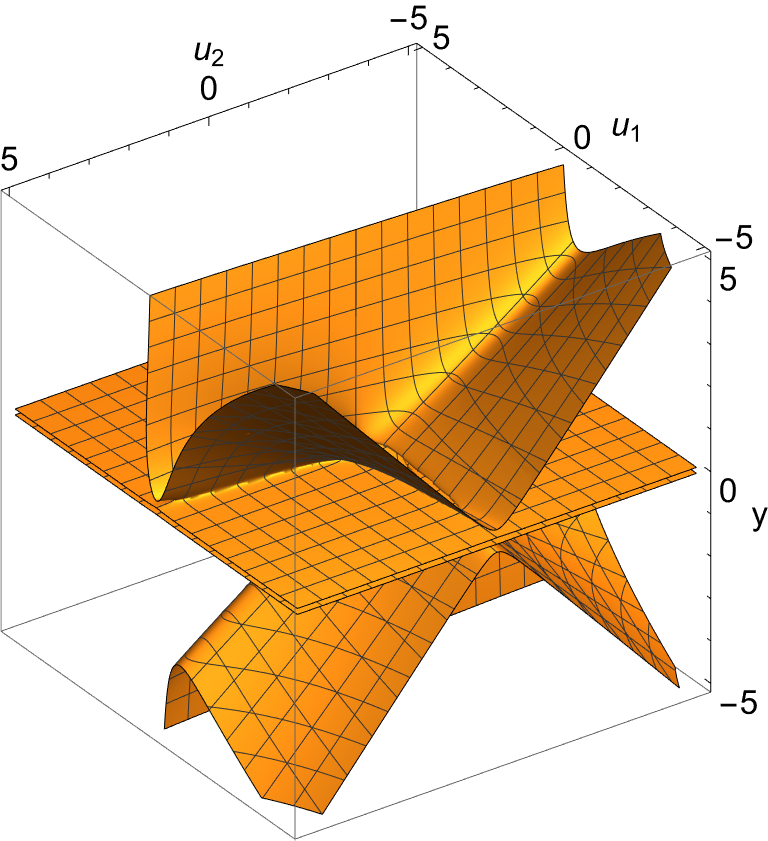}
\quad\,\,
\includegraphics[width=65mm]{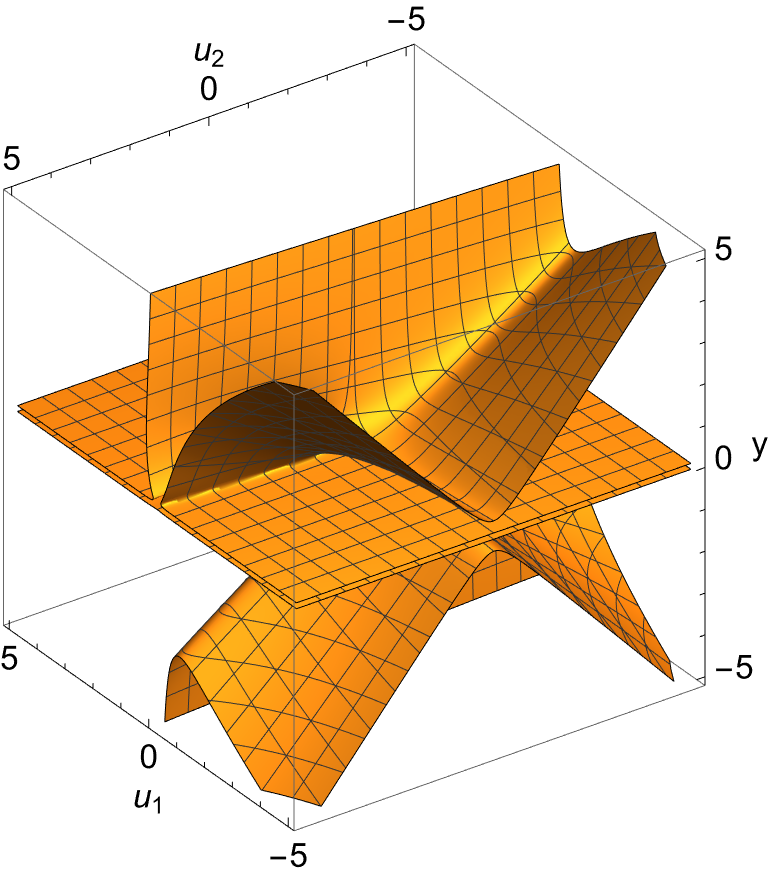}
\includegraphics[width=54mm]{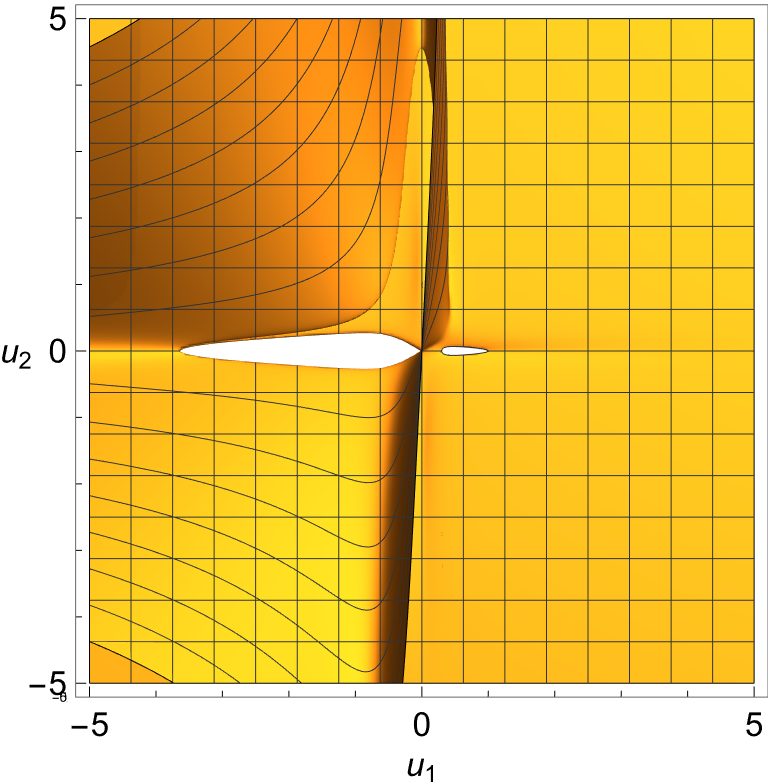}
\includegraphics[width=54mm]{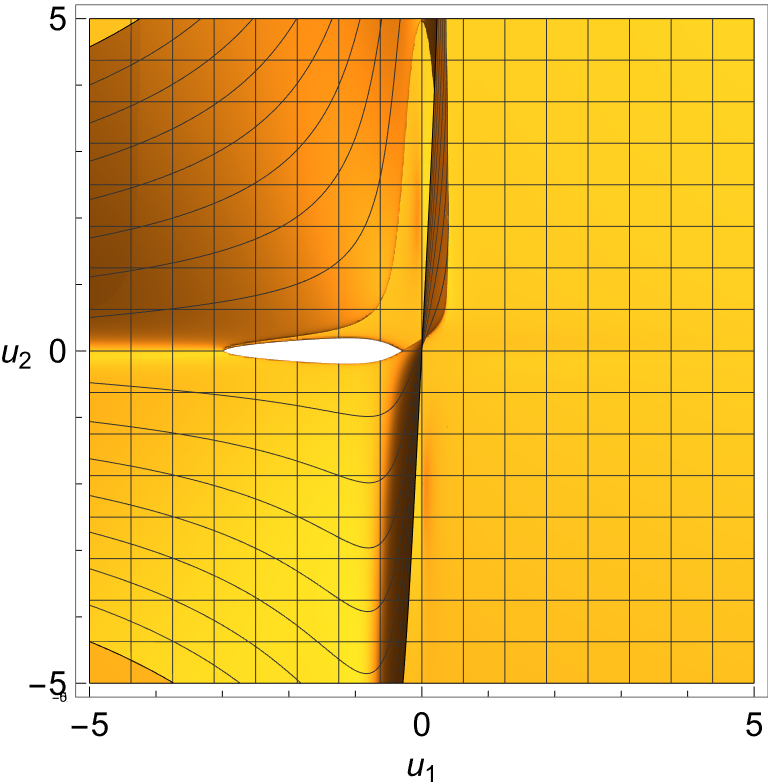}
\includegraphics[width=54mm]{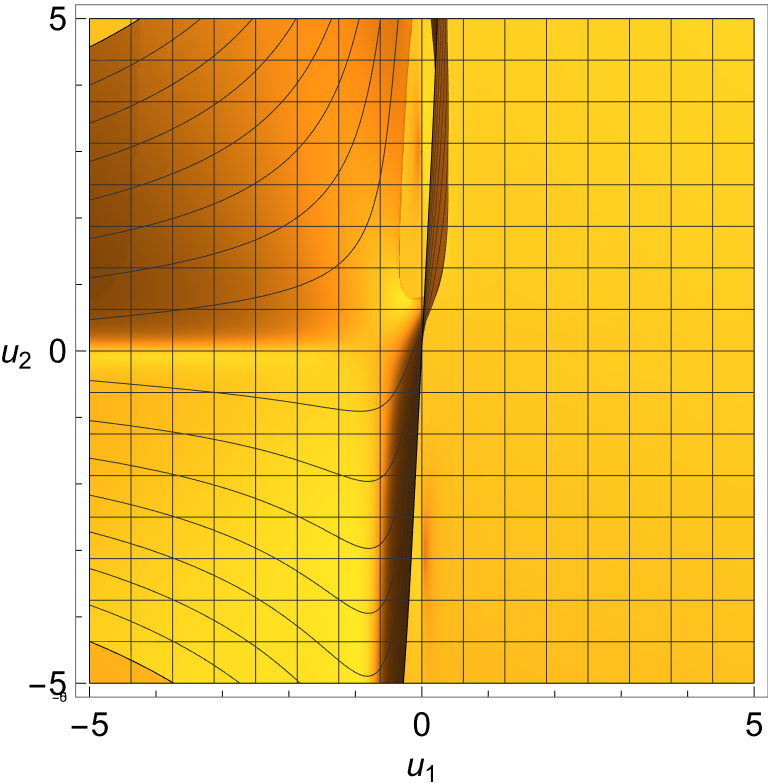}
\caption{\emph{The first row on the right shows a doubly connected topological space in the $(in)$ asymptotic state, which  in the process of evolution to the $(out)$ asymptotic state transits into two simply connected half-spaces with displaced holes. The second row shows three two-dimensional pictures-projections of the manifold $\Sigma^{(2)}_{\bf{u }}(t)$ onto the plane $\bigl(u_1,u_2\bigr)$, characterizing the evolution of the topological singularities of the manifold when transitioning between these asymptotic states defined by the data $(\Omega_0^- =\Omega_0(t)=1,\,\,\Omega_0(t)=2,\, \,\Omega_0 ^+=\Omega_0(t) =3 )$ and $ (\varepsilon^{(r)}=1,\,\,\varepsilon^{(i)}= 0.01)$ respectively.}
\label{overflow}}
\end{figure}

In the presence of strong elastic and weak inelastic processes in the environment, the calculations lead to formation of the following surfaces (see FIG 9).
An analysis of the pictures shows that the arising manifold is a two-connected topological space, which during evolution transit to the manifold consisting from
two sub-manifolds each of which is a simply connected topological space with spatially shifted typological singularities. Moreover, the shift between the topologies
of the upper and lower sub-manifolds is present already in the $(in)$ asymptotic state.  In the course of evolution, this shift only increases, while the gap
intersection decreases and already in the $(out)$ state becomes equal to zero. Moreover, the shift between the topologies of the upper
$\Sigma^{(2)+}_{\bf{u}}(t)$  and lower $\Sigma^{(2)-}_{\bf{u}}(t)$ sub-manifolds is present in the $(in)$ asymptotic state (this can be verified by analyzing three-dimensional graphs in FIG 9). In the course of evolution, this shift only increases, while the gap  intersection decreases and already in the $(out)$ state becomes equal to zero. In other words, the manifold $\Sigma^{(2)}_{\bf{u}}(t)$ can be represented as the union of two submanifolds:
\begin{equation}
\Sigma^{(2)}_{\bf{u}}(t)\cong  \Sigma^{(2)+}_{\bf{u}}(t) \sqcup \Sigma^{(2)-}_{\bf{u}}(t),
\label{9.n03}
\end{equation}
where $\Sigma^{(2)+}_{\bf{u}}(t) $ denotes a submanifold of the upper half-space, while $\Sigma^{(2)-}_{\bf{u}}(t)$ is a submanifold of the lower
half-space. As we will see below, this manifold  can evolve into a disjoint union of two submanifolds in the course of evolution.
 \begin{figure}[ht!]
\centering
\includegraphics[width=65mm]{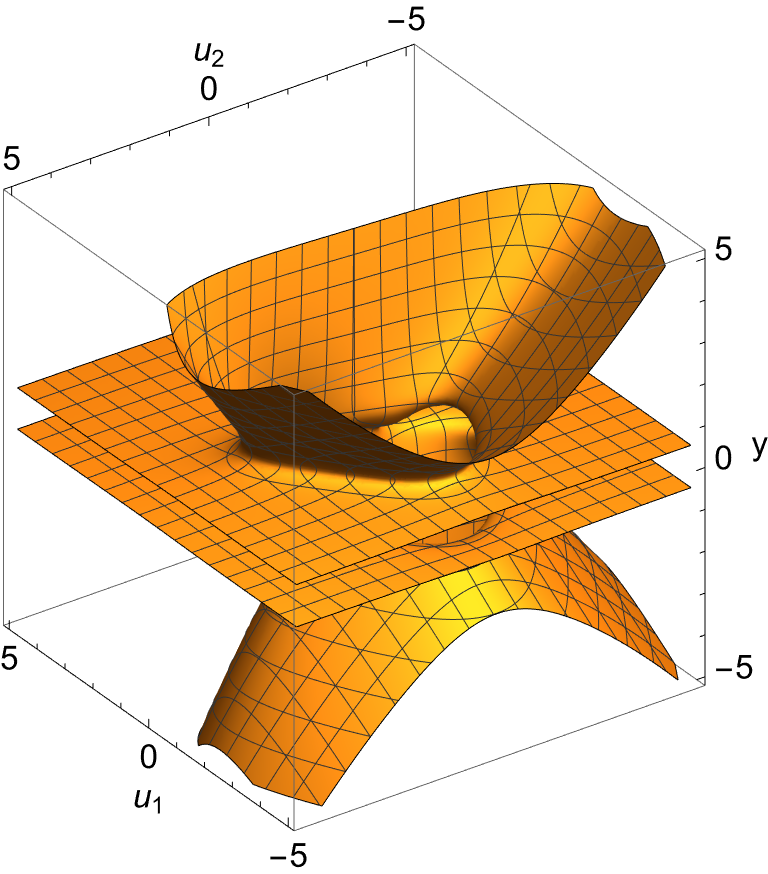}
\quad\,\,
\includegraphics[width=65mm]{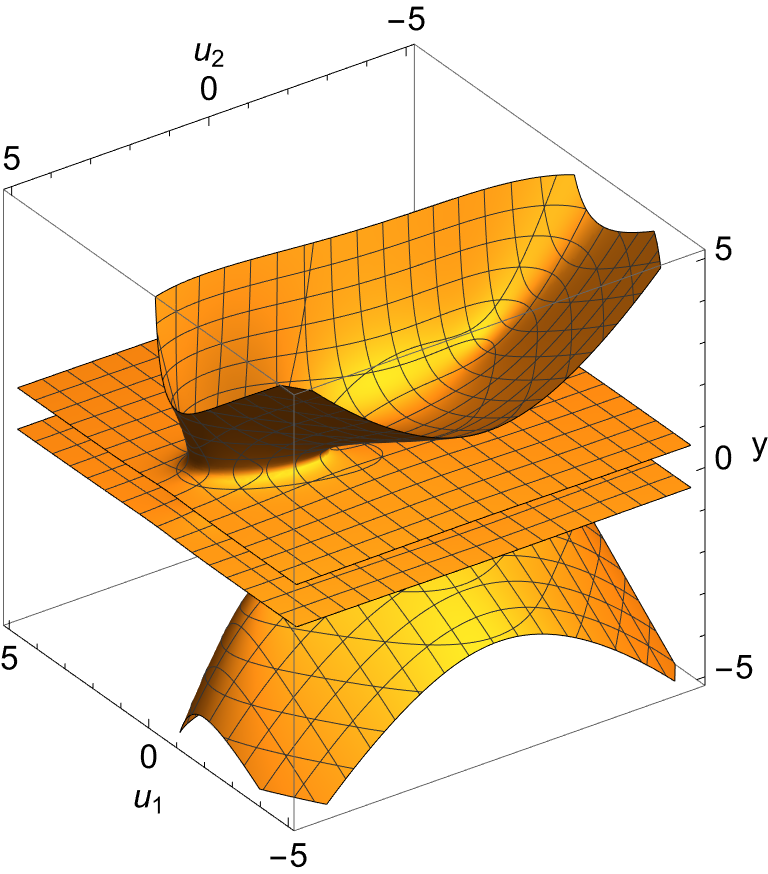}
\includegraphics[width=54mm]{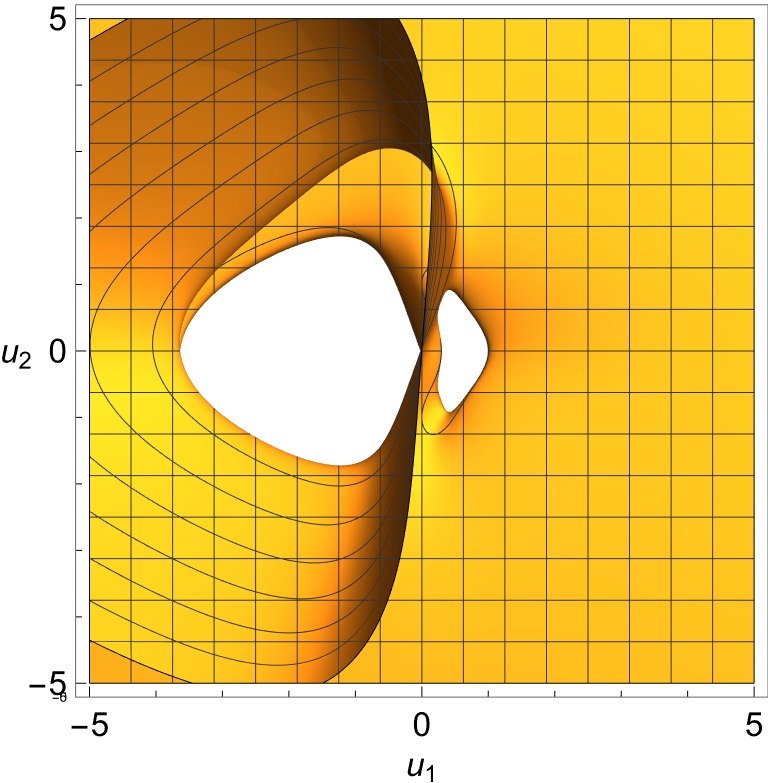}
\includegraphics[width=54mm]{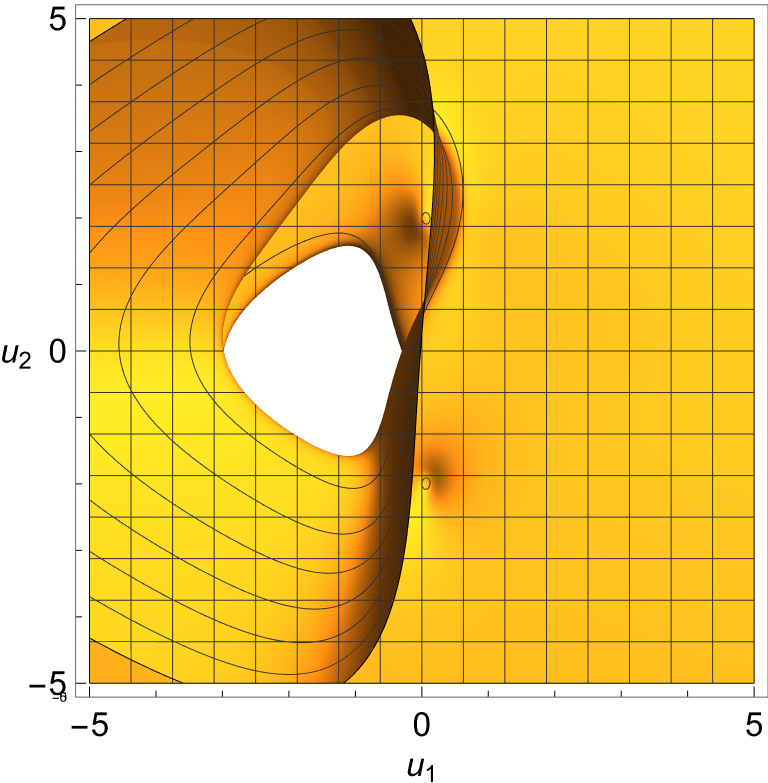}
\includegraphics[width=54mm]{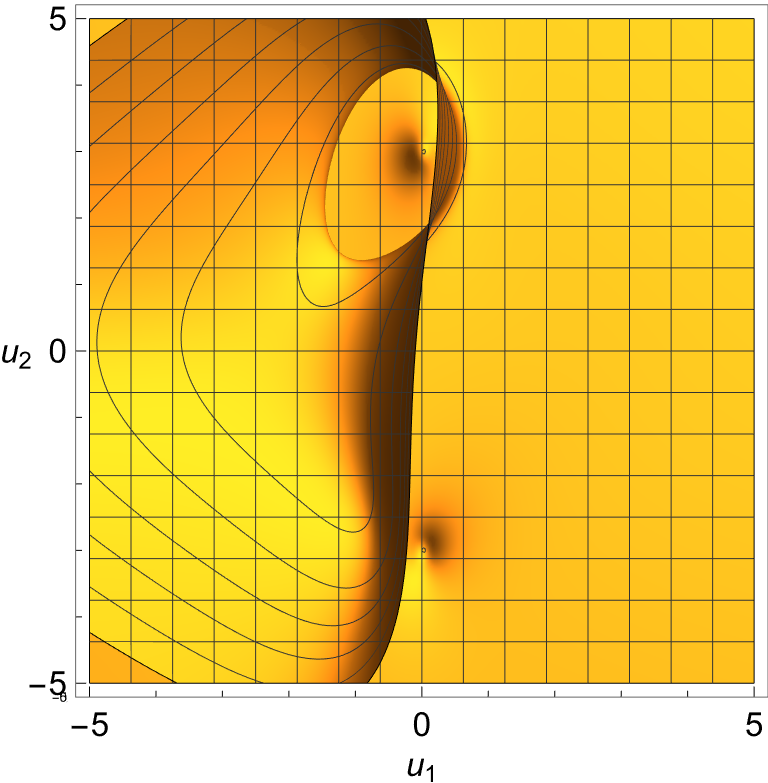}
\caption{\emph{The two figures in the first row (from left to right) show the evolution of the manifold  $\Sigma^{(2)}_{\bf{u}}(t)$ from the
asymptotic state $(in)$ with data $\bigl(\Omega_0 ^-=1,\,\,\varepsilon^{(r)}=1,\,\varepsilon^{(i)}=0.5\bigr)$ to the $(out )$ asymptotic state
with data $\bigl (\Omega_0^+=3,\,\,\varepsilon^{(r)}=1,\,\varepsilon^{(i)}=0.5\bigr)$. The second row represents two-dimensional projections
of this manifold onto the $(u_1,u_2)$ plane, showing the evolution of its topological features.}
\label{overflow}}
\end{figure}
\begin{figure}[ht!]
\centering
\includegraphics[width=54mm]{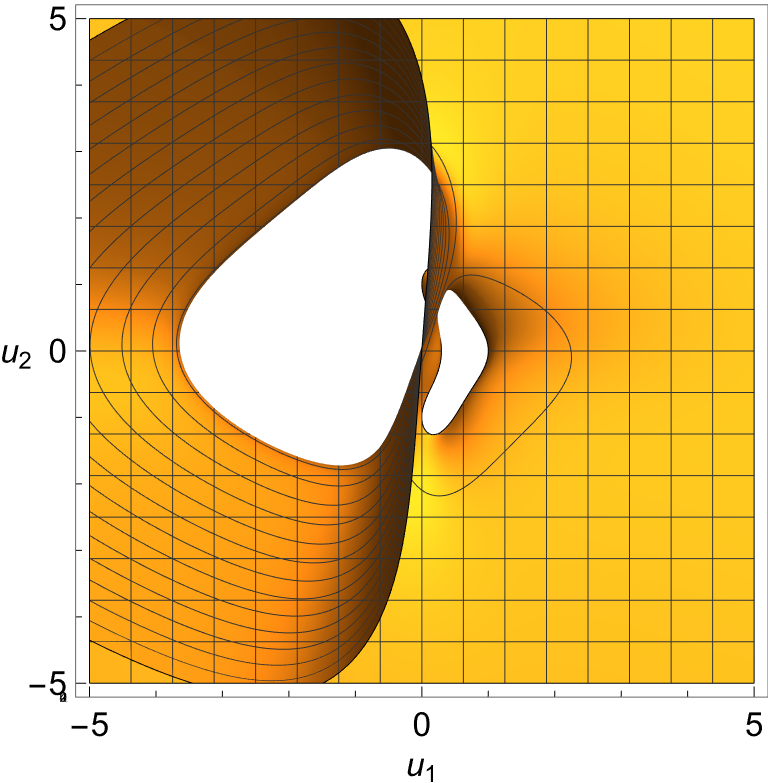}
\includegraphics[width=54mm]{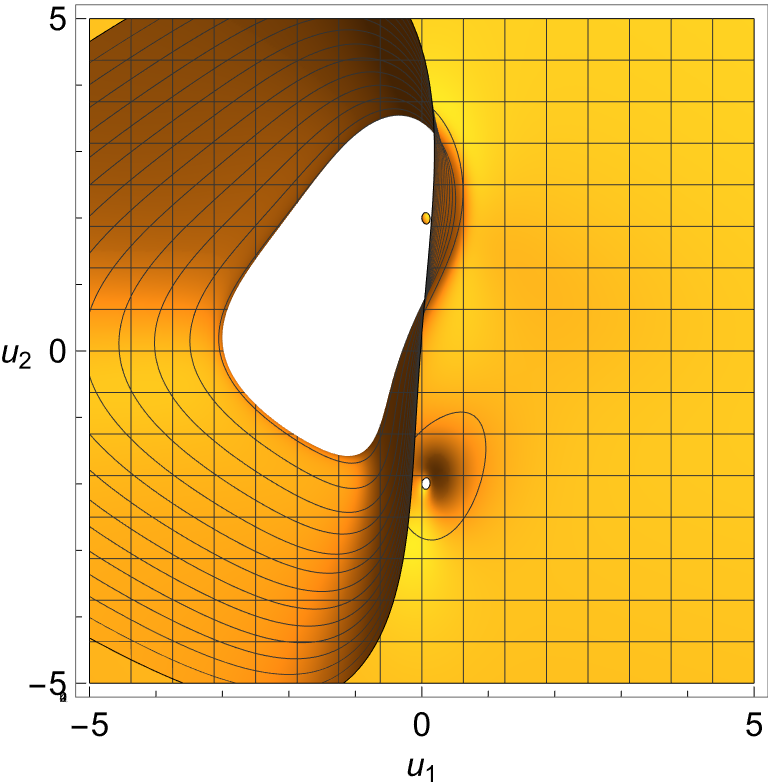}
\includegraphics[width=54mm]{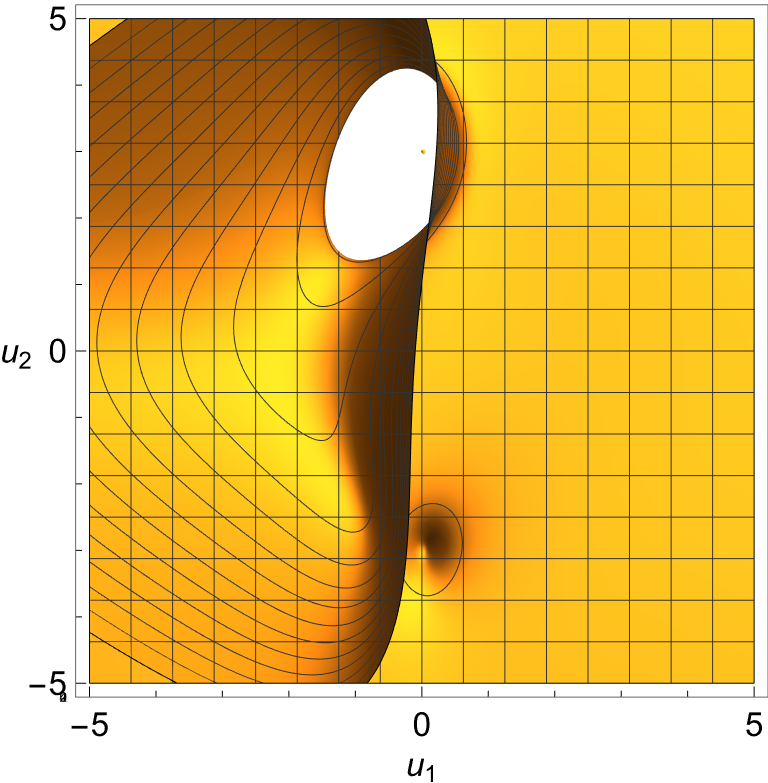}
\includegraphics[width=54mm]{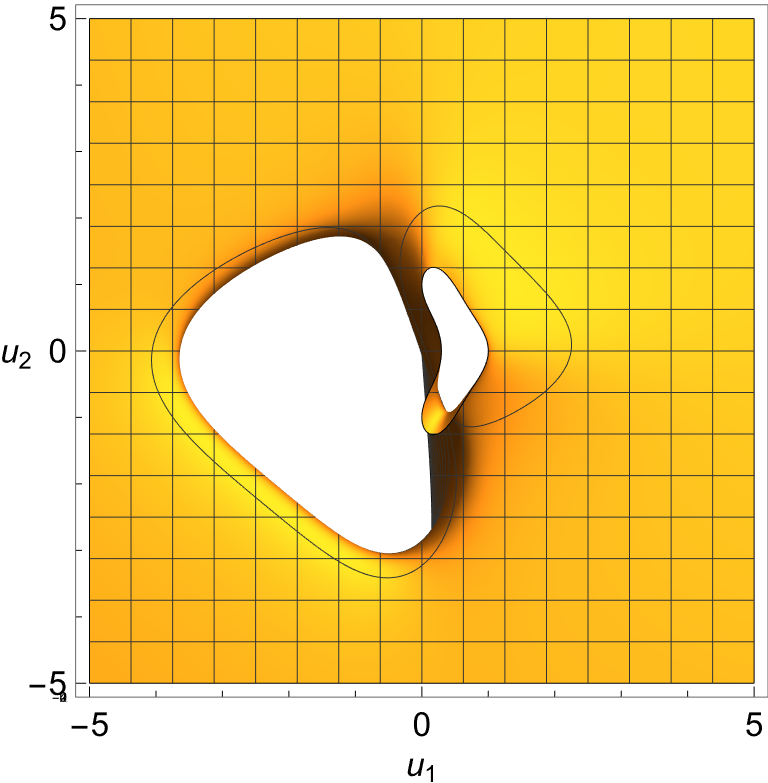}
\includegraphics[width=54mm]{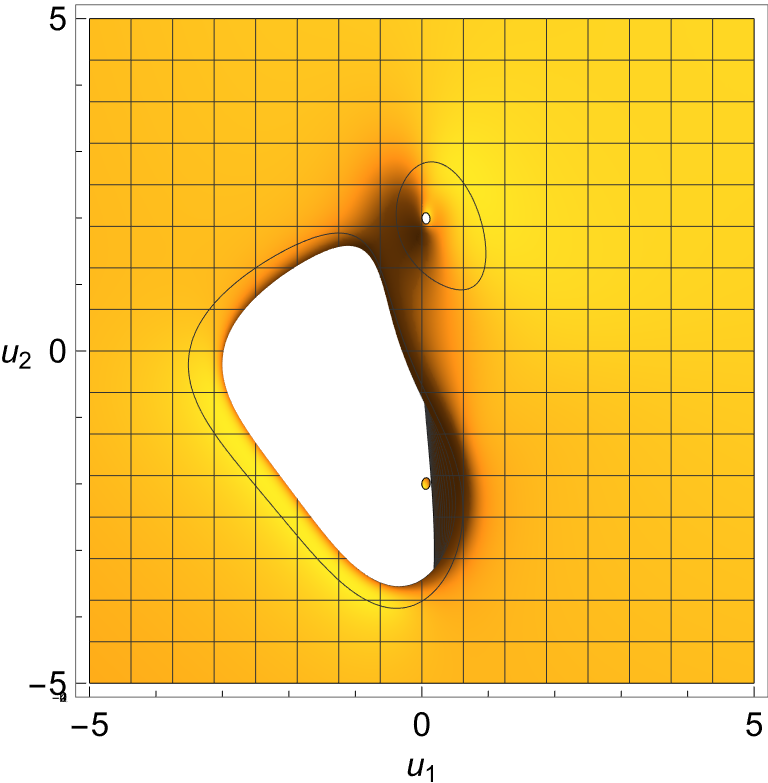}
\includegraphics[width=54mm]{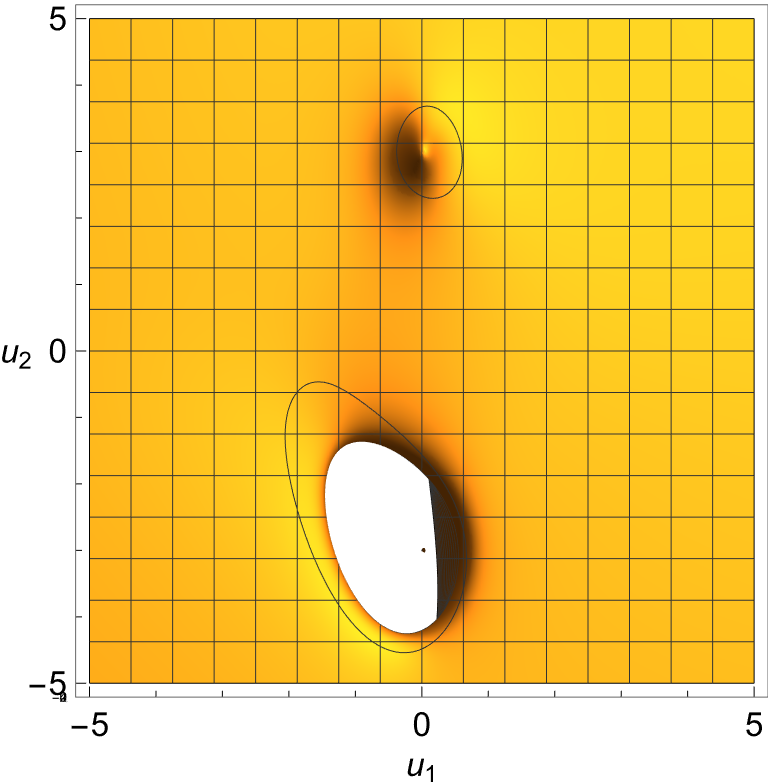}
\caption{\emph{The three figures in the first row (from left to right) show the evolution of the topological singularities of the submanifold $\Sigma^{(2)+}_{\bf{u}}(t)$ from the asymptotic state $(in)$ with data $\bigl(\Omega_0 ^-=1,\,\,\varepsilon^{(r)}=1,\,\varepsilon^{(i)}=0.5\bigr) $, to $(out )$ an
asymptotic state where the frequency is $\Omega_0^+=3$. In the second row (from left to right) there are figures showing a similar evolution for the submanifold $\Sigma^{(2)-}_{\bf{u}}(t)$.}
\label{overflow}}
\end{figure}

Finally, let us consider the case when strong elastic and nonweak inelastic processes occur in the environment. As calculations show (see FIG 10),
in this case  the manifold also has a doubly connected topology, but with different sizes and slot configurations.  In particular, in the first row in FIG
10 shows three-dimensional visualizations of the elements of the metric tensor $g^{12}(u_1,u_2,t)$ and $g^{21}(u_1,u_2,t )$ as functions of coordinates
$(u_1,u_2)$ in asymptotic states $ (in)$ and $(out)$, respectively. The second row (FIG 10) shows the graphs showing the evolution of the topological
features of the manifold $\Sigma^{(2)}_{\bf{u}}(t)$ in the transition from $(in)$ to the state $(out)$. As can be seen from the figures, the manifold
$\Sigma^{(2)}_{\bf{u}}(t)$ loses its topological features in the process of evolution. However, since we know that the manifold $\Sigma^{(2)}_{\bf{u}}(t)$
is the union of two submanifolds (see (\ref{9.n03})), a natural question arises: do these submanifolds retain any of their topological singularities during evolution? As the calculations and their two-dimensional visualizations (FIG 11) show, each of the submanifolds (see the first and second rows of the graphs) in the process of evolution passes from a doubly connected to a simply connected topology, however, these topologies are displaced and do not have a common hole (see graphs in the second row of FIG 10 obtained by combining the graphs of the first two rows of FIG 11).

It is important to note that,  depending on the parameters, a three-connected topological manifold also arises, which is a  union of two oriented three-connected topological submanifolds $\Sigma^{(2)+}_{\bf {u}}(t)$ and $\Sigma^{(2)-}_{\bf{u}}(t)$. In the end, we note that the manifold $\Sigma^{(2)}_{\bf {u}}(t)$, regardless of the state of the environment, at the end of its evolution in the state $(out)$, is a disjoint union of two submanifolds.

  \subsection{Features of entropy calculation}
The integrals in the equations for the ordinary (\ref{6.n01}) and generalized (\ref{6.n03}) entropies were calculated using the trapezoidal method.
 \begin{figure}[ht!]
\centering
\includegraphics[width=55mm]{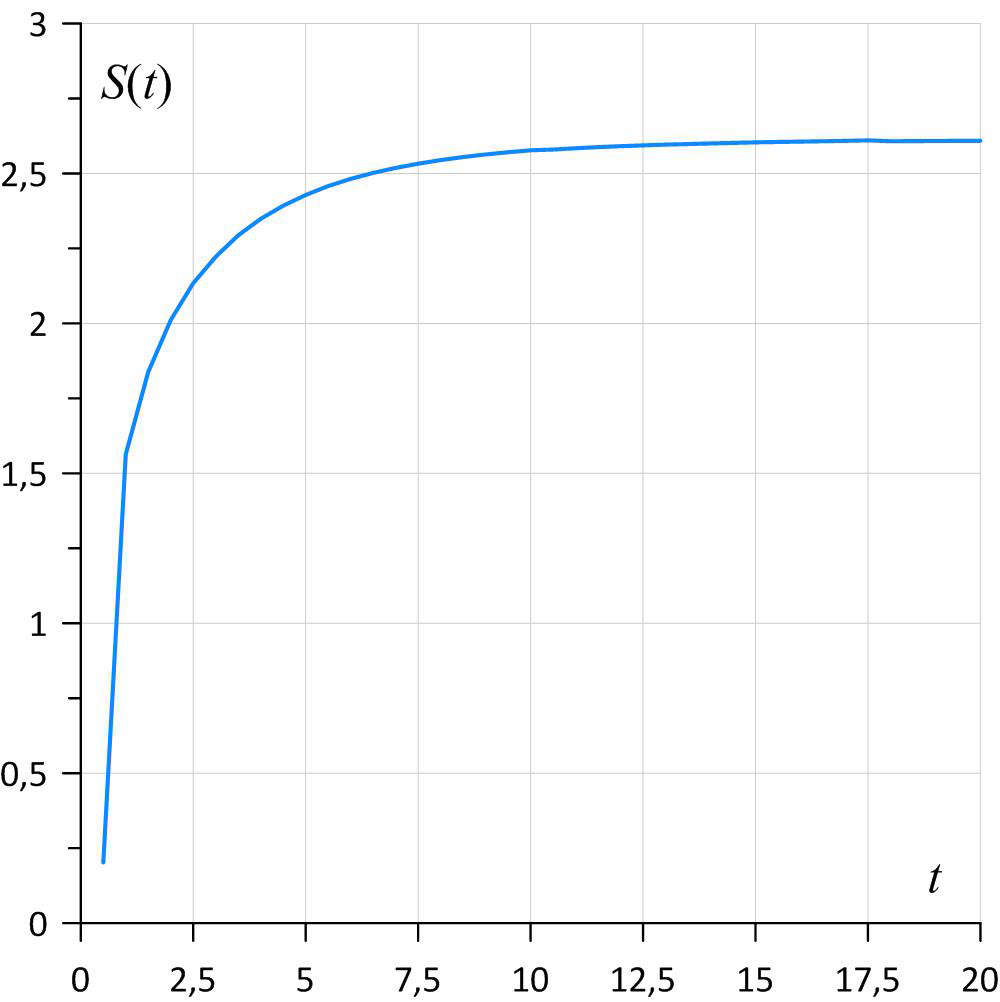}
\qquad
\includegraphics[width=55mm]{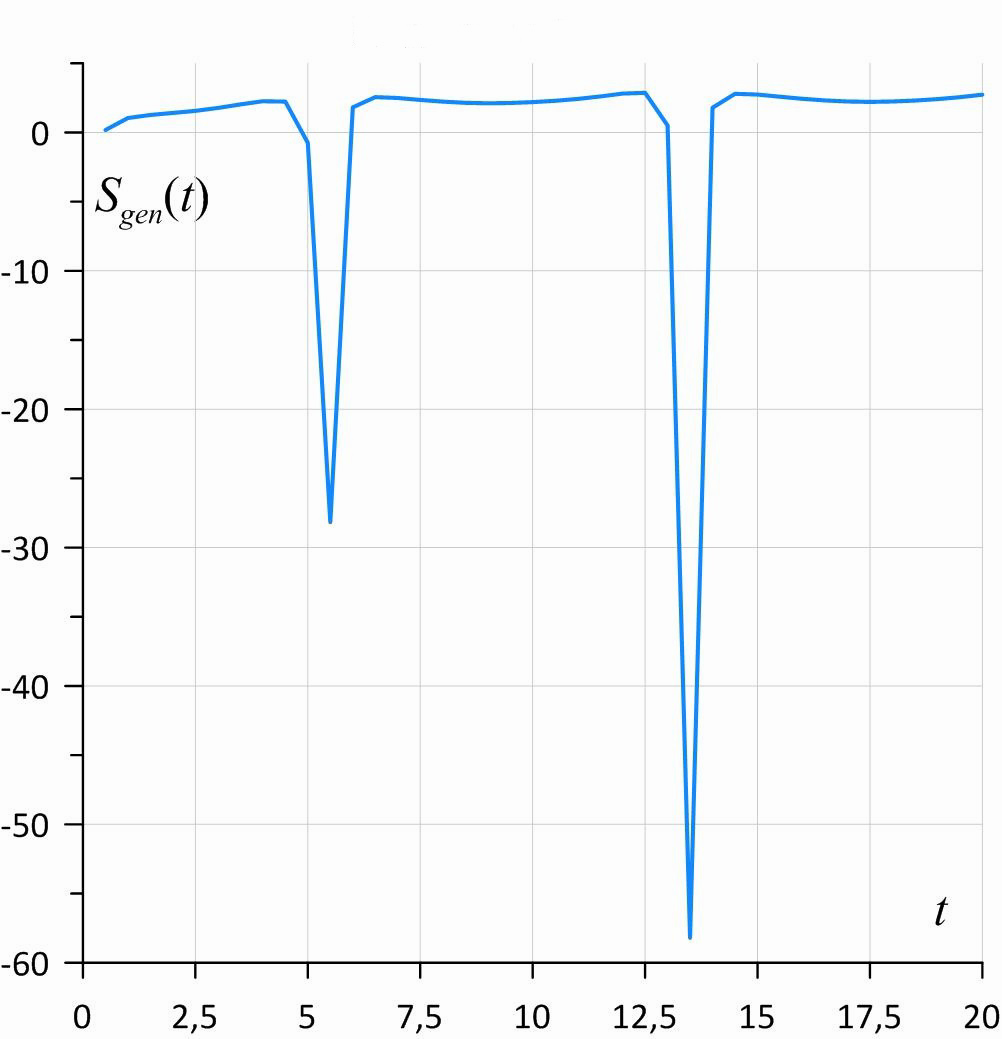}
\qquad
\includegraphics[width=55mm]{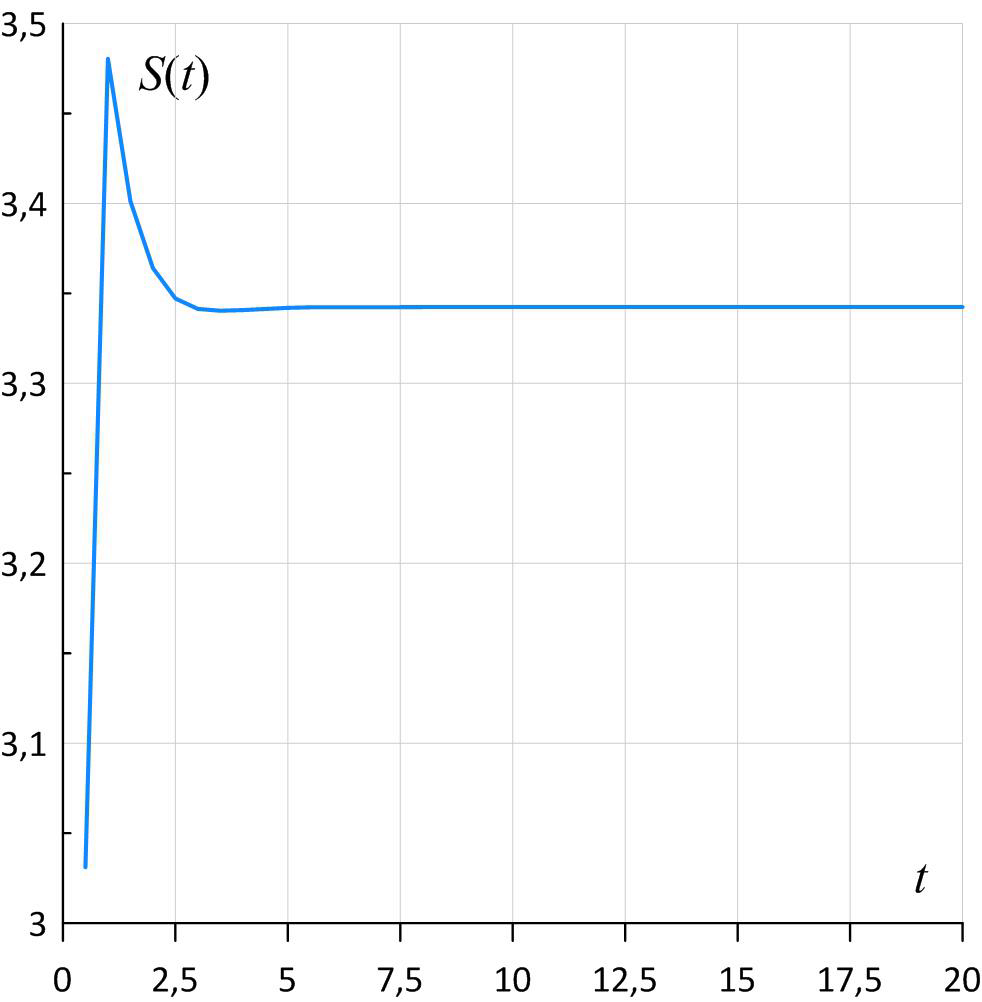}
\qquad
\includegraphics[width=55mm]{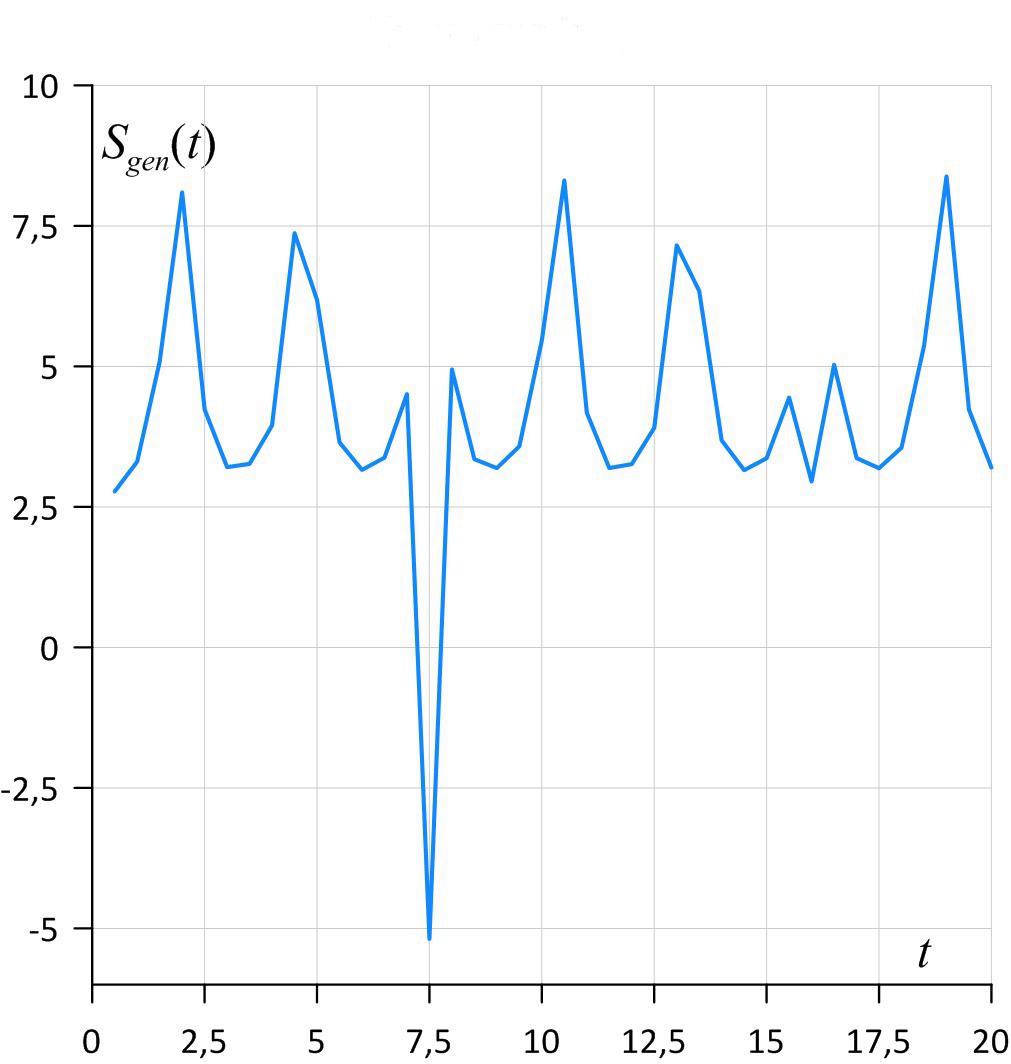}
\includegraphics[width=55mm]{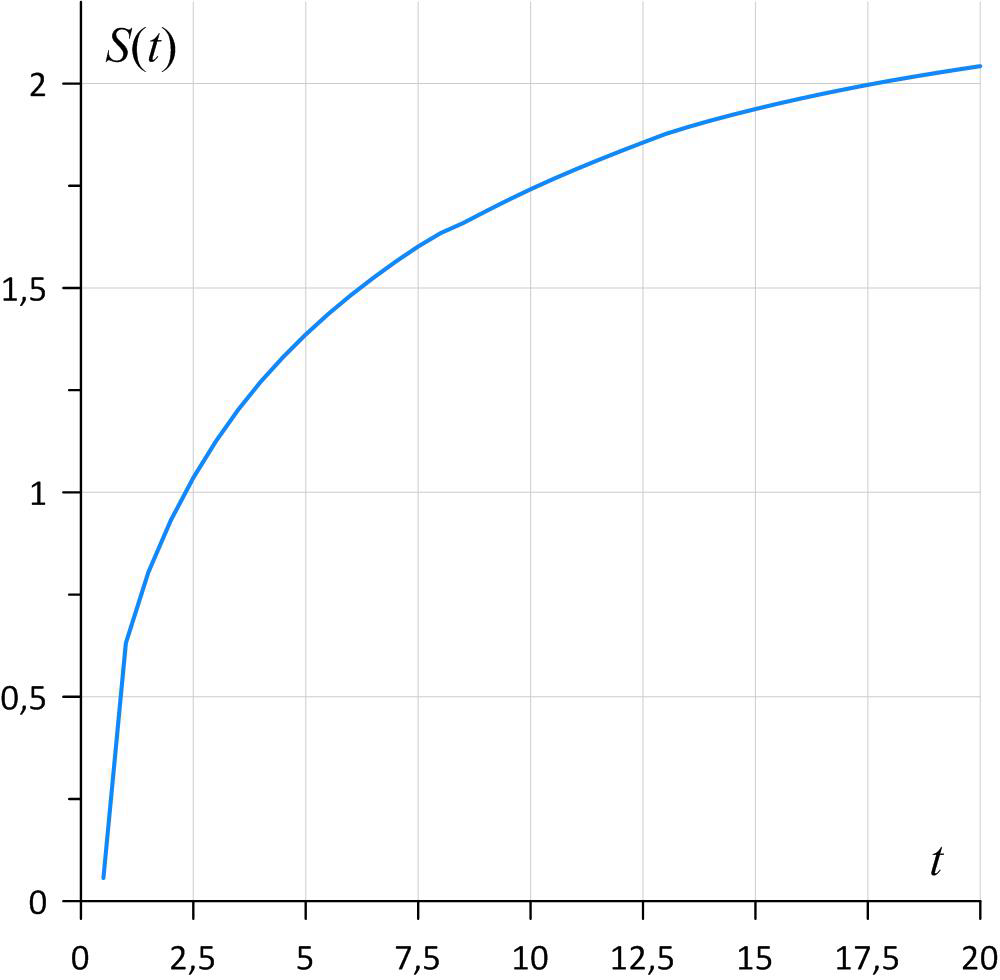}
\qquad
\includegraphics[width=55mm]{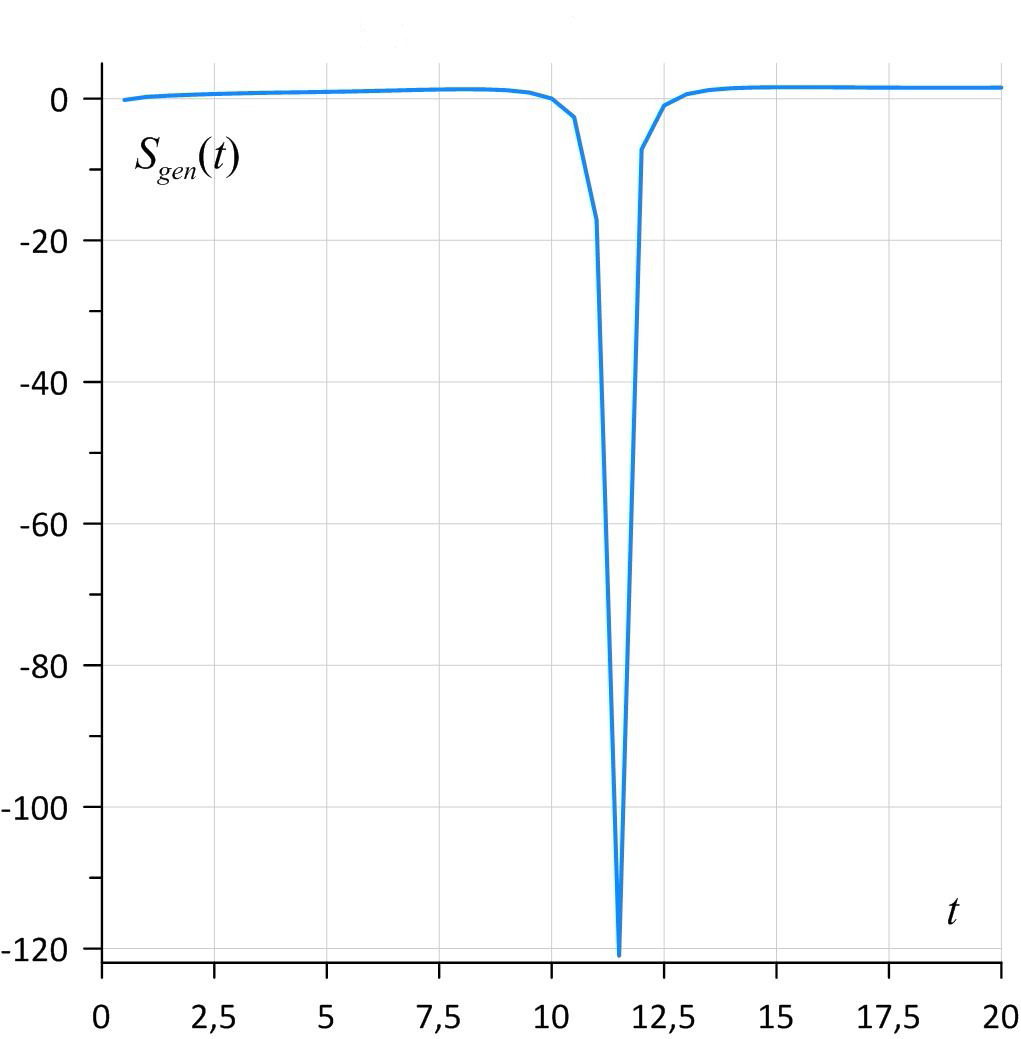}
\caption{\emph{In the figure, the left column of three graphs describes the Shannon entropy for three different environmental states while the
right column describes the generalized entropy describing the  joint system for the same environmental states.}
\label{overflow}}
\end{figure}
However, due to the fact that in the expressions for the entropy there is a term of type $A\ln(A)$, which has a singularity at $A=0$, the condition
$A\ln(A)=0$ is introduced, which makes it possible to eliminate this singularity. We performed entropy calculations using the formulas (\ref{6.n01}) and
 (\ref{6.n03}) for three different  states of the environment and visualized them (see FIG  12) taking into account the data of the \textbf{Table}.
As follows from these graphs, the usual entropy $\mathcal{S}(t)$ (left column) in the first two cases continuously increases with time and reaches a constant value in the $(out)$ state. In the third case, when inelastic precessions in the environment are strong enough, as we observe, from some moment the increase
in entropy passes into the stage of its decrease, and already at the long times, when the system goes into $(out)$ state, it takes constant value.
The right column contains graphs of the generalized entropy $\mathcal{S}_{gen}(t)$, which, in addition to being nonmonotonic, also contain areas
with negative values. This behavior of entropy is quite explainable for a closed self-organizing system, which at some point can generate negentropy to
stabilize the state of the joint system. In particular, the negative value of entropy can also be explained by the short-term capture or "swallowing" of a
small or oscillator subsystem by a large subsystem or thermostat.

As can be seen from the graphs of the right column, in the first two cases, the generalized entropy $\mathcal{S}_{gen}(t)$ relatively fast tends
a constant value, while  in the third case the process of stabilization of the combined system proceeds non-monotonically and takes a long time.


\section{C\lowercase{onclusion}}
The main achievement of this work is the development of a mathematically rigorous representation that
allows one to study the statistical properties of a classical oscillator with its random environment as a problem
of self-organization of a closed self-consistent system. Note that such a statement in the philosophical
sense corresponds to the consideration of the problem within the framework of Plato's concept, which excludes
the loss of information regarding the joint CORE system. The mathematical implementation of this idea is carried
out within the framework of a complex probabilistic process that satisfies a SDE of the Langevin-type.
The paper considers three typical scenarios of a random environment or thermostat, for which, in the limit
of statistical equilibrium, the kinetic equations for the distribution of the fields of the environment are
derived (see equations (\ref{3.05})-(\ref {3.n05}) and (\ref{3v.n05})-(\ref{3vt.n05})). With the help of
these equations, the measures of functional spaces are determined and the mathematical expectations of the
corresponding physical parameters are constructed. We use the generalized Feynman-Kac theorem
\cite{Agev,GevA} to compactify the infinite-dimensional functional integral describing the expectation
of the oscillator trajectory and reduce it to the two-dimensional integral representation, where the
integrand is the solution of the complex second-order PDE  given on a two-dimensional manifold
$\Sigma^{(2)}_{\bf{u}}(t)$.

The second most important result of the paper is the proof that the subspace $\Sigma^{(2)}_{\bf{u}}(t)$
is generally described by a noncommutative geometry, which also has topological singularities. In particular,
 in the case of nonintensive random processes in the environment, i.e. when
 $(\varepsilon^{(r)},\varepsilon^ {(i) })<<1$ the manifold $\Sigma^{(2)}_{\bf{u}}(t)\cong
\mathbb{E}^{(2)+}_{\bf{u}}(t) \sqcup\mathbb{E}^{(2)-}_{\bf{u}}(t)$, where $\mathbb{E}^{(2)+}_{\bf{u}}(t) $
 and $\mathbb{E}^{(2)-}_{\bf{u}}(t)$ are Euclidean subspaces with one singular boundary (see FIG 8).
As the power of random processes in the environment $(\varepsilon^{(r)},\varepsilon^ {(i) })\sim1$ increases,
both the geometric properties of the manifold $\Sigma^{(2)}_{\bf{u}}(t)\cong \Sigma^{(2)+}_{\bf{u}}(t)
\sqcup \Sigma^{(2)-}_{\bf{u}}(t)$ and their topological features change strongly. In particular, in this case,
as shown in FIG 10, the submanifolds $\Sigma^{(2)+}_{\bf{u}}(t)$ and $\Sigma^{(2)-}_{\bf{u}}(t)$ have a
non-Euclidean curvilinear geometry and doubly connected topologies, which in the $(out)$ state go over to a
simply connected topology. In other words, conformational transformations of an additional subspace of a
self-organizing classical system $\Sigma^{(2)}_{\bf{u}}(t)$, taking into account its geometric and typological
features, lead to radical differences in the description of the dynamics of a classical system without a medium
and when it is immersed in an environment.

In this work, an efficient mathematical algorithm for sequential and parallel calculation of various characteristics of
the problem is developed, taking into account that $\Sigma^{(2)}_{\bf{u}}(t)$ is Euclidean space. However, it is
obvious that it is more correct to perform calculations on the manifold $\Sigma^{(2)}_{\bf{u}}(t)$, which is the
union of two topological submanifolds $\Sigma^{( 2)+}_{\bf {u}}(t)$ and $\Sigma^{(2)-}_{\bf{u}}(t)$, respectively.
In the near future, it is planned to generalize the computational algorithm for performing calculations on just such a
manifold. Recall that in this case the distribution of the environmental fields will be described by the tensor equation
(\ref{nw3.01})-(\ref{nw3.02}) where the off-diagonal element $g^{12}(u_1,u_2,t)=-g^{21}(u_1,u_2,t)$ will be
determined by the algebraic equation of the 4\emph{th} degree (\ref{nw3z.0zf3}). The latter will allow numerical
methods to study important features of the dynamics of a classical system within the framework of an ideologically
more consistent and accurate Platonic concept.

In conclusion, it should be noted that the study of quantum analogues of the considered classical models, taking into
account the noncommutativity of the emerging geometries, will be extremely interesting and rich in new and unexpected results.

\section{A\lowercase{cknowledgments}}
Gevorkyan A.S. is grateful to grant N 21T-1B059 of the Science Committee of Armenia, which partially
funded this work. The research was carried out with partial financial support from the Ministry of Science
 and Higher Education of the Russian Federation within the framework of the program "World-Class Research
Center: Advanced Digital Technologies" (contract No. 075-15-2020-903 dated November 16, 2020).

\end{document}